\PassOptionsToPackage{hyphens}{url}
\PassOptionsToPackage{dvipsnames,svgnames,x11names}{xcolor}
\documentclass[12pt]{article}

\usepackage{amsmath,amssymb}
\usepackage{mathtools,amsthm}
\usepackage{tikz}
\usepackage{tikz-qtree}
\usepackage{enumitem}
\usepackage{multirow}
\usepackage{array}
\usepackage[T1]{fontenc}
\usepackage[utf8]{inputenc}
\usepackage{textcomp}
\usepackage{iftex}
\usepackage[comma]{natbib}
\usepackage{lmodern}
\IfFileExists{upquote.sty}{\usepackage{upquote}}{}
\IfFileExists{microtype.sty}{\usepackage{microtype}\UseMicrotypeSet[protrusion]{basicmath}}{}
\IfFileExists{parskip.sty}{\usepackage{parskip}}{\setlength{\parindent}{0pt}\setlength{\parskip}{6pt plus 2pt minus 1pt}}
\usepackage{xcolor}
\makeatletter
\ifx\paragraph\undefined\else
  \let\oldparagraph\paragraph
  \renewcommand{\paragraph}{\@ifstar\xxxParagraphStar\xxxParagraphNoStar}
  \newcommand{\xxxParagraphStar}[1]{\oldparagraph*{#1}\mbox{}}
  \newcommand{\xxxParagraphNoStar}[1]{\oldparagraph{#1}\mbox{}}
\fi
\ifx\subparagraph\undefined\else
  \let\oldsubparagraph\subparagraph
  \renewcommand{\subparagraph}{\@ifstar\xxxSubParagraphStar\xxxSubParagraphNoStar}
  \newcommand{\xxxSubParagraphStar}[1]{\oldsubparagraph*{#1}\mbox{}}
  \newcommand{\xxxSubParagraphNoStar}[1]{\oldsubparagraph{#1}\mbox{}}
\fi
\makeatother

\usepackage{longtable,booktabs}
\usepackage{calc}
\usepackage{etoolbox}
\makeatletter
\patchcmd\longtable{\par}{\if@noskipsec\mbox{}\fi\par}{}{}
\makeatother
\IfFileExists{footnotehyper.sty}{\usepackage{footnotehyper}}{\usepackage{footnote}}
\makesavenoteenv{longtable}
\usepackage{graphicx}
\makeatletter
\def\maxwidth{\ifdim\Gin@nat@width>\linewidth\linewidth\else\Gin@nat@width\fi}
\def\maxheight{\ifdim\Gin@nat@height>\textheight\textheight\else\Gin@nat@height\fi}
\makeatother
\setkeys{Gin}{width=\maxwidth,height=\maxheight,keepaspectratio}
\makeatletter
\def\fps@figure{htbp}
\makeatother

\usepackage{caption}
\usepackage{subcaption}
\usepackage{float}
\floatstyle{ruled}
\makeatletter
\@ifundefined{c@chapter}{\newfloat{codelisting}{h}{lop}}{\newfloat{codelisting}{h}{lop}[chapter]}
\makeatother
\floatname{codelisting}{Listing}

\usepackage[ruled,noend]{algorithm2e}

\PassOptionsToPackage{unicode}{hyperref}
\usepackage{bookmark}
\IfFileExists{xurl.sty}{\usepackage{xurl}}{}
\hypersetup{
  pdftitle={Average Treatment Effect Estimation with Non-binary Instrumental Variables},
  pdfauthor={Mei Dong; Lin Liu; Dingke Tang; Geoffrey Liu; Wei Xu; Linbo Wang},
  pdfkeywords={Mendelian randomization; Semiparametric methods; Unmeasured confounding},
  colorlinks=true,
  linkcolor=blue,
  filecolor=Maroon,
  citecolor=Blue,
  urlcolor=Blue,
  pdfcreator={LaTeX}
}

\graphicspath{{./figure/}}
\usetikzlibrary{shapes,decorations,arrows,calc,arrows.meta,fit,positioning,shapes.geometric,backgrounds,shapes.swigs}
\tikzset{
  -Latex,
  auto,
  node distance=1cm and 1cm,
  semithick,
  state/.style={ellipse,draw,minimum width=0.7cm},
  point/.style={circle,draw,inner sep=0.04cm,fill,node contents={}},
  bidirected/.style={Latex-Latex,dashed},
  el/.style={inner sep=2pt,align=left,sloped}
}

\newtheorem{theorem}{Theorem}
\newtheorem{lemma}{Lemma}

\newtheorem{remark}{Remark}
\newtheoremstyle{nodot}{3pt}{3pt}{\itshape}{}{\bfseries}{}{0.5em}{\thmname{#1}\thmnumber{ #2}\thmnote{ (#3)}}
\theoremstyle{nodot}
\newtheorem{assumption}{Assumption}

\newcommand{\independent}{\perp\!\!\!\perp}
\newcommand{\notindependent}{\not\!\perp\!\!\!\perp}

\DeclareMathOperator*{\argmin}{arg\,min}

\newcounter{cond}
\renewcommand{\thecond}{C\arabic{cond}}
\newcounter{deltaoneassumption}

\newcounter{supassumption}
\renewcommand{\thesupassumption}{S\arabic{supassumption}}
\newenvironment{cond}{\refstepcounter{cond}\par\thecond\ignorespaces}{\par\ignorespacesafterend}

\def\OP{\mathrm{OP}}
\def\bbR{\mathbb{R}}
\def\calZ{\mathcal{Z}}
\def\calX{\mathcal{X}}

\def\diff{\mathrm{d}}

\def\eff{\mathrm{eff}}

\def\np{\mathrm{np}}
\def\sp{\mathrm{sp}}
\def\calE{\mathcal{E}}
\def\var{\mathrm{Var}}
\def\cov{\mathrm{Cov}}
\def\expit{\mathrm{expit}}
\def\calM{\mathcal{M}}

\def\E{\mathbb{E}}
\def\opt{\mathrm{opt}}

\def\muyp{\dot{\mu}^Y}
\def\mudp{\dot{\mu}^D}

\def\ch{\mathrm{ch}}

\def\bbO{\boldsymbol{O}}
\def\bbo{\boldsymbol{o}}
\def\bbX{\boldsymbol{X}}
\def\bbx{\boldsymbol{x}}

\DeclareRobustCommand{\primeassumptionnumber}[1]{\texorpdfstring{\ensuremath{5^{\ifcase#1\or\prime\or\prime\prime\or\prime\prime\prime\fi}}}{5\ifcase#1\or'\or''\or'''\fi}}
\providecommand{\theHassumption}{\theassumption}
\newcommand{\setprimeassumption}[2]{\renewcommand{\theassumption}{\primeassumptionnumber{#1}}\renewcommand{\theHassumption}{#2}}

\begin{document}
\def\spacingset#1{\renewcommand{\baselinestretch}{#1}\small\normalsize}
\spacingset{1}
\title{\bf Average Treatment Effect Estimation with Non-binary Instrumental Variables}
\author{
Mei Dong\textsuperscript{1},
Lin Liu\textsuperscript{2},
Dingke Tang\textsuperscript{3},
Geoffrey Liu\textsuperscript{4},
Wei Xu\textsuperscript{1},
Linbo Wang\textsuperscript{5}\thanks{Correspondence: linbo.wang@utoronto.ca and wei.xu@uhn.ca}
}
\date{\vspace{0.6em}
  \textsuperscript{1}\small Division of Biostatistics, University of Toronto, Toronto, Ontario M5T 3M7, Canada\\
  \textsuperscript{2}Institute of Natural Sciences, Shanghai Jiao Tong University, Shanghai 200240, China\\
  \textsuperscript{3}Department of Mathematics and Statistics, University of Ottawa, Ottawa, Ontario K1N 9B4, Canada\\
  \textsuperscript{4}Princess Margaret Cancer Centre, University Health Network, Toronto, Ontario M5G 2C4, Canada\\
  \textsuperscript{5}Department of Statistical Sciences, University of Toronto, Toronto, Ontario M5G 1X6, Canada
}
{\maketitle\par}

\bigskip
\begin{abstract}
Non-binary instrumental variables, especially continuous ones, are common in practice. A binary recoding induces a Wald ratio but may discard useful variation and reduce efficiency. Although fully nonparametric approaches can in principle use the entire instrument, they often require high-dimensional nuisance estimation which can be unstable with rich covariates. We address this problem by developing a generalized Wald estimand for binary treatments that uses the full variation in a non-binary instrument. Under standard instrumental-variable assumptions and a homogeneity condition, the estimand yields a common identification formula for categorical and continuous instruments. We further develop its semiparametric efficiency theory and construct a locally efficient debiased estimator using risk-minimization reparameterizations and double cross-fitting to accommodate flexible machine learning while improving numerical stability. The central technical challenge is that the many Wald ratios generated by a non-binary instrument must agree, thereby imposing overidentifying restrictions on the observed-data law. In this setting, characterizing the tangent space is nonstandard: it requires a second-order parametric submodel, a construction that, to our knowledge, has not been standard in semiparametric efficiency theory. Simulations show stable performance across sample sizes and greater efficiency than estimators based on dichotomized instruments. In an application to the Princess Margaret Cancer Centre lung cancer cohort, associational analyses link excess body weight to lower two-year mortality, a seemingly protective pattern often called the obesity paradox. The proposed instrumental-variable analysis instead suggests increased mortality, pointing to residual confounding behind this paradox.
\end{abstract}


\vfill
\newpage
\spacingset{1.8}
\allowdisplaybreaks

\section{Introduction}

Obesity is typically associated with poorer oncology outcomes. Paradoxically, however, many observational studies have reported that patients with non-small cell lung cancer and higher body mass index, abbreviated as BMI, experience lower mortality, a phenomenon often referred to as the ``obesity paradox'' \citep{zhang2017obesity}. Unmeasured confounding likely contributes to this pattern \citep{park2018plausibility}. Potential confounders such as socioeconomic status, physical activity, and diet are difficult to measure accurately and are often unaccounted for in observational studies. To better understand this apparent contradiction, we aim to estimate the average treatment effect of being overweight at diagnosis, defined as BMI $\geq$ 25, on two-year mortality among patients with non-small cell lung cancer using data from the institutional lung cancer cohort at Princess Margaret Cancer Centre in Toronto, a major contributor to the International Lung Cancer Consortium, \url{https://ilcco.iarc.fr}.

Instrumental variable methods provide a powerful framework for addressing unmeasured confounding, a central challenge in observational studies like ours. A valid instrument must be correlated with the treatment but affects the outcome only through the treatment. In genetic epidemiology, genetic variants associated with the treatment are commonly used as instruments \citep{sanderson2022mendelian}. A single variant, coded as 0, 1, or 2 by the number of risk alleles, often only constitutes a weak instrument. To strengthen the instrument, researchers often construct a polygenic risk score (PRS), which aggregates the effects of many genetic variants into a weighted sum and can provide a stronger, continuous instrument \citep{davies2015many}. This motivates the setting studied here, in which the treatment is binary but the instrument may be non-binary. Non-binary instruments also appear in other substantive applications; for example, researchers have used distance to the nearest cycle path as an instrument for cycle-commuting behavior \citep{berrie2024does}, and cigarette prices as an instrument for smoking \citep{leigh2004instrumental}.

There has been extensive research on the identification and estimation of treatment effects with binary instruments. The literature can be organized according to the identification assumptions employed; see \citet{swanson2018partial} and \citet{levis2024nonparametric} for detailed reviews. Under a so-called monotonicity assumption, the local average treatment effect, namely the treatment effect among compliers whose treatment status would change with the instrument, can be nonparametrically identified \citep{angrist1996identification}. Alternatively, under a homogeneity assumption, the average treatment effect can be identified nonparametrically. Specifically, this assumption rules out unmeasured confounders that simultaneously modify both the effect of the instrument on the treatment and the effect of the treatment on the outcome \citep{wang2018bounded,cui2021semiparametric, hartwig2023average}. More recently, \citet{liu2025multiplicative} and \citet{lee2025inference} proposed a multiplicative instrumental variable model, which assumes no multiplicative interaction between the instrument and the unmeasured confounder in the treatment model, and showed that the treatment effect on the treated can be nonparametrically identified.

Non-binary instruments add a central challenge beyond the binary case: they generate many possible Wald ratios, one for each comparison of instrument values. If the goal is to identify a single average treatment effect, these ratios should agree rather than represent unrelated targets; if they do agree, the remaining statistical problem is how to use the additional variation efficiently. Existing approaches typically avoid this question by specifying regression or structural models, changing the estimand to a local or treated-population effect, or retaining a fully nonparametric formulation that requires estimating high-dimensional regressions and other nuisance components. Classical structural equation models identify the average treatment effect through linear regression systems \citep{wooldridge2010econometric}, structural mean models impose parametric mean restrictions \citep{robins1994correcting}, and semiparametric partially linear instrumental variable models typically focus on homogeneous treatment effects or low-dimensional treatment-effect heterogeneity \citep{okui2012doubly, young2024rose, scheidegger2026inference}. Methods based on monotonicity identify local effects, such as the local instrumental variable curve \citep{kennedy2019robust, zeng2025nonparametric}. \citet{tchetgen2013alternative} instead identify the average treatment effect on the treated under a homogeneous treatment-selection-bias assumption. Fully nonparametric instrumental variable formulations can avoid strong functional-form restrictions, but require accurate estimation of high-dimensional nuisance functions when many covariates are included, which can be unstable in finite samples \citep{newey2003instrumental, darolles2011nonparametric, hartford2017deep}.
In subsequent independent work, \citet{chen2025identification} study a related nonparametric instrumental variable equation and establish identification under an additive-instrument condition.
This is a useful existence result for a fully nonparametric equation. Our approach instead starts from the many Wald ratios generated by a non-binary instrument and unifies them through a generalized Wald estimand; we then develop the corresponding semiparametric theory under the resulting overidentified observed-data model and use that theory to obtain an estimator that is stable with flexible covariate adjustment. This distinction is practically important; in the simulation designs considered in Section~\ref{sec: sim}, estimators based on a fully nonparametric approach proposed by \citet{chen2025identification} were substantially less stable than the estimator proposed here.

We formalize this idea under standard instrumental variable assumptions and a homogeneity condition. Every conditional Wald ratio formed from two distinct instrument values then identifies the same conditional causal effect. The same principle extends to conditional generalized Wald ratios based on weighted pairwise differences in conditional means for categorical instruments or weighted average derivatives of the conditional mean functions for continuous instruments. Using the conditional ``Riesz representer'' \citep{chernozhukov2024conditional}, we express these ratios in a single residual form and define an average generalized Wald estimand. This estimand unifies binary, categorical and continuous instruments: different residual forms weight the conditional distribution of the instrument in different ways, but all identify the average treatment effect.

This unification reveals the main technical structure of the paper. With a non-binary instrument, equality of the many conditional generalized Wald ratios means that the observed distribution is overidentified. This overidentification imposes algebraic constraints on the observed-data model, yielding a semiparametric model with a nontrivial tangent space \citep{chen2018overidentification}. This is not merely a routine semiparametric projection problem: the overidentifying constraints make the tangent space itself nonstandard, and characterizing it becomes part of the efficient influence function derivation. The key technical obstacle is that regular one-dimensional paths based on first-order perturbations generally do not remain inside this constrained model. At the same time, forcing the first-order score itself to satisfy the observed-data constraint would exclude admissible score directions. We therefore characterize the tangent space by constructing higher-order paths that remain inside the constrained model, which gives the efficient influence function of the average generalized Wald estimand. To the best of our knowledge, this use of higher-order paths to characterize a first-order tangent space has not been standard in the semiparametric statistical literature, where higher-order paths have typically been used to extend the first-order differentiability of statistical functionals to higher orders \citep{pfanzagl1983asymptotic, van2014higher}. See the proof of Theorem~\ref{th:nonparametric} and Remark~\ref{S-rem:non-convex} in Supplementary Material~\ref{S-app:th:nonparametric} for more details.

The overidentified structure also shapes a practical estimator. Although the full efficient influence function provides the efficiency benchmark, direct implementation requires several complex nuisance functions. We therefore work with a locally efficient influence function obtained from a submodel in which the efficient influence function simplifies; at distributions outside this submodel, the resulting influence function remains valid for the full semiparametric model. The proposed debiased estimator permits flexible nuisance estimation and is consistent and asymptotically normal under regularity conditions. A central implementation challenge is that the influence function contains reciprocal and quotient nuisance terms, for which direct substitution can amplify estimation error. We instead target these nonlinear quantities directly through risk-minimization reparameterizations, improving finite-sample stability. Because the corresponding empirical risks use preliminary nuisance estimates, double cross-fitting separates preliminary nuisance estimation, risk minimization, and target-parameter estimation, avoiding Donsker-type empirical-process conditions on the nuisance estimators. Empirical results show that the proposed estimator performs reliably across a range of sample sizes and avoids the precision loss incurred by dichotomizing a continuous instrument. The real-data application using the proposed estimator suggests that being overweight at diagnosis increases two-year lung cancer mortality, offering a possible resolution to the obesity paradox.

\section{Preliminaries}\label{sec2}

\subsection{Framework and Notation}\label{sec: framework}
Suppose we observe $n$ independent and identically distributed observations $(\bbO_1, \ldots, \bbO_n)$, where $\bbO \coloneqq (Y, D, Z, \bbX)$ follows a common distribution $P$ with density or mass function $p(\bbo)$. Here, $\bbX \in \mathcal{X}$ is a vector of baseline covariates, $Z \in \mathcal{Z}$ is a categorical or continuous instrument, $D \in \{0, 1\}$ is a binary exposure or treatment, and $Y \in \bbR$ is the outcome of interest. The support of $Z$ is $\calZ = \{0,1,\ldots,J-1\}$ when categorical with $J$ levels, and $\calZ \subseteq \bbR$ when continuous. Let $U$ denote an unmeasured, possibly vector-valued, confounder.

A valid instrument satisfies the following assumptions~\citep{didelez2007mendelian}:
\begingroup
\renewcommand{\theassumption}{1}
\begin{assumption}[Instrument relevance]
\label{as:relevance}
     $Z \notindependent D \mid \bbX$ almost surely.
\end{assumption}
\endgroup
\begin{assumption}[Independence]
\label{as:ind}
     $Z \independent U \mid \bbX$.
\end{assumption}
\begin{assumption}[Exclusion restriction]
\label{as:ER}
     $Z \independent Y \mid (D, \bbX, U)$.
\end{assumption}

Figure~\ref{fig:fig1} presents graphical representations of the conditional instrumental variable model. As shown in Figure~\ref{DAG}, conditional on $\bbX$, $Z$ is associated with $Y$ only when $D$ has a causal effect on $Y$. The bi-directed arrow between $Z$ and $D$ indicates that $Z$ need not causally affect $D$.

\begin{figure}[!htbp]
    \centering
    \begin{subfigure}[b]{0.45\textwidth}
\begin{tikzpicture}[->, >=stealth', shorten >=1pt, auto, node distance=2.4cm, thick]
\tikzset{line width=1pt, outer sep=0pt, ell/.style={draw,fill=white, inner sep=2pt, line width=1pt}, swig vsplit={gap=3pt, line color right=red}}
\node[name = Z, ell, shape = ellipse]{$Z$};
\node[name = D, ell, right of = Z, shape = ellipse]{$D$};
\node[name = X, ell, above of = D, shape = ellipse, node distance = 2cm]{$\bbX$};
\node[name = Y, ell, right of = D, shape = ellipse]{$Y$};
\node[name = U, ell, fill = gray, below right of = D, shape = ellipse, node distance = 1.8cm] {$U$};
\draw[->,line width=1pt,>=stealth]
(X) edge[bend right] (Z)
(X) edge (D)
(X) edge[bend left] (Y)
(Z) edge (D)
(D) edge[bend left, <->] (Z)
(D) edge [red] (Y)
(U) edge (Y)
(U) edge (D);
\end{tikzpicture}
\caption{A causal graph with a bi-directed arrow.}
\label{DAG}
\end{subfigure}
\begin{subfigure}[b]{0.45\textwidth}
    \centering
\begin{tikzpicture}[->, >=stealth', shorten >=1pt, auto, node distance=3cm, thick]
\tikzset{line width=1pt, outer sep=0pt, ell/.style={draw,fill=white, inner sep=2pt, line width=1pt}, swig vsplit={gap=3pt, line color right=red}}
\node[name=Z,shape=swig vsplit]{
\nodepart{left}{$Z$}
\nodepart{right}{$z$} };
\node[name=D, right of = Z, shape=swig vsplit]{
\nodepart{left}{$D (z)$}
\nodepart{right}{$d$} };
\node[name = X, ell, above of = D, shape = ellipse, node distance = 1.9cm]{$\bbX$};
\node[name = Y, ell, right of = D, shape = ellipse]{$Y (d)$};
\node[name = U, ell, fill = gray, below right of = D, shape = ellipse, node distance = 2cm] {$U$};
\draw[->,line width=1pt,>=stealth]
(X) edge[bend right] (Z.120)
(X) edge (D.80)
(X) edge[bend left] (Y)
(Z) edge (D)
(D) edge[bend left, <->] (Z.230)
(D) edge [red] (Y)
(U) edge (Y)
(U) edge (D.250);
\end{tikzpicture}
    \caption{A SWIG with a bi-directed arrow.}
    \label{SWIG}
\end{subfigure}
\caption{Causal graphs for an instrumental variable model defined by Assumptions~\ref{as:relevance}--\ref{as:dsep}. The bi-directed arrow between $Z$ and $D$ represents potential unmeasured common causes; $\bbX$, $Z$, $D$, and $Y$ are observed, and $U$ is unobserved. The left panel shows a causal graph~\citep{pearl2009causality} with a bi-directed arrow; the right panel shows a single-world intervention graph, or SWIG~\citep{richardson2013single}.
}
\label{fig:fig1}
\end{figure}

Under the potential outcomes framework, let $D(z)$ denote the treatment that would be received if the instrument took value $z$, and let $Y(d)$ denote the potential outcome under treatment level $d$. Under the consistency assumption, we observe $D = D(Z)$ and $Y = Y(D) = DY(1) + (1 - D)Y(0)$. We are interested in estimating the population average treatment effect $\E\left[Y(1)-Y(0)\right]$. Assumptions~\ref{as:ind} and~\ref{as:ER} can be read off from the SWIG in Figure~\ref{SWIG} via $d$-separation~\citep{richardson2013single}. The following assumption is also implied by the SWIG:
\begin{assumption}[$d$-separation]
\label{as:dsep}
$Y(d) \independent (Z, D) \mid (\bbX, U)$.
\end{assumption}

Throughout, let $F$ denote cumulative distribution functions; for example, $F(z \mid \bbX) \coloneqq \Pr(Z \le z \mid \bbX)$. Define $\mu^Y(Z, \bbX) \coloneqq \E[Y \mid Z, \bbX]$ and $\mu^D(Z, \bbX) \coloneqq \E[D \mid Z, \bbX]$. We also define $\mu^Y(\bbX) \coloneqq \E[Y \mid \bbX]$, $\mu^D(\bbX) \coloneqq \E[D \mid \bbX]$, and $\mu^Z(\bbX) \coloneqq \E[Z \mid \bbX]$. Let
$\mathbb{P}_n f \coloneqq n^{-1}\sum_{i=1}^n f(\bbO_i)$ denote the empirical average of a measurable function $f$. For a random variable $W$, let $L^2(P_W)$ be the Hilbert space of square-integrable functions of $W$, equipped with inner product $\langle \ell, \rho \rangle \coloneqq \E[\ell(W)\rho(W)]$. We
assume $\mu^Y(Z,\bbX),\mu^D(Z,\bbX)\in L^2(P_{Z,\bbX})$. When weighted average derivatives of the conditional mean functions $\mu^D(Z, \bbX)$ and $\mu^Y(Z, \bbX)$ are considered for continuous $Z$, we write $\mudp(z,\bbx) \coloneqq \partial\mu^D(z,\bbx)/\partial z$ and $\muyp(z,\bbx) \coloneqq \partial\mu^Y(z,\bbx)/\partial z$, and assume these derivatives exist and belong to $L^2(P_{Z,\bbX})$.

\subsection{Identification with Binary Instrumental Variables}\label{sec: identification-binIV}

Under Assumptions~\ref{as:relevance}--\ref{as:dsep} alone, the average treatment effect is generally not point identifiable. With a binary instrument, point identification can be obtained by adding the following no unmeasured common effect modifier assumption~\citep{wang2018bounded,cui2021semiparametric, hartwig2023average}, which stipulates that no unmeasured confounder acts as a common effect modifier of both the additive effect of the instrument on the treatment and the additive treatment effect on the outcome:
\begingroup
\setprimeassumption{1}{assumption.5prime}
\begin{assumption}
\label{as:5c}
$\cov\{\E[D\mid Z=1, \bbX, U] - \E[D\mid Z=0, \bbX, U], \E[Y(1) - Y(0) \mid \bbX, U] \mid \bbX\}=0$ almost surely.
\end{assumption}
\endgroup

Assumption \ref{as:5c} allows either component to vary with $U$ separately, but rules out shared additive effect modification within levels of $\bbX$.

When $Z$ is binary, under Assumptions~\ref{as:relevance}--\ref{as:dsep} and~\ref{as:5c},
\[
    \E[Y(1)-Y(0)\mid\bbX]
    =
    \dfrac{\mu^Y(1,\bbX)-\mu^Y(0,\bbX)}
          {\mu^D(1,\bbX)-\mu^D(0,\bbX)}
\]
whenever the denominator is nonzero~\citep{wang2018bounded}. This ratio is formed from the only nonredundant pair of instrument values, $Z=1$ and $Z=0$, using the corresponding conditional mean differences for the outcome and treatment regressions.

\subsection{Weighted Linear Functionals and ``Riesz Representations''}\label{sec: Riesz Representer}
Moving beyond binary instruments, weights can combine information across multiple values of $Z$ to quantify IV--treatment and IV--outcome associations. For example, with categorical instruments,
the IV--treatment association can be summarized by a weighted average of pairwise conditional-mean differences, such as
\(
\E\left[ \sum_{0 \leq k < j \leq J-1} \omega_{jk}(\bbX)\{\mu^D(j,\bbX)-\mu^D(k,\bbX)\} \right],
\)
where $\omega_{jk}(\bbX)\in L^2(P_{\bbX})$ for $0 \leq k < j \leq J-1$. For continuous instruments, the analogous object is a weighted average derivative of the conditional mean functions,
$\E[\omega(Z,\bbX)\mudp(Z,\bbX)],$
where $\omega(Z,\bbX)\in L^2(P_{Z,\bbX})$. More generally, let $h \mapsto m_\omega(z,\bbx; h)$ denote a linear operator for each $(z, \bbx) \in \calZ \times \calX$ that is linear in the regression curve $z\mapsto h(z,\bbX)$. The corresponding IV--treatment association is the value of the linear functional $h\mapsto \E[m_\omega(Z, \bbX; h)]$ at $h=\mu^D$:
\[
    \E[m_\omega(Z, \bbX; \mu^D)],
\]
where $m_\omega(Z, \bbX; \mu^D)= \sum_{0 \leq k < j \leq J-1} \omega_{jk}(\bbX)\{\mu^D(j,\bbX)-\mu^D(k,\bbX)\}$ for categorical instruments and $m_\omega(Z, \bbX; \mu^D) = \omega(Z,\bbX)\mudp(Z,\bbX)$ for continuous instruments.

The Riesz type theorem~\citep[e.g.][Theorem~3.4]{conway2019course} allows bounded linear functionals of the regression curve to be written as $L^2(P_{Z,\bbX})$ inner products. Specifically, for any bounded linear functional $h\mapsto \E[m_\omega(Z, \bbX; h)]$, there exists a unique element $\gamma_\omega(Z, \bbX) \in L^2(P_{Z,\bbX})$, referred to as the ``Riesz representer'' in the recent statistics and econometrics literature, such that
\begin{equation*}
    \E[m_\omega(Z, \bbX; h)]
    =   \E[\gamma_\omega(Z,\bbX)h(Z,\bbX)]
\end{equation*}
for every $h \in L^2(P_{Z, \bbX})$. Similarly, a conditional ``Riesz representer'' of the linear operator $h\mapsto\E[m_\omega(Z, \bbX; h)\mid\bbX]$, coined in \citet{chernozhukov2024conditional}, is any square-integrable $\gamma_\omega(Z,\bbX)$ satisfying
\begin{equation}\label{eqn:conditional-rr}
    \E[m_\omega(Z, \bbX; h)\mid\bbX]   =\E[\gamma_\omega(Z,\bbX)h(Z,\bbX)\mid\bbX],
\end{equation}
for every admissible $h$. Explicit forms of $\gamma_{\omega}$, together with the corresponding positivity and smoothness conditions, are collected in Lemma~\ref{S-le:rr} in Supplementary Material~\ref{S-app:riesz-details}. Under those conditions, the same $\gamma_{\omega}$ also satisfies the conditional identity in \eqref{eqn:conditional-rr}.

For the weighted linear operators considered here, $m_\omega$ is unchanged if $h(z,\bbX)$ is shifted by a term depending only on $\bbX$. Thus the right-hand side of \eqref{eqn:conditional-rr} must also be invariant to such shifts, which forces $\E[\gamma_\omega(Z,\bbX)\mid\bbX]=0$. The converse also holds. For categorical instruments, any conditionally mean-zero function $\gamma(Z,\bbX)$ induces a weighted pairwise-difference representation, although the pairwise weights are not unique unless a difference basis is fixed. For continuous instruments, the analogous weighted derivative representation is unique: if $\gamma\in L^2(P_{Z,\bbX})$ satisfies $\E[\gamma(Z,\bbX)\mid\bbX]=0$, then a conditional version of~\citet[][Theorem 1]{hines2026parameterizing} gives the induced weight
\begin{equation}\label{omega}
   \omega(z,\bbx)
   =
   -\dfrac{F(z\mid\bbx)\E[\gamma(Z,\bbX)\mid Z\leq z,\bbX=\bbx]}
   {p(z\mid\bbx)},
\end{equation}
for which $\E[\gamma(Z,\bbX)h(Z,\bbX)\mid\bbX]
=\E[\omega(Z,\bbX)\dot h(Z,\bbX)\mid\bbX]$ for every admissible smooth $h$. These representation results generalize Stein's lemma; details are given in Supplementary Material~\ref{S-app:stein-connection}.

\section{Identification of the Average Treatment Effect}\label{sec: identification}
In this section, we identify the average treatment effect
$\E[Y(1)-Y(0)]$ with non-binary instruments. The binary identification result in Section~\ref{sec: identification-binIV} relies on a homogeneity condition that rules out unmeasured common effect modification. For non-binary instruments, the same idea should not be tied to a single pair of instrument values. We use the following version, which requires the additive instrument--treatment effect and the additive treatment--outcome effect do not share any unmeasured effect modifier, regardless of which instrument level is compared with the covariate-specific mean:
\begingroup
\renewcommand{\theassumption}{5}
\begin{assumption}[No unmeasured common effect modifier]
\phantomsection
\label{as:5d}
$\forall \, z  \in \calZ, \,\cov \{\E[D\mid Z=z, \bbX, U] - \E[D\mid  \bbX, U],  \E[Y(1) - Y(0) \mid \bbX, U] \mid \bbX\} =0$  almost surely.
\end{assumption}
\endgroup

Assumption \ref{as:5d} reduces to Assumption \ref{as:5c} when $Z$ is binary. The formal identification result is as follows.
\begin{theorem}\label{th:identification}
    Under Assumptions~\ref{as:relevance}--\ref{as:5d}, for any
$\gamma(Z, \bbX) \in L^2(P_{Z, \bbX})$ such that $\delta_\gamma^D(\bbX) \neq 0$ almost surely, the conditional average treatment effect equals the conditional generalized Wald estimand:
\begin{equation}\label{eq: ATE-generalIV}
 \E[Y(1) - Y(0) \mid \bbX ] =  \dfrac{\E[\{\gamma(Z, \bbX) - \E[\gamma(Z, \bbX) \mid \bbX]\}\mu^Y(Z, \bbX) \mid \bbX]}
    {\E[\{\gamma(Z, \bbX) - \E[\gamma(Z, \bbX) \mid \bbX]\} \mu^D(Z, \bbX) \mid \bbX]} \coloneqq \dfrac{\delta_\gamma^Y(\bbX)}{\delta_\gamma^D(\bbX)} \coloneqq \delta_\gamma(\bbX).
\end{equation}
Then the marginal average treatment effect is identified as $\Delta_\gamma = \E[\delta_\gamma(\bbX)]$.
\end{theorem}
In general, without the instrumental variable assumptions, $\delta_\gamma(\bbX)$ depends on the choice of the function $\gamma(Z, \bbX)$. Under those assumptions, however, whenever
$\delta_\gamma^D(\bbX) \neq 0$ almost surely, the quantity $\delta_\gamma(\bbX)$ does not depend on the
choice of $\gamma(Z, \bbX)$. In this case, we denote $\delta_\gamma(\bbX)$ by $\delta(\bbX)$ and its marginal
mean by $\Delta \coloneqq \E[\delta(\bbX)]$.

We next explain how the identification formula \eqref{eq: ATE-generalIV} arises. The main implication of Assumption~\ref{as:5d} is that all valid conditional Wald ratios based on two distinct instrument values identify the same conditional causal effect. Specifically, under Assumptions~\ref{as:relevance}--\ref{as:5d}, for any distinct $z_0,z_1\in\calZ$,
  whenever $\mu^D(z_1,\bbX)-\mu^D(z_0,\bbX)\neq0$,
\begin{equation}\label{identification: twozs}
     \E[Y(1)-Y(0)\mid\bbX]
     =
     \dfrac{\mu^Y(z_1,\bbX)-\mu^Y(z_0,\bbX)}
           {\mu^D(z_1,\bbX)-\mu^D(z_0,\bbX)}.
\end{equation}

For a continuous instrument, if $\mu^Y(z,\bbx)$ and $\mu^D(z,\bbx)$ are differentiable in $z$, taking the limit $z_1\to z_0=z$ in \eqref{identification: twozs} gives the local derivative ratio
$
     \E[Y(1)-Y(0)\mid\bbX]
     =
     \dfrac{\muyp(z,\bbX)}{\mudp(z,\bbX)}
$
at values where $\mudp(z,\bbX)\neq0$. One could then average these local Wald ratios across values of the instrument,
\(
     \E\left[
     \tilde{\omega}(Z,\bbX)
     \dfrac{\muyp(Z,\bbX)}{\mudp(Z,\bbX)}
     \,\middle|\,\bbX
     \right],
\)
for weights satisfying $\E[\tilde{\omega}(Z,\bbX)\mid\bbX]=1$. However, this form requires that $\mudp(Z,\bbX) \neq 0$ almost surely, which can be too strong: the treatment may respond to the instrument over some ranges of $Z$ but not others. Reparameterizing the weights as $\omega(Z,\bbX)=\tilde{\omega}(Z,\bbX)/\mudp(Z,\bbX)$ gives a weaker instrument relevance requirement
\begin{equation}\label{identification:weighted-derivative}
     \E[Y(1)-Y(0)\mid\bbX]
     =
     \dfrac{\E[\omega(Z,\bbX)\muyp(Z,\bbX)\mid\bbX]}
           {\E[\omega(Z,\bbX)\mudp(Z,\bbX)\mid\bbX]},
\end{equation}
provided that the denominator is nonzero.

For categorical instruments, differentiability is unavailable, but the same idea is implemented by averaging weighted pairwise conditional-mean differences rather than weighted derivatives:
\begin{equation}\label{identification:categorical-weighted}
     \E[Y(1)-Y(0)\mid\bbX]
     =
     \dfrac{\sum_{0\leq k<j\leq J-1}\omega_{jk}(\bbX)\{\mu^Y(j,\bbX)-\mu^Y(k,\bbX)\}}
           {\sum_{0\leq k<j\leq J-1}\omega_{jk}(\bbX)\{\mu^D(j,\bbX)-\mu^D(k,\bbX)\}},
\end{equation}
again provided that the denominator is nonzero. When $J=2$, \eqref{identification:categorical-weighted} reduces to the binary Wald estimand in Section~\ref{sec: identification-binIV}.

The conditional Riesz representations in Section~\ref{sec: Riesz Representer} put the conditional generalized Wald ratios \eqref{identification:weighted-derivative} and \eqref{identification:categorical-weighted} into a common residual form
\[
     \E[Y(1)-Y(0)\mid\bbX]  = \dfrac{\E[\gamma_\omega(Z,\bbX)\mu^Y(Z,\bbX)\mid\bbX]}      {\E[\gamma_\omega(Z,\bbX)\mu^D(Z,\bbX)\mid\bbX]},
\]
where $\gamma_\omega$ is the conditional ``Riesz representer'' defined in Section~\ref{sec: Riesz Representer}. The converse results in Section~\ref{sec: Riesz Representer} show that we can reverse this construction: instead of starting from a weight $\omega$ and deriving its representer $\gamma_\omega$, we may start from any conditionally mean-zero square-integrable function $\gamma(Z,\bbX)$ and view it as representing some weighted average derivatives or weighted average pairwise differences. Equivalently, for an arbitrary square-integrable $\gamma(Z,\bbX)$, we first residualize it as $\gamma(Z,\bbX)-\E[\gamma(Z,\bbX)\mid\bbX]$ and then use this mean-zero function in the generalized Wald ratio, which gives \eqref{eq: ATE-generalIV}. Thus \eqref{eq: ATE-generalIV} preserves the intuition of weighted aggregation over instrument values as in \eqref{identification:weighted-derivative} and \eqref{identification:categorical-weighted}, without imposing differentiability conditions.

\begin{remark}
   The following remarks are in order. First, when $Z$ is continuous,~\citet{hartwig2023average} show that the average treatment effect can be identified by the conventional Wald estimand $ \E \left[  {\cov(Z,  Y \mid \bbX)}/{\cov(Z,  D \mid \bbX)}\right].$ This is a special case of \eqref{eq: ATE-generalIV} by setting $\gamma(Z, \bbX) = Z - \mu^Z(\bbX)$. Second, practitioners sometimes dichotomize a continuous instrument at a predetermined threshold $z_0$. In this case, the conditional average treatment effect can be identified by
   $$\dfrac{\E[Y \mid Z \geq z_0, \bbX] - \E[Y \mid Z  < z_0, \bbX]}{\E[D \mid Z \geq z_0, \bbX] - \E[D \mid Z  < z_0, \bbX]},$$
   which is a special case of \eqref{eq: ATE-generalIV} obtained by taking $\gamma(Z, \bbX) = \{\mathbb{I}(Z \geq z_0) - (1-F(z_0 \mid \bbX))\} /\{F(z_0 \mid \bbX) (1-F(z_0 \mid \bbX))\}$. However, dichotomizing a continuous instrument generally leads to information loss, which we further investigate in Sections \ref{sec: sim} and \ref{sec: app}.
\end{remark}

\section{Estimation}\label{sec: estimation}

We now consider the estimation problem for the average treatment effect $\Delta$. We first characterize the observed-data constraints implied by our identification assumptions in Theorem \ref{th:identification}. In addition to the usual instrumental variable inequality constraints induced by Assumptions \ref{as:relevance}--\ref{as:dsep} \citep{kedagni2020generalized}, we have the constraint $\mu^Y(z_1, \bbX) - \mu^Y(z_0, \bbX) = \delta(\bbX) (\mu^D(z_1, \bbX) - \mu^D(z_0, \bbX))$, for all $z_0, z_1 \in \calZ$. Fixing $z_1=z$ and averaging this identity over $z_0$ with respect to the conditional distribution of $Z$ given $\bbX$ gives
\begin{equation}\label{constraint2}
    \mu^Y(z, \bbX) -  \mu^Y(\bbX)= \delta(\bbX)  \{\mu^D(z, \bbX) - \mu^D(\bbX)\} , \ \forall \ z \in \calZ.
\end{equation}
When $Z$ takes more than two values, the constraint~\eqref{constraint2} becomes nontrivial, arising from the requirement that any two distinct values of $Z$ yield the same $\delta(\bbX)$. The instrumental variable inequality constraints do not affect the tangent space and efficiency results when the true average treatment effect lies within the interior of its parameter space \citep{wang2018bounded}. Hence, we focus primarily on the constraint~\eqref{constraint2}.

We first derive the efficient influence function of $\Delta$, denoted by $\Psi_{\eff}$, under the semiparametric model $\calM_{\sp}$, which restricts the observed-data law only through constraint~\eqref{constraint2}. Because the resulting $\Psi_{\eff}$ involves nuisance functions that may be difficult to estimate in practice, we instead identify an influence function $\Psi_{\gamma_0}$ that is valid in $\calM_{\sp}$ and coincides with $\Psi_{\eff}$ on a simplifying submodel, thereby attaining local efficiency there. This leads to a double-cross-fitted debiased estimator that accommodates flexible machine learning for nuisance estimation.

\subsection{Semiparametric Efficiency Theory}\label{sec: semiparametric theory}

We now present the semiparametric theory for $\Delta$.
As a first step, we consider the estimation problem for the average generalized Wald estimand $\Delta_{\gamma}$ under the nonparametric model $\mathcal{M}_{\mathrm{np}}$ for a given function $\gamma(Z, \bbX)$. In this case, under the model $\mathcal{M}_{\mathrm{np}}$, $\Delta_\gamma$ has a unique influence function, also known as the efficient influence function, which we denote by $\Psi_{\gamma}$.
We then turn to the semiparametric model $\mathcal{M}_{\mathrm{sp}}$, under which all $\Delta_\gamma$ reduce to $\Delta$. Hence, any $\Psi_{\gamma}$ remains a valid, though not necessarily efficient, influence function of $\Delta$. Finally, we derive the efficient influence function of $\Delta$ in $\mathcal{M}_{\mathrm{sp}}$, denoted by $\Psi_{\mathrm{eff}}$.

\begin{theorem}\label{th:nonparametric}
The following hold.
   \begin{itemize}
    \item[(1)] For a given function $\gamma(Z, \bbX)$ such that $\delta_\gamma^D(\bbX) \neq 0$ almost surely, the unique influence function, also the efficient influence function, of $\Delta_{\gamma}$ in $\calM_{\np}$ is
    \begin{align*}
     \Psi_\gamma(\bbO; P) = \dfrac{\gamma(Z, \bbX) - \E[\gamma(Z, \bbX) \mid \bbX]}{\delta_{\gamma}^D(\bbX)}\calE_\gamma +\delta_{\gamma}(\bbX)  - \Delta_\gamma,
    \end{align*}
    where $\calE_\gamma \coloneqq Y - \mu^Y(\bbX) - \delta_\gamma(\bbX) \{ D -\mu^D(\bbX)\}$.
    \item[(2)] Under the semiparametric model $\calM_{\sp}$, $\delta_\gamma(\bbX) = \delta(\bbX)$, $\Delta_\gamma = \Delta $, and  $\calE_\gamma = \calE \coloneqq Y - \mu^Y(\bbX) - \delta(\bbX) \{ D -\mu^D(\bbX)\}$. Then
    \begin{equation*}
     \Psi_\gamma(\bbO; P) = \dfrac{\gamma(Z, \bbX) - \E[\gamma(Z, \bbX) \mid \bbX]}{\delta_{\gamma}^D(\bbX)} \calE +\delta(\bbX)  - \Delta,
    \end{equation*}
    and $\{ \Psi_\gamma(\bbO; P): \delta_\gamma^D(\bbX) \neq 0 \ \text{almost surely}\}$ is a class of IFs of $\Delta$.
    \item[(3)] Let $\sigma^2(Z, \bbX) \coloneqq \var[\calE \mid Z, \bbX] \equiv \E[\calE^{2} \mid Z, \bbX]$. Then the efficient influence function of $\Delta$ under the semiparametric model $\calM_{\sp}$ is
    \begin{equation*}
    \Psi_{\eff}(\bbO; P) = \gamma_{\opt}(Z, \bbX)  \calE + \delta(\bbX) - \Delta,
    \end{equation*}
    where
    \begin{equation*}
    \gamma_{\opt} (Z, \bbX) =  \dfrac{\E\left[\dfrac{1}{\sigma^2(Z, \bbX)} \,\middle\vert\, \bbX \right] \dfrac{\mu^D (Z, \bbX)}{\sigma^2(Z, \bbX)} - \dfrac{1}{\sigma^2(Z, \bbX)} \E\left[\dfrac{\mu^D(Z, \bbX)}{\sigma^2(Z, \bbX)} \,\middle\vert\, \bbX \right]}{\E \left[\dfrac{1}{\sigma^2(Z, \bbX)} \,\middle\vert\, \bbX \right] \E \left[\dfrac{\mu^D(Z, \bbX)^2 }{\sigma^2(Z, \bbX)} \,\middle\vert\, \bbX \right] - \E^2 \left[\dfrac{\mu^D(Z, \bbX)}{\sigma^2(Z, \bbX)} \,\middle\vert\, \bbX\right]}.
    \end{equation*}
    Furthermore, $\Psi_{\eff}$ belongs to the class $\{ \Psi_\gamma(\bbO; P): \delta_\gamma^D(\bbX) \neq 0 \ \text{almost surely}\}$ since $\E[\gamma_{\opt} (Z, \bbX) \mid \bbX] =0 $ and   $\delta_{\gamma_{\mathrm{opt}}}^D(\bbX)
    = \E \big[\gamma_{\opt}(Z,\bbX)\,\mu^D(Z, \bbX) \mid \bbX\big] = 1$.
\end{itemize}
\end{theorem}

The proof of Theorem~\ref{th:nonparametric} is provided in Supplementary Material~\ref{S-app:th:nonparametric}. Statements (1) and (2) follow from standard semiparametric efficiency theory for a given $\gamma(Z, \bbX)$. The nonstandard contribution is Statement (3): deriving the efficient influence function under $\calM_{\sp}$ requires characterizing the tangent space induced by the constraint~\eqref{constraint2}. In particular, the efficient influence function is obtained by projecting any influence function onto this tangent space, denoted $\Lambda_{\sp}$. To characterize $\Lambda_{\sp}$, we must show the existence of a parametric submodel in $\calM_{\sp}$ whose score lies in $\Lambda_{\sp}$. The main challenge is that standard first-order perturbations of the form $p_t(\bbo) = p(\bbo)\{1 + t s(\bbo)\}$, with $t \in \bbR$ and score function $s(\bbo)$, may generate parametric submodels outside $\calM_{\sp}$. Restricting $s(\bbo)$ so that this first-order path remains inside the model would exclude many directions that should belong to the tangent space. To address this, we construct a second-order perturbation that ensures the resulting parametric submodel remains within $\calM_{\sp}$ without imposing unnecessary restrictions on $s(\bbo)$. While second- or higher-order paths are typically introduced in the study of higher-order tangent spaces~\citep[e.g.][]{pfanzagl1983asymptotic}, here they are required even for deriving the first-order tangent space. Further details and rationale for this construction are provided in Remark~\ref{S-rem:non-convex} of the Supplementary Material.

\begin{remark}
\label{rem:bin}
Provided that $\sigma^2(Z,\bbX)>0$, the denominator of $\gamma_{\rm opt}(Z,\bbX)$ is nonzero almost surely under Assumption~\ref{as:relevance}. By the Cauchy--Schwarz inequality, the denominator is zero if and only if $\mu^D(Z, \bbX)$ does not depend on $Z$ given $\bbX$, namely, the instrument is not relevant.
 \end{remark}

Theorem \ref{th:nonparametric} suggests that one may estimate $\Delta$ based on $\Psi_{\eff}(\bbO; P)$ by solving the estimating equation $\mathbb{P}_n\Psi_{\eff}(\bbO; \widehat{P})=0$, where $\widehat{P}$ includes estimates of the nuisance functions that appear in $\Psi_{\eff}(\bbO;P)$.
However, the nuisance functions $\sigma^2(Z, \bbX)$, $\E[1/\sigma^2(Z, \bbX) \mid \bbX]$, $\E [\mu^D(Z, \bbX)/\sigma^2(Z, \bbX) \mid \bbX]$, and $\E\left[{\{\mu^D(Z, \bbX)\}^2 }/{\sigma^2(Z, \bbX)} \,\middle\vert\, \bbX \right]$ can be difficult to estimate in practice because their functional forms are typically unknown. Moreover, several of these nuisance functions depend intricately on other nuisance functions and can be complicated even when the preliminary regression functions are very simple \citep[e.g.][]{young2024rose}. These difficulties make direct estimation based on $\Psi_{\eff}(\bbO;P)$ less attractive as a practical strategy, even though $\Psi_{\eff}$ characterizes the semiparametric efficiency bound.

For completeness, in Supplementary Material~\ref{S-sec:debiasedestimator_eff}, we construct a semiparametric efficient debiased estimator $\widehat{\Delta}_{\,\eff}$ using standard cross-fitting, where the nuisance functions and the target parameter are estimated using different samples~\citep{chernozhukov_doubledebiased_2018}.
Although this estimator is asymptotically normal and semiparametrically efficient under suitable regularity conditions and high-level rate conditions, our simulation results show that it can have poor finite-sample performance; see Tables~\ref{S-tab:datagen-binary-eif} and~\ref{S-tab:datagen2-binary-eif} for details. This suggests that, even under relatively simple data-generating processes, the nuisance functions in $\Psi_{\eff}(\bbO; P)$ may be complex enough that estimators based on the efficient influence function can be outperformed by estimators based on simpler influence functions. We turn to such alternatives in the following section.

\subsection{A Locally Efficient Influence Function}\label{sec: localeff}
We now use the efficient influence function derived above to construct an influence function that is easier to model and compute. The key idea is localization: we identify submodels of $\calM_{\sp}$ on which the evaluation of $\Psi_{\eff}(\bbO; P)$ becomes tractable, and then use the resulting simplified form to guide the construction of a valid influence function in the full semiparametric model. We achieve this goal by introducing the following condition that enables evaluation of $\Psi_{\eff}(\bbO; P)$ without requiring knowledge of $\sigma^2(Z, \bbX)$ or $\E[1/\sigma^2(Z, \bbX) \mid \bbX]$.
\begin{cond}\label{cond:homoskedastic}
  \emph{Conditional homoskedasticity}: The conditional variance of the residual $\calE$ given $(Z, \bbX)$, $\sigma^2(Z, \bbX) $, does not depend on $Z$.
\end{cond}
\begin{remark}
Condition~\ref{cond:homoskedastic} is similar in spirit to, but much weaker than, the homoskedasticity assumption, namely that $\sigma^2(Z, \bbX)$ is constant, commonly imposed in partially linear instrumental variable models \citep{okui2012doubly, scheidegger2026inference}.
\end{remark}
Under Condition~\ref{cond:homoskedastic}, the efficient influence function in
$\calM_{\sp}$ simplifies to
\begin{equation}\label{eq:eif_conhom}
   \dfrac{\mu^D(Z, \bbX) - \mu^D(\bbX)}{\var\{\mu^D(Z, \bbX) \mid \bbX\}}\, \calE + \delta(\bbX) - \Delta.
\end{equation}
It follows that the influence function in \eqref{eq:eif_conhom} is locally efficient in the submodel
\[
\mathcal{M}_{\ch} \coloneqq \{ P \in \mathcal{M}_{\sp}: P \text{ satisfies Condition~\ref{cond:homoskedastic} } \},
\]
a subset of $\mathcal{M}_{\sp}$. However, this condition may not hold in practice. It is therefore important that the validity of
\eqref{eq:eif_conhom} does not rely on Condition~\ref{cond:homoskedastic}. The class $\{ \Psi_\gamma(\bbO; P): \delta_\gamma^D(\bbX) \neq 0 \ \text{almost surely}\}$ in Theorem~\ref{th:nonparametric} provides the bridge back to the original semiparametric model: every member of this class is a valid influence function of $\Delta$ in $\calM_{\sp}$, regardless of whether Condition~\ref{cond:homoskedastic} holds. Comparing~\eqref{eq:eif_conhom} with this class suggests the weight $\gamma_0(Z,\bbX)=\mu^D(Z,\bbX)$. The corresponding indexed influence function is the localization of $\Psi_{\eff}$ that we use below: if $\gamma_0$ were fixed at its population value, $\Psi_{\gamma_0}(\bbO;P)$ would coincide with the locally efficient form in \eqref{eq:eif_conhom} on $\calM_{\ch}$ while remaining a member of the valid influence-function class in $\calM_{\sp}$. The following theorem demonstrates that this localized influence function is valid for $\Delta$ in $\calM_{\sp}$ and is locally efficient at $\calM_{\ch}$.
The proof is in Supplementary Material~\ref{S-sec:prooflocaleff}.

\begin{theorem}\label{th:localeff}
$\Psi_{\gamma_0}(\bbO; P)$ is locally efficient for $\Delta$ in $\mathcal M_{\sp}$ at the submodel $\mathcal M_{\ch}$; consequently, the corresponding semiparametric efficiency bound is $\E[\Psi_{\gamma_0}^2(\bbO;P)]$. When Condition~\ref{cond:homoskedastic} fails, $\Psi_{\gamma_0}(\bbO; P)$ remains a valid influence function of $\Delta$ under $\mathcal{M}_{\sp}$.
\end{theorem}
\begin{remark}
The construction of $\Psi_{\gamma_0}$ is not merely a formal substitution into \eqref{eq:eif_conhom}. The simplifying submodel identifies the weight $\gamma_0(Z,\bbX)=\mu^D(Z,\bbX)$, for which $\delta_{\gamma_0}^D(\bbX)=\var\{\mu^D(Z,\bbX)\mid\bbX\}$, so that $\Psi_{\gamma_0}$ coincides with the efficient influence function in \eqref{eq:eif_conhom} on $\mathcal M_{\ch}$. Since this weight depends on the observed-data law, validity in the full model does not follow simply by treating $\gamma_0$ as a fixed function. In the proof, $\gamma_0$ is allowed to vary as the nuisance function $\mu^D(Z,\bbX)$ when differentiating $\Delta_{\gamma_0}$; the corresponding nonparametric efficient influence function is simplified to \eqref{eq:eif_conhom} under the semiparametric constraint defining $\mathcal M_{\sp}$. Thus the same $\Psi_{\gamma_0}$ remains a valid influence function in $\mathcal M_{\sp}$ while retaining local efficiency at $\mathcal M_{\ch}$.
\end{remark}
\begin{remark}
Condition \ref{cond:homoskedastic} is compatible with our instrumental variable assumptions; see Supplementary Material~\ref{S-example:Homoskedastic} for an example where it holds. This condition is required only for $\Psi_{\gamma_0}$ to be locally efficient when the instrument is non-binary.  When the instrument is binary, $\Psi_{\gamma_0}$ coincides with the efficient influence function of $\Delta$ under $\calM_{\np}$. Therefore, if Condition \ref{cond:homoskedastic} fails, $\Psi_{\gamma_0}$ is a valid influence function of $\Delta$ with a non-binary instrument and equals the efficient influence function of $\Delta$ with a binary instrument.
\end{remark}

In practice, $\gamma_0(Z,\bbX)=\mu^D(Z,\bbX)$ is typically unknown and must be estimated. The next subsection constructs an estimator based on $\Psi_{\gamma_0}$ that accounts for this estimation step.

\subsection{A Semiparametric Locally Efficient Debiased Estimator}\label{sec: debiasedestimator}

We now propose a semiparametric locally efficient debiased estimator based on $\Psi_{\gamma_0}(\bbO;P)$. The construction has two ingredients. The first is a risk-minimization reparameterization of two \emph{composite nuisance functions}, defined below, which improves stability when the influence function contains reciprocal or quotient terms.
The second is \emph{double cross-fitting}, introduced because some \emph{composite nuisance functions} are fitted using preliminary regression estimates.

Constructing estimators of $\Delta$ based on $\Psi_{\gamma_0}(\bbO;P)$ requires estimating the following nuisance functions: $\mu^D(Z,\bbX),\mu^D(\bbX),\mu^Y(\bbX),\delta_{\gamma_0}^D(\bbX)$, and $\delta(\bbX)$. We call the last two \emph{composite nuisance functions}, because they are conditional expectations of some other nuisance parameters. Specifically, $\delta_{\gamma_0}^D(\bbX)=\E\left[\{\mu^D(Z,\bbX)-\mu^D(\bbX)\}\mu^D(Z, \bbX) \mid \bbX\right]$, and $\delta(\bbX) = {\delta_{\gamma_0}^Y(\bbX)}/{\delta_{\gamma_0}^D(\bbX)}$ with $ \delta_{\gamma_0}^Y(\bbX) =\E\left[\{\mu^D(Z,\bbX)-\mu^D(\bbX)\}\mu^Y(Z, \bbX)\mid\bbX\right]$. In the influence function, $\delta_{\gamma_0}^D(\bbX)$ appears through two nonlinear targets: its reciprocal $\{\delta_{\gamma_0}^D(\bbX)\}^{-1}$ and the quotient $\delta_{\gamma_0}^Y(\bbX)/\delta_{\gamma_0}^D(\bbX)$. A direct-substitution strategy would estimate $\delta_{\gamma_0}^D(\bbX)$ and $\delta_{\gamma_0}^Y(\bbX)$ separately and then form these two targets by division.
Such a strategy can amplify the estimation error, especially when the estimated $\delta_{\gamma_0}^D(\bbX)$ is close to zero. Instead, we estimate the reciprocal $\{\delta_{\gamma_0}^D(\bbX)\}^{-1}$ and the quotient $\delta(\bbX)$ directly through weighted risk-minimization reparameterizations. For notational simplicity, write
$
\tau(\bbX)=\left\{\delta_{\gamma_0}^D(\bbX)\right\}^{-1},
\xi^D (Z,\bbX)=\mu^D(Z,\bbX)-\mu^D(\bbX).
$
Specifically, the key identity is that, for any random variable $T$, the minimizer of $\E[\{\xi^D(Z,\bbX)\}^2\{T-g(\bbX)\}^2]$ over $g\in L^2(P_{\bbX})$ is
\begin{equation*}
 \dfrac{\E[\{\xi^D(Z,\bbX)\}^2T\mid\bbX]}{\E[\{\xi^D(Z,\bbX)\}^2\mid\bbX]}  ,
\end{equation*}
provided that the denominator is nonzero. Taking $T=\{\xi^D(Z,\bbX)\}^{-2}$ and $T=\{Y-\mu^Y(\bbX)\}/\xi^D(Z,\bbX)$ gives
\begin{align}
\tau(\bbX)
&=\left\{\E\left[\{\xi^D(Z,\bbX)\}^2 \mid \bbX\right]\right\}^{-1} \notag\\
&=\argmin_{g_1 \in L^2(P_{\bbX})}\E\left[\xi^D (Z, \bbX)^2\left\{\dfrac{1}{\xi^D (Z, \bbX)^2}-g_1(\bbX)\right\}^2\right], \label{g1}\\
\delta(\bbX)
&=\argmin_{g_2 \in L^2(P_{\bbX})}\E\left[\xi^D (Z, \bbX)^2\left\{\dfrac{Y-\mu^Y(\bbX)}{\xi^D (Z, \bbX)}-g_2(\bbX)\right\}^2\right]. \label{g2}
\end{align}
Thus $\tau(\bbX)$ and $\delta(\bbX)$ can be estimated by minimizing empirical analogues of \eqref{g1} and \eqref{g2}, with $\xi^D (Z, \bbX)$ and $\mu^Y(\bbX)$ replaced by preliminary estimates. Similar reparameterization strategies have also been used by \citet[Lemma 5.1]{hines2026parameterizing}.

A further complication is that the empirical risks of \eqref{g1} and \eqref{g2} are built using preliminary estimates of $\xi^D(Z,\bbX)$ and $\mu^Y(\bbX)$. Standard cross-fitting separates nuisance estimation from target-parameter estimation, but it does not by itself separate the preliminary regressions from the induced regressions used to estimate $\tau(\bbX)$ and $\delta(\bbX)$. If the same observations are used for both stages, additional Donsker-type empirical-process conditions are often needed for consistent estimation of $\tau(\bbX)$ and $\delta(\bbX)$; see Remark~\ref{rem:inner-cross-fitting} and Supplementary~\ref{S-sec:debiasedestimator_eff}. Such conditions may be difficult to justify when the covariates are high-dimensional. We therefore use an additional layer of sample splitting. For each held-out fold $\mathcal I_k, k = 1, \ldots, K$, its complement $\mathcal I_k^c$ is split into two subsamples: one is used to estimate $\mu^D(Z,\bbX)$, $\mu^D(\bbX)$ and $\mu^Y(\bbX)$, and the other is used to estimate $\tau(\bbX)$ and $\delta(\bbX)$. The roles of the two subsamples are then swapped to make full use of the data, and the resulting fits are averaged.
For a held-out observation $i\in \mathcal I_k$, define
\begin{align}
\widehat{\calE}_{i,(-k)}
&=Y_i-\widehat{\mu}^Y_{(-k)}(\bbX_i)
-\widehat{\delta}_{(-k)}(\bbX_i)\left\{D_i-\widehat{\mu}^D_{(-k)}(\bbX_i)\right\}, \notag\\
\widehat\psi_{i,k}
&=\left\{\widehat{\mu}^D_{(-k)}(Z_i,\bbX_i)-\widehat{\mu}^D_{(-k)}(\bbX_i)\right\}
\widehat{\tau}_{(-k)}(\bbX_i)\widehat{\calE}_{i,(-k)}
  +\widehat{\delta}_{(-k)}(\bbX_i).
\label{eq:estimated-local-score}
\end{align}
The resulting algorithm is as follows.

\begin{algorithm}[H]
\caption{Double-cross-fitted locally efficient debiased estimator}
\label{alg:dcf-estimator}
\DontPrintSemicolon
\KwIn{Observed data $\{\bbO_i=(Y_i,D_i,Z_i,\bbX_i):i=1,\ldots,n\}$ and number of folds $K$.}
\KwOut{$\widehat\Delta_{\mathrm{RM}}$.}
Randomly split $\mathcal I=\{1,\ldots,n\}$ into $K$ disjoint folds $\mathcal I_1,\ldots,\mathcal I_K$.\;
\For{$k=1,\ldots,K$}{
Set $\mathcal I_k^c=\mathcal I\setminus\mathcal I_k$, and split $\mathcal I_k^c$ into two non-overlapping folds $\mathcal I_k^{c_1}$ and $\mathcal I_k^{c_2}$.\;
\For{$j=1,2$}{
Set $j'=3-j$.\;
Using observations in $\mathcal I_k^{c_j}$, estimate $\mu^D(Z,\bbX)$, $\mu^D(\bbX)$ and $\mu^Y(\bbX)$; denote the fitted functions by $\widehat\mu_{(-k,c_j)}^D(Z,\bbX)$, $\widehat\mu_{(-k,c_j)}^D(\bbX)$ and $\widehat\mu_{(-k,c_j)}^Y(\bbX)$.\;
Set $\widehat\xi^D_{(-k,c_j)}(Z,\bbX)=\widehat\mu^D_{(-k,c_j)}(Z,\bbX)-\widehat\mu^D_{(-k,c_j)}(\bbX)$.\;
Using observations in $\mathcal I_k^{c_{j'}}$, fit the weighted regressions in \eqref{g1} and \eqref{g2}, with $\xi^D$ and $\mu^Y$ replaced by $\widehat\xi^D_{(-k,c_j)}$ and $\widehat\mu_{(-k,c_j)}^Y(\bbX)$; denote the fitted functions by $\widehat\tau_{(-k,c_{j'})}(\bbX)$ and $\widehat\delta_{(-k,c_{j'})}(\bbX)$.\;
}
Average the two role-swapped fits pointwise to obtain $\widehat\mu^D_{(-k)}(Z,\bbX)$, $\widehat\mu^D_{(-k)}(\bbX)$, $\widehat\mu^Y_{(-k)}(\bbX)$, $\widehat\tau_{(-k)}(\bbX)$ and $\widehat\delta_{(-k)}(\bbX)$.\;
Evaluate $\widehat\psi_{i,k}$ in \eqref{eq:estimated-local-score} for each $i\in\mathcal I_k$.\;
}
Return $\widehat{\Delta}_{\mathrm{RM}}=n^{-1}\sum_{k=1}^K\sum_{i\in\mathcal I_k}\widehat\psi_{i,k}$.\;
\end{algorithm}

For the asymptotic analysis, let $\widehat P_{(-k)}$ collect the fold-specific nuisance estimators constructed in Algorithm~\ref{alg:dcf-estimator}. Then $\widehat\psi_{i,k}=\psi_{\gamma_0}(\bbO_i;\widehat P_{(-k)})$, where $\psi_{\gamma_0}$ is the uncentered estimating function defined by $\Psi_{\gamma_0}(\bbO;P)=\psi_{\gamma_0}(\bbO;P)-\Delta$.
For $k=1,\ldots,K$, define the following $L^2(P)$ estimation errors:
\begin{align*}
r_Y^{(-k)} &= \|\widehat{\mu}^Y_{(-k)}(\bbX)-\mu^Y(\bbX)\|_{P,2},  r_D^{(-k)} = \|\widehat{\mu}^D_{(-k)}(\bbX)-\mu^D(\bbX)\|_{P,2},  r_\tau^{(-k)} = \|\widehat{\tau}_{(-k)}(\bbX)-\tau(\bbX)\|_{P,2},\\
r_{D,Z}^{(-k)} &= \|\widehat{\mu}^D_{(-k)}(Z,\bbX)-\mu^D(Z,\bbX)\|_{P,2}, \quad
r_\delta^{(-k)} = \|\widehat{\delta}_{(-k)}(\bbX)-\delta(\bbX)\|_{P,2}.
\end{align*}
We impose the following conditions for all $k=1,\ldots,K$:
\begin{enumerate}[label={D\arabic*}, ref={D\arabic*}, leftmargin=*]
    \item
    \label{as:Boundedness} \emph{Boundedness}. There exists a constant $C<\infty$ such that $\|Y\|_{\infty} \leq C$, $\|\mu^Y(\bbX)\|_{\infty} \leq C$, $\|\delta(\bbX)\|_{\infty} \leq C$, and $\|\tau(\bbX)\|_{\infty} \leq C$; moreover, $\|\widehat{\delta}_{(-k)}(\bbX)\|_{\infty}$, $\|\widehat{\tau}_{(-k)}(\bbX)\|_{\infty}$, $\|\widehat{\mu}_{(-k)}^D(Z,\bbX)\|_{\infty}$, $\|\widehat{\mu}_{(-k)}^D(\bbX)\|_{\infty}$, and $\|\widehat{\mu}_{(-k)}^Y(\bbX)\|_{\infty}$ are $O_p(1)$.
    \item
    \label{as:Consistency} \emph{Consistency}. $r_{D,Z}^{(-k)}+r_D^{(-k)}+r_Y^{(-k)}+r_\tau^{(-k)}+r_\delta^{(-k)}=o_p(1)$.
    \item
    \label{as:Rate} \emph{Rate conditions}. $\{r_{D,Z}^{(-k)}+r_D^{(-k)}\}\{r_D^{(-k)}+r_Y^{(-k)}+r_\delta^{(-k)}\}+r_\tau^{(-k)}r_\delta^{(-k)}=o_p(n^{-1/2})$.
\end{enumerate}
Condition~\ref{as:Boundedness} requires that the outcome, the nuisance functions, and their estimators be uniformly bounded. In particular, uniform boundedness of $\tau(\bbX)$ strengthens the nonzero IV--treatment association used for identification: it requires $\delta_{\gamma_0}^D(\bbX)$ to be bounded away from zero, which controls the reciprocal in the asymptotic analysis. Condition~\ref{as:Consistency} requires consistency of all nuisance estimators in the $L^2(P)$-norm. Condition~\ref{as:Rate} is a second-order product-rate condition. Since $\mu^D(\bbX)$ appears in both factors of the first product, this condition requires that $r_D^{(-k)}=o_p(n^{-1/4})$, which is weaker than requiring the parametric rate. If $\mu^D(\bbX)$ is estimated at this rate or faster, then $\mu^D(Z,\bbX)$, $\mu^Y(\bbX)$ and $\delta(\bbX)$ may converge more slowly, provided the product rate remains $o_p(n^{-1/2})$.

The following theorem establishes that $\widehat{\Delta}_{\mathrm{RM}}$ is asymptotically normal in $\calM_{\sp}$ and locally efficient at $\calM_{\ch}$ under Conditions~\ref{as:Boundedness}--\ref{as:Rate}.
\begin{theorem}\label{th:Delta_1}
Suppose Conditions~\ref{as:Boundedness}--\ref{as:Rate} hold. Then, under $\calM_{\sp}$, $\sqrt{n}(\widehat{\Delta}_{\mathrm{RM}} - \Delta) \to_d N(0, \mathcal{V})$, where $\mathcal{V} := \var\{\Psi_{\gamma_0}(\bbO;P)\}$ attains the semiparametric local efficiency bound for $\Delta$ at $\calM_{\ch}$. Moreover, a consistent estimator of $\mathcal{V}$ is
\begin{align*}
    \widehat{\mathcal{V}} = \dfrac{1}{n} \sum_{k=1}^K \sum_{i \in \mathcal{I}_k} \left\{ \psi_{\gamma_0}(\bbO_i; \widehat{P}_{(-k)}) - \widehat{\Delta}_{\mathrm{RM}}\right\}^2.
\end{align*}
\end{theorem}

\begin{remark}\label{rem:inner-cross-fitting}
Standard cross-fitting separates the observations used for nuisance estimation from those used to estimate the target parameter, thereby avoiding Donsker-type conditions when controlling the final empirical-process term involving $\psi_{\gamma_0}(\cdot;\widehat P)$ \citep{chernozhukov_doubledebiased_2018}. The additional split used here has a different purpose: it allows the estimation error of $\widehat\tau$ and $\widehat\delta$ to be controlled by the estimation error for the preliminary nuisance estimates and the oracle learning error for the second-stage learner. A detailed error decomposition illustrating this point is provided in Supplementary Material~\ref{S-app:samplesplit}.
\end{remark}
\begin{remark}
    A closely related way to motivate debiased estimators is through Neyman-orthogonal moment equations \citep{chernozhukov_doubledebiased_2018}. Our construction is instead based on the influence function from semiparametric theory. Under specific conditions studied by \citet{chen2026equivalence}, such an influence-function-based estimating equation may also be interpreted as Neyman-orthogonal.
\end{remark}

\section{Simulation Studies}\label{sec: sim}
In this section, we evaluate the finite-sample performance of the proposed estimator. We then generate three binary covariates and one continuous covariate. Let $\bbX=(X_1, X_2, X_3, X_4)$ denote baseline covariates, where $X_1 \sim \textrm{Bernoulli}(0.5)$, $X_2 \sim \textrm{Bernoulli}(0.9)$, $X_3 \sim \textrm{Bernoulli}(0.4)$ and $X_4 \sim \textrm{N}(0, 1)$. The unmeasured confounder is generated as $U \sim \textrm{Unif}(0, 1)$. Throughout, let $\expit(t) \coloneqq \{1+\exp(-t)\}^{-1}$. Given $\bbX$, the instrument is generated from
\begin{align*}
Z=\boldsymbol{\zeta}^\top\bbX+\epsilon_Z,\qquad \epsilon_Z\sim\textrm{N}(0,1).
\end{align*}
Conditional on $(Z, \bbX, U)$, we consider two treatment mechanisms:
\begin{align*}
\text{without interaction:}\quad
\Pr(D=1\mid Z,\bbX,U)
&=\expit\{-0.75+1.5Z+\boldsymbol{\iota}^\top\bbX+2U\},\\
\text{with interaction:}\quad
\Pr(D=1\mid Z,\bbX,U)
&=\expit\{-0.9+1.5Z+0.5ZX_4+\boldsymbol{\iota}^\top\bbX+2U\}.
\end{align*}
The second mechanism includes a $ZX_4$ interaction term, so that instrument strength varies with the continuous covariate $X_4$. The second mechanism creates a more challenging setting for nuisance-function estimation.

To mimic our real-data application, we generate a binary outcome $Y$ conditional on $(D, \bbX, U)$ using the odds product model of \citet{richardson2017modeling}. Specifically, $Y$ is generated as follows:
\begin{align*}
    &  \delta(\bbX, U)= \tanh(\boldsymbol{\alpha}^\top \bbX), \quad  \OP(\bbX, U) = \exp(0.1 + \boldsymbol{\theta}^\top \bbX + 0.5U);\\
   & \E[Y \mid D, \bbX, U] =  \delta(\bbX, U) D + \E[Y \mid D=0, \bbX, U],
\end{align*}
where $\E[Y \mid D=0, \bbX, U] = \{2(\OP(\bbX, U) - 1)\}^{-1} \{ \OP(\bbX, U)(2 - \delta(\bbX, U)) + \delta(\bbX, U) - \sqrt{\{\OP(\bbX, U) (\delta(\bbX, U) -2 ) - \delta(\bbX, U)\}^2 + 4 \OP(\bbX, U) ( 1- \delta(\bbX, U))(1-\OP(\bbX, U))\}}$.

The parameter values are $\boldsymbol{\alpha}= (0.2, 0.4, -0.5, 0.5)^\top$,  $\boldsymbol{\zeta} = (-1, 0.5, 0.5, 0.1)^\top$, $\boldsymbol{\iota} =(-0.5, 0.5, -0.5, -0.4)^\top$, $\boldsymbol{\theta} = (-1, 0.5, -0.5, -0.5)^\top$. We are interested in estimating the average generalized Wald estimand $\Delta=\E[\delta(\bbX)]$, whose true value is 0.202.

For nuisance-function estimation, we consider using Super Learner, abbreviated as SL, implemented using the \texttt{SuperLearner} package in \textsf{R} \citep{van2007super}. The SL library includes generalized linear models, generalized additive models, penalized generalized linear models, random forests, and XGBoost. To assess the role of risk minimization in estimating the reciprocal $\tau(\bbX)$ and the quotient $\delta(\bbX)$, we compare two algorithms. The first is $\widehat{\Delta}_{\mathrm{RM}}$, defined in Algorithm~\ref{alg:dcf-estimator}, which estimates $\tau(\bbX)$ and $\delta(\bbX)$ by risk minimization. The second is $\widehat{\Delta}_{\mathrm{DS}}$, which replaces the risk-minimization steps for $\tau(\bbX)$ and $\delta(\bbX)$ in Algorithm~\ref{alg:dcf-estimator} by direct substitution; see \ref{S-sec: alg-DS} for details.

We also evaluated the estimators proposed by \citet{chen2025identification}. Their estimation procedure uses standard cross-fitting, with generalized additive models to estimate nuisance functions and relies on direct substitution for estimating \emph{composite nuisance functions}, making it similar in spirit to $\widehat{\Delta}_{\mathrm{GAM,DS}}$ defined in Supplementary \ref{S-sec:debiasedestimator_eff}. However, their estimators are constructed from influence functions derived under the nonparametric model $\calM_{\np}$, rather than from the corresponding influence function under the semiparametric model $\calM_{\sp}$, which we have shown admits a simpler form. This difference may affect finite-sample performance in three ways. First, the nonparametric influence function contains additional terms, which may increase variability. Second, it involves more complicated nuisance functions, making nuisance estimation more involved. Finally, direct substitution can lead to unstable estimates, as shown below. In our simulation designs, these estimators had substantially poorer finite-sample performance than the proposed estimator $\widehat{\Delta}_{\mathrm{RM}}$. We therefore do not report the corresponding results.

For the treatment mechanism without interaction, we further dichotomize the continuous instrument at its 20th, 50th and 80th percentiles and assess the performance of the proposed estimator. All simulation scenarios are evaluated using 2000 Monte Carlo replicates at sample sizes $n=500, 1000, 2000$, and $5000$ with $K=5$ folds. The reproduction code for this study is publicly available at \url{https://github.com/MeiDongstat/nbiv}.

\begin{figure}[!ht]
    \centering
  \noindent\makebox[\textwidth][c]{
    \includegraphics[width=1\linewidth, height=0.5\textheight, keepaspectratio,
      trim=6pt 6pt 6pt 6pt, clip]{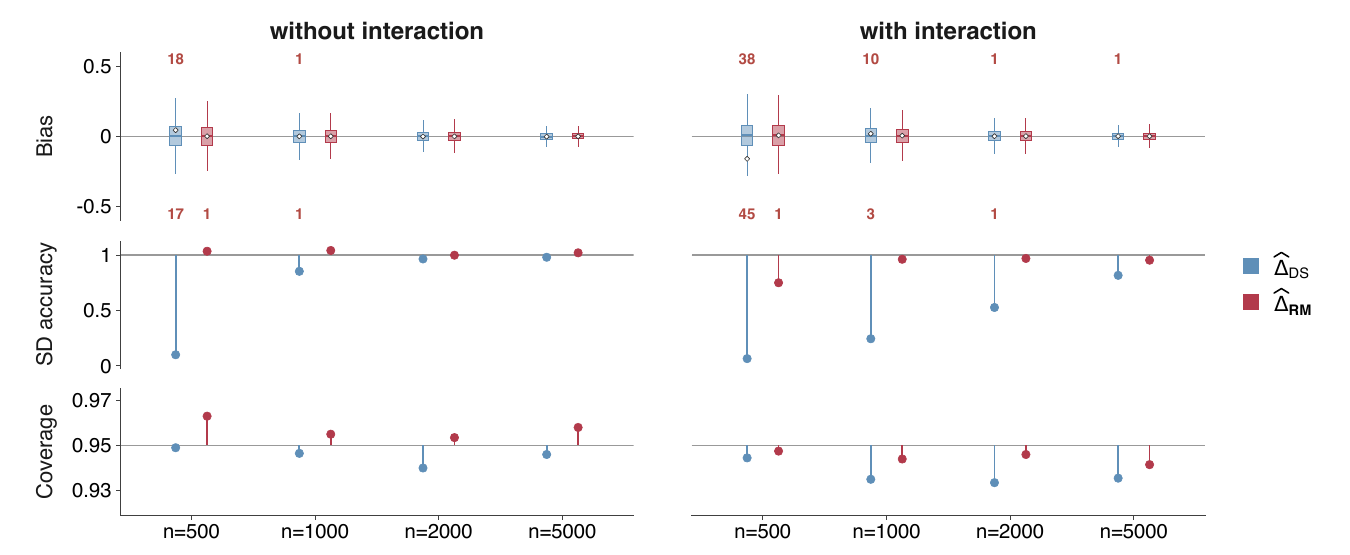}
  }
    \caption{Simulation results for $\widehat{\Delta}_{\mathrm{RM}}$ and $\widehat{\Delta}_{\mathrm{DS}}$ under the treatment mechanism without interaction in the left panel and with interaction in the right panel. The boldface legend entry highlights $\widehat{\Delta}_{\mathrm{RM}}$, our recommended implementation. In the upper panel, red numbers indicate the number of estimates outside the displayed range, either above 0.50 or below $-0.50$. In the middle panel, SD accuracy is defined as the ratio of the average estimated standard error to the Monte Carlo standard deviation. The nominal confidence-interval coverage level is 95\%.}
    \label{fig}
\end{figure}

Figure~\ref{fig} displays the finite-sample performance of the proposed estimator and its implementation variant under the two treatment mechanisms. Numerical summaries are reported in Tables~\ref{S-tab:datagen-binary-if} and~\ref{S-tab:datagen2-binary-if}. Across both settings, the risk-minimization estimator $\widehat{\Delta}_{\mathrm{RM}}$ is stable, with negligible bias, root mean squared error decreasing with sample size, and empirical coverage close to the nominal 95\% level. The SD accuracy is close to one, indicating good agreement between the average estimated standard error and the Monte Carlo standard deviation, apart from some underestimation at $n=500$ under the interaction mechanism.

The direct-substitution estimator $\widehat{\Delta}_{\mathrm{DS}}$ is more sensitive to the treatment mechanism. When the treatment model has no interaction, it performs comparably to $\widehat{\Delta}_{\mathrm{RM}}$ for moderate to large sample sizes, although its root mean squared error is substantially inflated at $n=500$. Under the interaction mechanism, this instability is more pronounced: the mean bias and root mean squared error are large at small sample sizes, and the standard errors substantially underestimate the Monte Carlo variability. In contrast, the median bias remains small, suggesting that the deterioration is driven by a small number of extreme estimates. This behaviour is expected from the data-generating mechanism. The term $ZX_4$ makes the instrument--treatment association more heterogeneous across covariate strata, which in turn makes the denominator nuisance $\delta_{\gamma_0}^{D}(\bbX)$ more variable. Direct substitution can therefore be unstable when the estimated denominator is close to zero. The risk-minimization estimator avoids explicit reciprocal or quotient formation for the composite nuisance and hence is more stable in finite samples.

Simulation results using dichotomized instruments are reported in Tables~\ref{S-tab:datagen-quantile-binary-q20-if1}--\ref{S-tab:datagen-quantile-binary-q80-if1}. Dichotomizing the continuous instrument increases variability relative to using the continuous instrument, especially at the 20th and 80th percentiles, where the induced binary instrument is more imbalanced. The median split is less harmful because it yields a more balanced binary instrument, but it still discards information and does not recover the precision obtained by retaining the full continuous instrument. Across thresholds, the risk-minimization estimators remain more stable than their direct-substitution counterparts, while the latter continue to exhibit occasional extreme behaviour. These findings are consistent with the information loss induced by hard thresholding of a continuous instrument.

\section{Application: the Princess Margaret Cancer Centre Lung Cancer Dataset}\label{sec: app}

We now investigate the ``obesity paradox'', the phenomenon that motivates this work. We are interested in estimating the causal effect of being overweight at diagnosis on two-year mortality among patients with non-small cell lung cancer, a clinically important outcome. This analysis draws on a retrospective cohort from the Princess Margaret Cancer Centre, including patients diagnosed between 1974 and 2013 under a standardized protocol. The sample in the analysis is restricted to individuals with available BMI at diagnosis, stage at diagnosis, vital status at last follow-up, date of death where applicable, and genotyping data. In total, 1395 patients meet these criteria. The germline genotyping is performed using the Illumina Infinium OncoArray-500K BeadChip. During follow-up, we observe 1008 deaths (72.3\%). The median follow-up time among censored patients is 8.8 years, and the median overall survival time is 3.5 years.

We perform standard genotype quality control and exclude 48 patients for any of the following: (1) discordance between genetic and self-reported sex, (2) poor genotyping quality with heterozygosity or missingness $> 5\%$, and (3) excess first- or second-degree relatedness. We also remove 18 patients with missing smoking status and 47 patients lost to follow-up within two years, resulting in 1282 patients for the final analysis. At baseline, 640 participants (49.9\%) are female and 642 (50.1\%) are male; the median age at diagnosis is 66 years, with range 23-95. The disease stage at diagnosis is stage I in 488 patients (38.1\%), stage II in 128 (10.0\%), stage III in 315 (24.6\%), and stage IV in 351 (27.4\%). A total of 118 individuals (9.2\%) are never smokers; the rest are former or current smokers. The final analytic cohort is also highly homogeneous in self-reported ancestry: 1279 patients (99.8\%) self-identify as Caucasian. We use a binary exposure rather than continuous BMI to target a conventional clinical contrast and to avoid imposing a linear relation between BMI and mortality, which may be thresholded or otherwise nonlinear in oncology cohorts. Because only 16.4\% of patients have obesity at diagnosis, defined as BMI $\ge$ 30, we define the exposure $D$ as a binary indicator of being overweight at diagnosis, with BMI $\ge$ 25 if and only if $D = 1$. The outcome $Y$ is two-year mortality, defined as death within two years of diagnosis. In the final cohort, the prevalence of being overweight at diagnosis is 50.6\%, and the two-year mortality rate is 36.0\%. The baseline covariates $\bbX$ used in our analysis include age, sex, smoking status, and disease stage, dichotomized as stage I-II versus stage III-IV. These covariates are selected for their potential to modify the effect of the instrument on the exposure or the effect of the exposure on the outcome. The disease stage is dichotomized because patients with early-stage I-II and late-stage III-IV disease typically receive different treatments.

To answer our causal question, we use BMI-related genetic variants as instruments. To strengthen instrument relevance, we construct a PRS for BMI using summary statistics from the Genetic Investigation of ANthropometric Traits consortium~\citep{yengo2018meta}. We apply the standard clumping-and-thresholding approach with linkage-disequilibrium clumping at $r^2=0.01$ within a 10 Mb window and set the P-value threshold to $1 \times 10^{-8}$ to match the significance level reported in~\citet{yengo2018meta}. One caveat to using the PRS as an instrument is horizontal pleiotropy, which means that genetic variants influence other exposures that may be related to the outcome, thus violating Assumption~\ref{as:ER}.
Since smoking is a well-established prognostic factor for lung cancer survival, we mitigate potential pleiotropy via smoking by excluding variants reported to be significantly associated with any smoking phenotype, including smoking initiation, smoking cessation, cigarettes per day, and age of smoking initiation, among samples of European ancestry in Phase II of the GWAS \& Sequencing Consortium of Alcohol and Nicotine use study \citep{saunders2022genetic}. After filtering and matching with our individual-level data, 860 single-nucleotide polymorphisms are used to construct the PRS, which we standardize for analysis. The F-statistic for the standardized PRS is 29.99, indicating a strong instrument \citep{burgess2011avoiding}.

Alongside the proposed estimator $\widehat{\Delta}_{\mathrm{RM}}$, we report four other estimators: (1) the crude risk difference, $\E[ Y \mid D=1]-\E[ Y \mid D=0]$, which does not account for confounders; (2) the covariate-adjusted risk difference, $\E[\E[ Y \mid D=1, \bbX]-\E[ Y \mid D=0, \bbX]]$, estimated via regression using the R package \texttt{brm} \citep{richardson2017modeling}; (3) the two-stage least squares estimate, with the PRS as a single instrument using the \texttt{AER} package; and (4) the proposed estimator using the PRS dichotomized at the median as an instrument. We apply two-stage least squares with linear models to obtain approximate estimates of the causal risk difference, irrespective of the natural constraint of a binary outcome. We also include age, sex, smoking status, and dichotomized disease stage, in the last three estimators. We repeat the proposed estimator 20 times, each with $K=5$ folds and a different random seed, and report the median point estimate and median standard error to reduce variability from sample splitting \citep{chernozhukov_doubledebiased_2018}. All nuisance models are fitted using SL, with generalized linear models, generalized additive models, penalized generalized linear models, random forests, and XGBoost as candidate learners.

Table~\ref{tab:rd-estimates} reports the estimated risk difference for being overweight at diagnosis on two-year mortality among patients with non-small cell lung cancer. The crude and covariate-adjusted analyses gave negative estimates, consistent with observational reports of an inverse association between BMI and mortality in lung cancer~\citep{zhang2017obesity}. In contrast, the proposed estimator $\widehat{\Delta}_{\mathrm{RM}}$ estimated an increase in two-year mortality risk of 0.300, with a 95\% confidence interval 0.182 to 0.419. This discrepancy suggests residual confounding in associational analyses. Recent studies \citep[e.g.][]{shepshelovich2019body} show that, among never-smokers, overweight status is associated with poorer survival than normal weight, while the opposite is observed among smokers. Because smokers comprise most of the cohort, the overall inverse association is largely driven by the smoker subgroup, suggesting that adjustment only for smoking status or pack-years may not adequately address residual confounding. The median-dichotomized instrument analysis was much less precise, with a substantially wider confidence interval, in line with the simulation evidence in Section~\ref{sec: sim}.

\begin{table}[!ht]
\centering
\caption{Risk-difference estimates for two-year mortality among patients with non-small cell lung cancer}
\label{tab:rd-estimates}
\begin{tabular}{lcc}
Estimator & Estimate & 95\% confidence interval \\
Crude risk difference & $-0.050$ & ($-0.102$, 0.003) \\
Covariate-adjusted risk difference & $-0.040$ & ($-0.085$, 0.006) \\
Two-stage least squares & 0.155 & ($-0.171$, 0.480) \\
$\widehat{\Delta}_{\mathrm{RM}}$ & 0.300 & (0.182, 0.419) \\
$\widehat{\Delta}_{\mathrm{RM}}$, dichotomized instrument & 0.059 & ($-2.667$, 2.784) \\
\end{tabular}
\end{table}

\section{Discussion}\label{sec: discussion}

Motivated by the ``obesity paradox'' in lung cancer, we study the identification and estimation of the average treatment effect for a binary treatment with non-binary instruments. We introduce an average generalized Wald estimand and show that it identifies the average treatment effect for both categorical and continuous instruments. We then characterize the semiparametric tangent space of the observed-data model induced by the instrumental variable assumptions and develop the semiparametric theory for the average generalized Wald estimand. Utilizing an influence function in the semiparametric model that coincides with a locally efficient influence function within the submodel $\calM_{\ch}$, we propose a semiparametric locally efficient debiased estimator whose nuisance functions can be estimated with flexible, data-adaptive machine learning algorithms. The risk-minimization reparameterization of the reciprocal and quotient nuisance terms in the influence function also substantially improves numerical stability.

Our estimates in the real-data application should be interpreted with caution. Despite efforts to exclude smoking-related genetic variants before constructing the PRS, the remaining variants may still be associated with the outcome through alternative pathways, leading to an invalid instrument. Methods for handling invalid instruments will be developed in a separate article.

In a recent paper, \citet{tchetgen2024nudge} establishes nonparametric identification of the conditional counterfactual distribution function for the nudgeable subgroup, $\Pr(Y(d) \leq y \mid D(z=1) \neq D(z=0), \boldsymbol{V})$, where $\boldsymbol{V} \subseteq \bbX$, with a binary instrument. We show that the conditional counterfactual distribution function for the whole population is identifiable with both continuous and categorical instruments; see Supplementary Material~\ref{S-sec: counterfactual_distribution} for details. This enables estimation of a broad class of estimands via appropriate choices of functional. For example, an extension to time-to-event outcomes is promising \citep[e.g.][]{wang2023instrumental}.

\section*{Supplementary material}

The Supplementary Material gives explicit Riesz representers and their connection to Stein's lemma, details of the semiparametrically efficient debiased estimator, an example satisfying the conditional homoskedasticity condition, identification of conditional counterfactual distribution functions, proofs, and additional simulation results.

\section*{Declaration of the use of generative AI and AI-assisted technologies}

During the preparation of this work, the authors used ChatGPT for code organization and grammar checking. After using this tool/service, the authors reviewed and edited the content as necessary and take full responsibility for the content of the publication.

\phantomsection\label{supplementary-material-document}
\bigskip
\clearpage
\begingroup
\spacingset{1}
\begin{center}
{\large\bf Supplementary Material for ``Average Treatment Effect Estimation with Non-binary Instrumental Variables''}

\medskip
Mei Dong\textsuperscript{1}, Lin Liu\textsuperscript{2}, Dingke Tang\textsuperscript{3}, Geoffrey Liu\textsuperscript{4}, Wei Xu\textsuperscript{1} and Linbo Wang\textsuperscript{5}

\medskip
\textsuperscript{1}\small Division of Biostatistics, University of Toronto, Toronto, Ontario M5T 3M7, Canada\\
\textsuperscript{2}Institute of Natural Sciences, Shanghai Jiao Tong University, Shanghai 200240, China\\
\textsuperscript{3}Department of Mathematics and Statistics, University of Ottawa, Ottawa, Ontario K1N 9B4, Canada\\
\textsuperscript{4}Princess Margaret Cancer Centre, University Health Network, Toronto, Ontario M5G 2C4, Canada\\
\textsuperscript{5}Department of Statistical Sciences, University of Toronto, Toronto, Ontario M5G 1X6, Canada
\end{center}
\endgroup

\setcounter{proposition}{0}
\renewcommand\theproposition{S.\arabic{proposition}}
\setcounter{equation}{0}
\renewcommand\theequation{S.\arabic{equation}}
\setcounter{table}{0}
\renewcommand\thetable{S.\arabic{table}}
\setcounter{figure}{0}
\renewcommand\thefigure{S.\arabic{figure}}
\setcounter{algocf}{0}
\renewcommand\thealgocf{S.\arabic{algocf}}
\setcounter{section}{0}
\renewcommand\thesection{S.\arabic{section}}
\setcounter{theorem}{0}
\renewcommand\thetheorem{S.\arabic{theorem}}
\setcounter{remark}{0}
\renewcommand\theremark{S.\arabic{remark}}
\setcounter{lemma}{0}
\renewcommand\thelemma{S.\arabic{lemma}}
\makeatletter
\renewcommand{\theHsection}{S\arabic{section}}
\renewcommand{\theHsubsection}{S\arabic{section}.\arabic{subsection}}
\renewcommand{\theHsubsubsection}{S\arabic{section}.\arabic{subsection}.\arabic{subsubsection}}
\renewcommand{\theHequation}{S\arabic{equation}}
\providecommand{\theHtheorem}{}
\providecommand{\theHlemma}{}
\providecommand{\theHremark}{}
\providecommand{\theHproposition}{}
\renewcommand{\theHtheorem}{S\arabic{theorem}}
\renewcommand{\theHlemma}{S\arabic{lemma}}
\renewcommand{\theHremark}{S\arabic{remark}}
\renewcommand{\theHproposition}{S\arabic{proposition}}
\providecommand{\theHalgocf}{}
\renewcommand{\theHalgocf}{S\arabic{algocf}}
\makeatother

\section{Details of Results in Section~\ref{sec: Riesz Representer}}\label{S-app:riesz-details}

We collect the explicit Riesz representers for the weighted average of pairwise conditional-mean differences and weighted average derivative of conditional mean functions discussed in Section~\ref{sec: Riesz Representer}.

\begin{lemma}\label{S-le:rr}
For categorical $Z$, let $m_\omega(Z, \bbX; h)= \sum_{0 \leq k < j \leq J-1} \omega_{jk}(\bbX)\{h(j,\bbX)-h(k,\bbX)\}$, where $\omega_{jk}(\bbX)\in L^2(P_{\bbX})$ for $0 \leq k < j \leq J-1$. For continuous $Z$, let $m_\omega(Z, \bbX; h) = \omega(Z,\bbX)\dot h(Z,\bbX)$, where $\omega(Z,\bbX)\in L^2(P_{Z,\bbX})$ and $\dot h(z,\bbx) = \partial h(z,\bbx)/ \partial z$ exists and $\dot{h}(Z, \bbX) \in L^2(P_{Z,\bbX})$.  Suppose the following assumptions hold.
\begin{enumerate}[label={A\arabic*}, leftmargin=*, labelsep=0.5em]
\item \emph{Categorical $Z$.}
There exists $\varepsilon>0$ such that $p(Z=j\mid\bbX)\in(\varepsilon,1-\varepsilon)$ almost surely for each $j\in\{0,1,\ldots,J-1\}$.
\label{S-A1}
\item \emph{Continuous $Z$.}
(a) $\omega(z,\bbx)p(z\mid\bbx)$ is continuously differentiable in $z$; (b) $\omega(z, \bbx)p (z \mid \bbx)=0$ for $z$ on the boundary of support $\Omega(\bbx) \coloneqq \{z: p(z \mid \bbx) >0\}$; (c) $\partial_z\omega(Z,\bbX) + \omega(Z,\bbX)\partial_z\log p(Z\mid\bbX)\in L^2(P_{Z,\bbX})$.
\label{S-A2}
\end{enumerate}
Then for every bounded $h(Z, \bbX) \in L^2(P_{Z,\bbX})$, $ \E[m_\omega(Z, \bbX; h)] = \E[\gamma_\omega(Z, \bbX) h(Z, \bbX)]$ and $ \E[m_\omega(Z, \bbX; h) \mid \bbX] = \E[\gamma_\omega(Z, \bbX) h(Z, \bbX) \mid \bbX]$ hold. Specifically,
\begin{itemize}
\item[(1)] when $Z$ is categorical, under Assumption~\ref{S-A1},
\begin{equation*}
    \gamma_{\omega}(z,\bbx)
    =
    \dfrac{\sum_{k < z} \omega_{zk}(\bbx)-\sum_{k > z}\omega_{kz}(\bbx)}
    {p(z\mid\bbx)}.
\end{equation*}
\item[(2)] when $Z$ is continuous, under Assumption~\ref{S-A2},
\begin{equation*}
    \gamma_{\omega}(z,\bbx)
    =
    -\dfrac{\partial\omega(z,\bbx)}{\partial z}
    -\omega(z,\bbx)\dfrac{\partial\log p(z\mid\bbx)}{\partial z}.
\end{equation*}
\end{itemize}
\end{lemma}

\subsection{Connection to Stein's lemma}\label{S-app:stein-connection}
The representation results above can be viewed as conditional versions of Stein's lemma. Conditional on $\bbX$, any square-integrable function $\gamma(Z,\bbX)$ satisfying $\E[\gamma(Z,\bbX)\mid\bbX]=0$ defines a linear operator of the regression curve $h(z,\bbx)$ through
\[
    h \mapsto \E[\gamma(Z,\bbX)h(Z,\bbX)\mid \bbX].
\]
For categorical instruments, this functional annihilates constants and can therefore be written as a weighted sum of pairwise conditional-mean differences. For example, fixing category $0$ as a baseline and setting
\[
    \omega_{j0}(\bbx)=p(j\mid\bbx)\gamma(j,\bbx),\quad j=1,\ldots,J-1,
\]
with all other pairwise weights equal to zero gives
\[
    \E[\gamma(Z,\bbX)h(Z,\bbX) \mid \bbX=\bbx]
    = \sum_{j=1}^{J-1}\omega_{j0}(\bbx)\{h(j,\bbx)-h(0,\bbx)\} .
\]
This representation is not unique when all pairwise weights are allowed: different collections of pairwise weights can induce the same net coefficient on each instrument level. Uniqueness is recovered only after fixing a difference basis, such as baseline or adjacent differences.

For continuous instruments, the analogous representation is through a derivative operator. A conditional analogue of \citet[][Theorem 1]{S-hines2026parameterizing} implies that a conditionally mean-zero $\gamma(Z, \bbX)$, that is, any $\gamma \in L^2(P_{Z, \bbX})$ satisfying $\E[\gamma(Z,\bbX)\mid \bbX]=0$, induces the weight in \eqref{omega} such that
\begin{equation}\label{S-eqn: conditionallf}
   \E[\gamma(Z,\bbX)h(Z,\bbX) \mid \bbX]
    =    \E[\omega(Z,\bbX)\dot h(Z,\bbX) \mid \bbX].
\end{equation}
Proof of \eqref{S-eqn: conditionallf} is provided in Supplementary \ref{S-app:le:crrr}. The weight satisfies the zero-boundary condition, that is, $\lim_{z \to  \inf\Omega(\bbx)}\omega(z,\bbx)p(z\mid\bbx) = \lim_{z \to \sup\Omega(\bbx)}\omega(z,\bbx)p(z\mid\bbx)=0$. Within the class of weights satisfying this zero-boundary condition, the induced weight is unique: if two weights generate the same derivative operator for every smooth $h$, then their difference multiplied by $p(z\mid\bbx)$ must be constant in $z$, and the boundary condition forces that constant to be zero.

This also clarifies the connection with Stein's lemma. Recall that Stein's lemma states that a random variable $W$ is
normally distributed with mean $\E[W]$ and variance $\var(W)$ if and only if, for every absolutely
continuous $g:\mathbb{R}\to\mathbb{R}$ with $\E[|\dot{g}(W)|]<\infty$,
\[
    \E\!\left[\dfrac{W-\E(W)}{\var(W)}g(W) \right]
    =
    \E[\dot{g}(W)],
\]
where $\dot{g}(w)=\partial g(w)/\partial w$. A conditional version of
Stein's lemma is the Gaussian special case of \eqref{S-eqn: conditionallf} \citep{S-stein1972bound}. Taking $\omega(z,\bbx)=1$ gives $\gamma(z,\bbx)=-\partial\log p(z\mid\bbx)/\partial z$. If $Z\mid\bbX=\bbx$ is normal with mean $m(\bbx)$ and variance $\sigma^2(\bbx)$, then $\gamma(z,\bbx)=\{z-m(\bbx)\}/\sigma^2(\bbx)$. Therefore, \eqref{S-eqn: conditionallf} becomes
\[
    \E\left[\dfrac{Z-m(\bbX)}{\sigma^2(\bbX)}h(Z,\bbX)\mid\bbX\right]
    =
    \E[\dot h(Z,\bbX)\mid\bbX],
\]
or equivalently $\E[\{Z-m(\bbX)\} h(Z,\bbX)\mid\bbX]=\sigma^2(\bbX)\E[\dot h(Z,\bbX)\mid\bbX]$. Conversely, taking $\gamma(z, \bbx) = \{z-m(\bbx)\}/\sigma^2(\bbx)$ in the Gaussian case, one can verify that the induced weight $\omega(z, \bbx) =1$, reducing to Stein's identity.

\newpage

\section{Semiparametrically Efficient Debiased Estimator}\label{S-sec:debiasedestimator_eff}
In this section, we construct a semiparametric efficient debiased estimator for the average treatment effect $\Delta$, denoted by $\widehat{\Delta}_{\,\eff}$, based on the efficient influence function $\Psi_{\eff}(\bbO; P)$.

Unlike the double cross-fitting procedure considered in Section~\ref{sec: debiasedestimator}, we use standard cross-fitting \citep{S-chernozhukov_doubledebiased_2018}: for each $k=1,\ldots,K$, all nuisance functions in $\Psi_{\eff}(\bbO;P)$ are estimated using observations in the complement $\mathcal{I}_k^c$ of the held-out fold, and the estimating equation is evaluated on the held-out fold $\mathcal{I}_k$. This choice is motivated mainly by practical considerations. Because the nuisance functions in $\Psi_{\eff}(\bbO;P)$ are nested through several layers of generated quantities, fully avoiding sample reuse would require a four-layer cross-fitting scheme, which is impractical. As a starting point, we estimate all nuisance functions involving reciprocals and quotients by direct substitution and establish the asymptotic theory of the corresponding estimator. At the end of this section, we discuss how some of these terms can instead be estimated by risk minimization. Finally, we compare the finite-sample performance of estimators based on $\Psi_{\eff}(\bbO;P)$ with that of estimators based on $\Psi_{\gamma_0}(\bbO;P)$.

The following algorithm describes the constructed semiparametric efficient debiased estimator using standard cross-fitting.

\begin{algorithm}[H]
\caption{Cross-fitted semiparametric efficient debiased estimator with direct-substitution}
\label{S-alg:eff-estimator}
\DontPrintSemicolon
\KwIn{Observed data $\{\bbO_i=(Y_i,D_i,Z_i,\bbX_i):i=1,\ldots,n\}$ and number of folds $K$.}
\KwOut{$\widehat{\Delta}^{\eff}_{\mathrm{DS}}$.}

Randomly split $\mathcal I=\{1,\ldots,n\}$ into $K$ disjoint folds $\mathcal I_1,\ldots,\mathcal I_K$.\;

\For{$k=1,\ldots,K$}{
Set $\mathcal I_k^c=\mathcal I\setminus\mathcal I_k$.\;

Using observations in $\mathcal I_k^c$, estimate $\mu^D(Z,\bbX)$, $\mu^Y(\bbX)$, and $\mu^D(\bbX)$; denote the fitted functions by $\widehat{\mu}^D_{(-k)}(Z,\bbX)$, $\widehat{\mu}^Y_{(-k)}(\bbX)$, and $\widehat{\mu}^D_{(-k)}(\bbX)$.\;

Using observations in $\mathcal I_k^c$, regress
$\{\widehat{\mu}^D_{(-k)}(Z,\bbX)-\widehat{\mu}^D_{(-k)}(\bbX)\}
\{D-\widehat{\mu}^D_{(-k)}(\bbX)\}$
on $\bbX$ and denote the fitted function by $\widehat{\delta}^D_{(-k)}(\bbX)$.\;

Using observations in $\mathcal I_k^c$, regress
$\{\widehat{\mu}^D_{(-k)}(Z,\bbX)-\widehat{\mu}^D_{(-k)}(\bbX)\}
\{Y-\widehat{\mu}^Y_{(-k)}(\bbX)\}$
on $\bbX$ and denote the fitted function by $\widehat{\delta}^Y_{(-k)}(\bbX)$.\;

Set
$\widehat{\delta}_{(-k)}(\bbX)
=
\widehat{\delta}^Y_{(-k)}(\bbX)/
\widehat{\delta}^D_{(-k)}(\bbX)$.\;

Define
$\widehat{\calE}_{(-k)}
=
Y-\widehat{\mu}^{Y}_{(-k)}(\bbX)
-\widehat{\delta}_{(-k)}(\bbX)
\{D-\widehat{\mu}^{D}_{(-k)}(\bbX)\}$.\;

Using observations in $\mathcal I_k^c$, regress $\widehat{\calE}_{(-k)}^2$ on $(Z,\bbX)$ and denote the fitted function by $\widehat{\sigma}^2_{(-k)}(Z,\bbX)$.\;

Using observations in $\mathcal I_k^c$, regress
$1/\widehat{\sigma}^2_{(-k)}(Z,\bbX)$,
$\widehat{\mu}^D_{(-k)}(Z,\bbX)/\widehat{\sigma}^2_{(-k)}(Z,\bbX)$, and
$\{\widehat{\mu}^D_{(-k)}(Z,\bbX)\}^2/\widehat{\sigma}^2_{(-k)}(Z,\bbX)$
on $\bbX$; denote the fitted functions by
$\widehat{\kappa}_{1,(-k)}(\bbX)$,
$\widehat{\kappa}_{2,(-k)}(\bbX)$, and
$\widehat{\kappa}_{3,(-k)}(\bbX)$.\;

Evaluate, for each $i\in\mathcal I_k$,
\[
\widehat{\psi}_{i,k}=
\dfrac{
\widehat{\kappa}_{1,(-k)}(\bbX_i)
\dfrac{\widehat{\mu}^D_{(-k)}(Z_i,\bbX_i)}
{\widehat{\sigma}^2_{(-k)}(Z_i,\bbX_i)}
-
\dfrac{1}{\widehat{\sigma}^2_{(-k)}(Z_i,\bbX_i)}
\widehat{\kappa}_{2,(-k)}(\bbX_i)}
{
\widehat{\kappa}_{1,(-k)}(\bbX_i)
\widehat{\kappa}_{3,(-k)}(\bbX_i)
-
\{\widehat{\kappa}_{2,(-k)}(\bbX_i)\}^2
}
\widehat{\calE}_{i,(-k)}
+
\widehat{\delta}_{(-k)}(\bbX_i).
\]
}

Return
$\widehat{\Delta}^{\,\eff}_{\mathrm{DS}} = n^{-1}\sum_{k=1}^K\sum_{i\in\mathcal I_k}\widehat{\psi}_{i,k}$.\;
\end{algorithm}
Equivalently, if $\widehat{P}_{(-k)}$ denotes the collection of fold-specific nuisance estimators constructed in Algorithm~\ref{S-alg:eff-estimator}, then $\widehat{\psi}_{i,k} = \psi(\bbO_i; \widehat{P}_{(-k)})$, where $\psi$ is the uncentered efficient influence function satisfying $\Psi_{\eff} (\bbO; P) = \psi(\bbO; P)  - \Delta$.
Again, before stating the asymptotic result of $\widehat{\Delta}^{\eff}_{\mathrm{DS}}$, we first state the assumptions needed to establish the asymptotic theory of the proposed estimator.
\setcounter{supassumption}{0}
\begin{itemize}
    \item [\refstepcounter{supassumption}\label{S-as:eif-positivity}\thesupassumption.] \emph{Positivity}. There exist constants $c_1, c_2, c_3 >0 $ such that $\sigma^2(Z, \bbX) > c_1$, $|\kappa_1(\bbX) \kappa_3(\bbX) - (\kappa_2(\bbX))^2 |>c_2$, and $|\delta^D(\bbX)|>c_3$ for all $\bbX \in \calX$. Moreover, $\widehat{\sigma}^2_{(-k)}(Z, \bbX)>c_1$, $|\widehat{\kappa}_{1, (-k)}(\bbX) \widehat{\kappa}_{3, (-k)}(\bbX) - (\widehat{\kappa}_{2, (-k)}(\bbX))^2 |>c_2$ , and $|\widehat{\delta}^D_{(-k)}(\bbX)|>c_3$ for all $\bbX \in \calX$ almost surely.
    \item [\refstepcounter{supassumption}\label{S-as:eif-boundedness}\thesupassumption.] \emph{Boundedness}. $\|Y\|_{\infty} < \infty$, $\|\mu^Y(\bbX)\|_{\infty} < \infty$, $\|\delta(\bbX)\|_{\infty} < \infty$, $\|\kappa_j(\bbX)\|_{\infty} < \infty$ for $j=1, 2, 3$; $\|\widehat{\delta}_{(-k)}(\bbX)\|_{\infty} = O_p(1)$, $\|\widehat{\mu}^D_{(-k)}(Z,\bbX)\|_{\infty}=O_p(1)$, $\|\widehat{\mu}^D_{(-k)}(\bbX)\|_{\infty}=O_p(1)$, $\|\widehat{\mu}^Y_{(-k)}(\bbX)\|_{\infty}=O_p(1)$, and $\|\widehat{\kappa}_{j, (-k)}(\bbX)\|_{\infty} = O_p(1)$ for $j=1, 2, 3$.
    \item [\refstepcounter{supassumption}\label{S-as:eif-consistency}\thesupassumption.] \emph{Consistency}. $\|\widehat{\mu}^D_{(-k)}(Z, \bbX) - \mu^D(Z, \bbX)\|_{P, 2} + \|\widehat{\kappa}_{1, (-k)}(\bbX) - \kappa_1(\bbX)\|_{P, 2} + \|\widehat{\kappa}_{2, (-k)}(\bbX) - \kappa_2(\bbX)\|_{P, 2} + \|\widehat{\kappa}_{3, (-k)}(\bbX) - \kappa_3(\bbX)\|_{P, 2} + \|\widehat{\mu}^D_{(-k)}(\bbX) - \mu^D(\bbX)\|_{P, 2} + \|\widehat{\mu}^Y_{(-k)}(\bbX) - \mu^Y(\bbX)\|_{P, 2} + \|\widehat{\delta}_{(-k)}(\bbX) - \delta(\bbX)\|_{P, 2} + \|\widehat{\sigma}^2_{(-k)}(Z, \bbX) - \sigma^2(Z, \bbX) \|_{P, 2} = o_p(1)$.
    \item [\refstepcounter{supassumption}\label{S-as:eif-rate}\thesupassumption.] \emph{Rate conditions}. $\big(\|\widehat{\mu}^D_{(-k)}(Z, \bbX) - \mu^D(Z, \bbX)\|_{P, 2} + \|\widehat{\kappa}_{1, (-k)}(\bbX) - \kappa_1(\bbX)\|_{P, 2} + \|\widehat{\kappa}_{2, (-k)}(\bbX) - \kappa_2(\bbX)\|_{P, 2} + \|\widehat{\sigma}^2_{(-k)}(Z, \bbX) - \sigma^2(Z, \bbX) \|_{P, 2} \big)\big(\|\widehat{\mu}^D_{(-k)}(\bbX) - \mu^D(\bbX)\|_{P, 2} + \|\widehat{\mu}^Y_{(-k)}(\bbX) - \mu^Y(\bbX)\|_{P, 2} + \|\widehat{\delta}_{(-k)}(\bbX) - \delta(\bbX)\|_{P, 2}\big) = o_p(n^{-1/2})$ and $\big(\|\widehat{\kappa}_{1, (-k)}(\bbX) - \kappa_1(\bbX)\|_{P, 2} + \|\widehat{\kappa}_{2, (-k)}(\bbX) - \kappa_2(\bbX)\|_{P, 2} + \|\widehat{\kappa}_{3, (-k)}(\bbX) - \kappa_3(\bbX)\|_{P, 2}\big)\|\widehat{\delta}_{(-k)}(\bbX) - \delta(\bbX)\|_{P, 2} = o_p(n^{-1/2})$.
\end{itemize}
Condition~\ref{S-as:eif-positivity} ensures that all nuisance functions and their estimators that appear in denominators are bounded away from zero. Condition~\ref{S-as:eif-boundedness} requires that the outcome, the nuisance functions, and their estimators be uniformly bounded. Condition~\ref{S-as:eif-consistency} requires consistency of all nuisance estimators in the $L^2(P)$ norm. Finally, condition~\ref{S-as:eif-rate} is a second-order product-rate condition. Both condition~\ref{S-as:eif-consistency} and condition~\ref{S-as:eif-rate} are high-level conditions because several nuisance functions are generated regressions built from preliminary nuisance estimates. For example, $\delta(\bbX)$ is constructed from regressions involving $\widehat\mu^D(Z,\bbX)$, $\widehat\mu^D(\bbX)$ and $\widehat\mu^Y(\bbX)$; $\widehat\sigma^2(Z,\bbX)$ is then constructed from residuals involving $\widehat\delta(\bbX)$; and the nuisance functions $\widehat\kappa_1(\bbX)$, $\widehat\kappa_2(\bbX)$ and $\widehat\kappa_3(\bbX)$ are further constructed from $\widehat\sigma^2(Z,\bbX)$ and $\widehat\mu^D(Z,\bbX)$. With standard cross-fitting, these \emph{composite nuisance functions} are fitted on the same training sample. Consequently, verifying consistency and rate requirements in Conditions~\ref{S-as:eif-consistency} and~\ref{S-as:eif-rate} requires controlling empirical-process terms induced by reusing the same observations across these generated regression stages. One sufficient route is to impose Donsker-type conditions on the nuisance classes and the induced generated-outcome loss classes, together with suitable oracle convergence rates for the corresponding regression procedures; see Supplementary Material~\ref{S-app:samplesplit} for a detailed example.

The following theorem establishes that $\widehat{\Delta}^{\eff}_{\mathrm{DS}}$ is asymptotically normal and efficient in $\calM_{\sp}$ under Conditions~\ref{S-as:eif-positivity}--\ref{S-as:eif-rate}.
\begin{theorem}\label{S-th:Delta_eff}
    Suppose Conditions~\ref{S-as:eif-positivity}--\ref{S-as:eif-rate} hold. Then, under $\calM_{\sp}$, $\sqrt{n}(\widehat{\Delta}^{\eff}_{\mathrm{DS}} - \Delta) \to_d N(0, \mathcal{V}_{\eff})$, where $\mathcal{V}_{\eff} := \var\{\Psi_{\eff}(\bbO;P)\}$ is the semiparametric efficiency bound. Moreover, a consistent estimator of $\mathcal{V}_{\eff}$ is given by
\begin{align*}
    \widehat{\mathcal{V}}_{\eff} = \dfrac{1}{n} \sum_{k=1}^K \sum_{i \in \mathcal{I}_k} \left\{ \psi(\bbO_i; \widehat{P}_{(-k)}) - \widehat{\Delta}^{\eff}_{\mathrm{DS}}\right\}^2.
\end{align*}
\end{theorem}

In Algorithm~\ref{S-alg:eff-estimator}, the reciprocal and quotient nuisance functions are obtained by direct substitution. Specifically, $\delta(\bbX)$ is formed as a ratio of two fitted conditional expectations, and $\{\sigma^2(Z,\bbX)\}^{-1}$ is obtained by inverting a fitted conditional variance. To improve numerical stability, we also consider a risk-minimization version of the same efficient-influence-function-based estimator. In particular, the reciprocal of $\sigma^2(Z,\bbX) = \E[\calE^2 \mid Z, \bbX]$ admits a direct risk-minimization characterization. Define
\begin{align}\label{S-eqn:riskminimization3}
    g_3^*= \argmin_{g_3 \in L^2(P_{Z,\bbX})} \E \left[  \calE^2 \left\{\frac{1}{\calE^2} - g_3(Z,\bbX)\right\}^2 \right].
\end{align}
Then $ g_3^*(Z,\bbX)=\{\sigma^2(Z,\bbX)\}^{-1}$. Therefore, $\{\sigma^2(Z,  \bbX)\}^{-1}$ can be estimated by minimizing the empirical analogue of \eqref{S-eqn:riskminimization3}, with $\calE$ replaced by its fitted residual. The quotient $\delta(\bbX)$ is estimated analogously using the risk-minimization reparameterization described in Section~\ref{sec: debiasedestimator}. We denote the resulting risk-minimization variant by $\widehat{\Delta}_{\mathrm{RM}}^{\eff}$, which replaces the direct-substitution steps for $\delta(\bbX)$ and $\{\sigma^2(Z,\bbX)\}^{-1}$ by the corresponding risk-minimization regressions in Algorithm~\ref{S-alg:eff-estimator}.

\subsection{Simulation Results}
We assess the finite-sample performance of the proposed semiparametrically efficient debiased estimators $\widehat{\Delta}_{\mathrm{DS}}^{\eff}$ and $\widehat{\Delta}_{\mathrm{RM}}^{\eff}$ using the simulation settings described in Section~\ref{sec: sim}. To compare their performance with that of estimators based on  the locally efficient influence function $\Psi_{\gamma_0}(\bbO;P)$, we further construct two estimators based on $\Psi_{\gamma_0}(\bbO;P)$, both using standard cross-fitting. Specifically, all nuisance functions $\widehat{\mu}^D_{(-k)}(Z, \bbX)$, $\widehat{\mu}^Y_{(-k)}(\bbX)$, $\widehat{\mu}^D_{(-k)}(\bbX)$, $\widehat{\tau}_{(-k)}(\bbX)$, and $\widehat{\delta}_{(-k)}(\bbX)$ are estimated using observations in the complement of the held-out fold $\mathcal{I}_{k}^c$ for $k=1, \ldots, K$. The first estimator is $\widehat{\Delta}_{\mathrm{SCF,DS}}$, which replaces the double cross-fitting procedure in Algorithm~\ref{S-alg:dcf-estimator-ds} with standard cross-fitting.  The second is $\widehat{\Delta}_{\mathrm{SCF,RM}}$, which replaces the double cross-fitting procedure in Algorithm~\ref{alg:dcf-estimator} with standard cross-fitting. The subscript ``SCF'' denotes standard cross-fitting and is used to distinguish these estimators from their double-cross-fitted counterparts.

All nuisance functions are estimated with generalized additive models, abbreviated as GAMs~\citep{S-hastie1990generalized}. The GAMs are fitted using the \texttt{gam()} function in the \texttt{mgcv} package in \textsf{R}. All simulation scenarios are evaluated using 2000 Monte Carlo replicates at sample sizes $n=500, 1000, 2000$, and $5000$ with $K=5$ folds.

Tables~\ref{S-tab:datagen-binary-eif} and~\ref{S-tab:datagen2-binary-eif} report the numerical results. In the relatively simple setting where the treatment mechanism does not include an interaction term, the locally efficient risk-minimization estimator $\widehat{\Delta}_{\mathrm{SCF,RM}}$ performs well across sample sizes, while the direct-substitution estimator $\widehat{\Delta}_{\mathrm{SCF,DS}}$ performs comparably only at larger sample sizes. By contrast, the semiparametrically efficient estimators perform poorly even in this setting. Although both $\widehat{\Delta}_{\mathrm{DS}}^{\eff}$ and $\widehat{\Delta}_{\mathrm{RM}}^{\eff}$ are generally median-centred, their finite-sample distributions are heavy-tailed, as reflected by the discrepancies between mean and median bias and by the large root mean squared errors. The risk-minimization version, $\widehat{\Delta}_{\mathrm{RM}}^{\eff}$, reduces some of this instability relative to direct substitution, but substantial tail behaviour remains. Both $\widehat{\Delta}_{\mathrm{DS}}^{\eff}$ and $\widehat{\Delta}_{\mathrm{RM}}^{\eff}$ have large average estimated standard errors and empirical coverage above the nominal 95\% level. However, the SD accuracies are far below one, suggesting that the fitted intervals are wide and the inferences are not reliable.

In the more challenging setting with an interaction in the treatment mechanism, all four standard-cross-fitted estimators deteriorate, although $\widehat{\Delta}_{\mathrm{SCF,RM}}$ remains the most stable among them. However, its coverage probability decreases as the sample size increases, indicating that standard cross-fitting with generalized additive models does not adequately control the error induced by the generated nuisance regressions in this setting. These simulation results show that the estimators constructed based on the efficient influence function may not perform well even under relative simple data-generating process.

\begin{table}[ht]
\centering
\caption{Simulation results of the semiparametric efficient and locally efficient debiased estimators under the treatment mechanism without interaction. Nuisance functions are all fitted with generalized additive models}
\label{S-tab:datagen-binary-eif}
\begin{tabular}{@{}c l r r r r r r r@{}}
n & Estimator & Bias & Median bias & MCSD & Avg. SE & SD acc.$^*$ & RMSE & Coverage \\[2pt]
\multirow{4}{*}{\centering $500$} & $\widehat{\Delta}_{\mathrm{DS}}^{\mathrm{eff}}$ & -1.443 & 0.025 & 1.587 & 5.145 & 0.072 & 70.989 & 0.973 \\
 & $\widehat{\Delta}_{\mathrm{RM}}^{\mathrm{eff}}$ & -0.956 & 0.005 & 0.979 & 3.400 & 0.078 & 43.768 & 0.973 \\
 & $\widehat{\Delta}_{\mathrm{SCF,DS}}$ & 32.961 & 0.014 & 29.778 & 61.473 & 0.046 & 1332 & 0.968 \\
 & $\widehat{\Delta}_{\mathrm{SCF,RM}}$ & 0.005 & 0.003 & 0.002 & 0.096 & 1.035 & 0.093 & 0.957 \\[4pt]
\multirow{4}{*}{\centering $1000$} & $\widehat{\Delta}_{\mathrm{DS}}^{\mathrm{eff}}$ & $0.311$ & $-0.038$ & $1.032$ & $4.187$ & $0.091$ & $46.121$ & $0.974$ \\
 & $\widehat{\Delta}_{\mathrm{RM}}^{\mathrm{eff}}$ & $-0.013$ & $0.011$ & $0.199$ & $1.928$ & $0.216$ & $8.911$ & $0.984$ \\
 & $\widehat{\Delta}_{\mathrm{SCF, DS}}$ & $0.198$ & $0.005$ & $0.321$ & $0.694$ & $0.048$ & $14.365$ & $0.955$ \\
 & $\widehat{\Delta}_{\mathrm{SCF, RM}}$ & $0.003$ & $0.003$ & $0.001$ & $0.064$ & $1.019$ & $0.063$ & $0.952$ \\[4pt]
\multirow{4}{*}{\centering $2000$} & $\widehat{\Delta}_{\mathrm{DS}}^{\mathrm{eff}}$ & $0.195$ & $-0.005$ & $0.290$ & $2.576$ & $0.199$ & $12.973$ & $0.975$ \\
 & $\widehat{\Delta}_{\mathrm{RM}}^{\mathrm{eff}}$ & $0.122$ & $-0.031$ & $0.259$ & $2.264$ & $0.195$ & $11.596$ & $0.983$ \\
 & $\widehat{\Delta}_{\mathrm{SCF, DS}}$ & $0.763$ & $0.001$ & $0.757$ & $0.813$ & $0.024$ & $33.858$ & $0.944$ \\
 & $\widehat{\Delta}_{\mathrm{SCF, RM}}$ & $0.001$ & $0.002$ & $0.001$ & $0.044$ & $0.987$ & $0.045$ & $0.943$ \\[4pt]
\multirow{4}{*}{\centering $5000$} & $\widehat{\Delta}_{\mathrm{DS}}^{\mathrm{eff}}$ & $0.632$ & $0.002$ & $0.481$ & $3.776$ & $0.175$ & $21.534$ & $0.981$ \\
 & $\widehat{\Delta}_{\mathrm{RM}}^{\mathrm{eff}}$ & $0.182$ & $-0.014$ & $0.250$ & $2.499$ & $0.224$ & $11.160$ & $0.982$ \\
 & $\widehat{\Delta}_{\mathrm{SCF, DS}}$ & $-0.002$ & $-0.002$ & $0.001$ & $0.028$ & $0.990$ & $0.028$ & $0.952$ \\
 & $\widehat{\Delta}_{\mathrm{SCF, RM}}$ & $0.001$ & $0.000$ & $0.001$ & $0.028$ & $0.988$ & $0.028$ & $0.953$ \\[2pt]
\multicolumn{9}{@{}p{0.92\textwidth}@{}}{\footnotesize $^*$ SD accuracy is the ratio of the average estimated standard error to the Monte Carlo standard deviation.}
\end{tabular}
\end{table}

\begin{table}[ht]
\centering
\caption{Simulation results of the semiparametric efficient and locally efficient debiased estimators under the treatment mechanism with interaction. Nuisance functions are all fitted with generalized additive models}
\label{S-tab:datagen2-binary-eif}
\begin{tabular}{@{}c l r r r r r r r@{}}
n & Estimator & Bias & Median bias & MCSD & Avg. SE & SD acc.$^*$ & RMSE & Coverage \\[2pt]
\multirow{4}{*}{\centering $500$} & $\widehat{\Delta}_{\mathrm{DS}}^{\mathrm{eff}}$ & -0.233 & 0.038 & 1.482 & 8.266 & 0.125 & 66.254 & 0.975 \\
 & $\widehat{\Delta}_{\mathrm{RM}}^{\mathrm{eff}}$ & -0.249 & 0.034 & 0.966 & 3.361 & 0.078 & 43.212 & 0.975 \\
 & $\widehat{\Delta}_{\mathrm{SCF, DS}}$ & 761 & 0.078 & 2207 & 3290 & 0.033 & 98680 & 0.972 \\
 & $\widehat{\Delta}_{\mathrm{SCF, RM}}$ & 0.028 & 0.026 & 0.002 & 0.096 & 0.930 & 0.106 & 0.929 \\[4pt]
\multirow{4}{*}{\centering $1000$} & $\widehat{\Delta}_{\mathrm{DS}}^{\mathrm{eff}}$ & $-1.159$ & $0.054$ & $2.766$ & $11.054$ & $0.089$ & $124$ & $0.978$ \\
 & $\widehat{\Delta}_{\mathrm{RM}}^{\mathrm{eff}}$ & $-0.177$ & $0.003$ & $0.205$ & $2.082$ & $0.227$ & $9.176$ & $0.973$ \\
 & $\widehat{\Delta}_{\mathrm{SCF, DS}}$ & $-221576$ & $0.072$ & $221513$ & $221841$ & $0.022$ & $9906381$ & $0.975$ \\
 & $\widehat{\Delta}_{\mathrm{SCF, RM}}$ & $0.030$ & $0.030$ & $0.002$ & $0.063$ & $0.881$ & $0.078$ & $0.888$ \\[4pt]
\multirow{4}{*}{\centering $2000$} & $\widehat{\Delta}_{\mathrm{DS}}^{\mathrm{eff}}$ & $-0.540$ & $0.006$ & $1.001$ & $8.542$ & $0.191$ & $44.779$ & $0.977$ \\
 & $\widehat{\Delta}_{\mathrm{RM}}^{\mathrm{eff}}$ & $3648$ & $0.030$ & $3651$ & $3656$ & $0.022$ & $163299$ & $0.975$ \\
 & $\widehat{\Delta}_{\mathrm{SCF, DS}}$ & $-656683$ & $0.069$ & $660694$ & $676714$ & $0.023$ & $29547043$ & $0.964$ \\
 & $\widehat{\Delta}_{\mathrm{SCF, RM}}$ & $0.035$ & $0.034$ & $0.001$ & $0.043$ & $0.845$ & $0.062$ & $0.822$ \\[4pt]
\multirow{4}{*}{\centering $5000$} & $\widehat{\Delta}_{\mathrm{DS}}^{\mathrm{eff}}$ & $5.097$ & $-0.002$ & $2.493$ & $16.721$ & $0.150$ & $112$ & $0.977$ \\
 & $\widehat{\Delta}_{\mathrm{RM}}^{\mathrm{eff}}$ & $-47.564$ & $0.012$ & $46.314$ & $49.828$ & $0.024$ & $2071$ & $0.978$ \\
 & $\widehat{\Delta}_{\mathrm{SCF, DS}}$ & $-181$ & $0.064$ & $2091$ & $5176$ & $0.055$ & $93503$ & $0.972$ \\
 & $\widehat{\Delta}_{\mathrm{SCF, RM}}$ & $0.042$ & $0.041$ & $0.001$ & $0.027$ & $0.831$ & $0.053$ & $0.626$ \\[2pt]
\multicolumn{9}{@{}p{0.92\textwidth}@{}}{\footnotesize $^*$ SD accuracy is the ratio of the average estimated standard error to the Monte Carlo standard deviation.}
\end{tabular}
\end{table}

\clearpage

\section[Algorithm of Delta_DS]{Algorithm of $\widehat{\Delta}_{\mathrm{DS}}$} \label{S-sec: alg-DS}
This section collects the algorithm for $\widehat{\Delta}_{\mathrm{DS}}$ that estimates the reciprocal $\tau(\bbX)$ and the quotient $\delta(\bbX)$ by direct substitutions.

\begin{algorithm}[H]
\caption{Double-cross-fitted locally efficient debiased estimator with direct substitution}
\label{S-alg:dcf-estimator-ds}
\DontPrintSemicolon
\KwIn{Observed data $\{\bbO_i=(Y_i,D_i,Z_i,\bbX_i):i=1,\ldots,n\}$ and number of folds $K$.}
\KwOut{$\widehat\Delta_{\mathrm{DS}}$.}
Randomly split $\mathcal I=\{1,\ldots,n\}$ into $K$ disjoint folds $\mathcal I_1,\ldots,\mathcal I_K$.\;
\For{$k=1,\ldots,K$}{
Set $\mathcal I_k^c=\mathcal I\setminus\mathcal I_k$, and split $\mathcal I_k^c$ into two non-overlapping folds $\mathcal I_k^{c_1}$ and $\mathcal I_k^{c_2}$.\;
\For{$j=1,2$}{
Set $j'=3-j$.\;
Using observations in $\mathcal I_k^{c_j}$, estimate $\mu^D(Z,\bbX)$, $\mu^D(\bbX)$ and $\mu^Y(\bbX)$; denote the fitted functions by $\widehat\mu_{(-k,c_j)}^D(Z,\bbX)$, $\widehat\mu_{(-k,c_j)}^D(\bbX)$ and $\widehat\mu_{(-k,c_j)}^Y(\bbX)$.\;
Set $\widehat\xi^D_{(-k,c_j)}(Z,\bbX)=\widehat\mu^D_{(-k,c_j)}(Z,\bbX)-\widehat\mu^D_{(-k,c_j)}(\bbX)$.\;
Using observations in $\mathcal I_k^{c_{j'}}$, regress $\{\widehat\xi^D_{(-k,c_j)}(Z,\bbX)\}^2$ on $\bbX$; denote the fitted functions by $\widehat\delta^D_{(-k,c_{j'})}(\bbX)$.\;
Using observations in $\mathcal I_k^{c_{j'}}$, regress $\widehat\xi^D_{(-k,c_j)}(Z,\bbX) \{Y-\widehat\mu^Y_{(-k,c_j)}(\bbX)\}$ on $\bbX$; denote the fitted function by $\widehat\delta^Y_{(-k,c_{j'})}(\bbX)$.\;
Set $\widehat\tau_{(-k,c_{j'})}(\bbX) = \{\widehat\delta^D_{(-k,c_{j'})}(\bbX)\}^{-1}$, $\widehat\delta_{(-k,c_{j'})}(\bbX) = { \widehat\delta^Y_{(-k,c_{j'})}(\bbX) }/{ \widehat\delta^D_{(-k,c_{j'})}(\bbX) }$.
}
Average the two role-swapped fits pointwise to obtain $\widehat\mu^D_{(-k)}(Z,\bbX)$, $\widehat\mu^D_{(-k)}(\bbX)$, $\widehat\mu^Y_{(-k)}(\bbX)$, $\widehat\tau_{(-k)}(\bbX)$ and $\widehat\delta_{(-k)}(\bbX)$.\;
Evaluate $\widehat\psi_{i,k}$ in \eqref{eq:estimated-local-score} for each $i\in\mathcal I_k$.\;
}
Return $\widehat{\Delta}_{\mathrm{DS}}=n^{-1}\sum_{k=1}^K\sum_{i\in\mathcal I_k}\widehat\psi_{i,k}$.\;
\end{algorithm}

\clearpage

\section{An Example of Conditionally Homoskedastic Errors}
\phantomsection
\label{S-example:Homoskedastic}

Suppose that Assumptions \ref{as:relevance}--\ref{as:5d} hold. Consider the following structural equation model:
\begin{align*}
    Z & = f_{\bbX}(\bbX) + \epsilon_Z, \quad \E[\epsilon_Z \mid \bbX] =0, \\
    D &= g_{Z, \bbX, U}(Z, \bbX, U) + \epsilon_D, \quad \E[\epsilon_D \mid Z, \bbX, U]=0, \\
    Y &= \delta(\bbX) D + h_{\bbX, U}(\bbX, U) + \epsilon_Y, \quad \E[\epsilon_Y \mid D, \bbX, U]=0,
\end{align*}
where $f_{\bbX}(\bbX), g_{Z, \bbX, U}(Z, \bbX, U)$, and $h_{\bbX, U}(\bbX, U)$ are arbitrary measurable functions.

Under the additional condition that the outcome error is conditionally homoskedastic with respect to $D$, namely, $\var(\epsilon_Y \mid D, \bbX, U)= \var(\epsilon_Y \mid \bbX, U)$, then,
\begin{align*}
    \sigma^2(Z, \bbX) & = \E\left[\left\{Y - \mu^Y(\bbX) - \delta(\bbX)(D -\mu^D(\bbX))\right\}^2 \mid Z, \bbX \right] \\
    &=\E\left[\left\{ h_{\bbX, U}(\bbX, U) - \E[ h_{\bbX, U}(\bbX, U) \mid \bbX] +  \epsilon_Y \right\}^2 \mid Z, \bbX \right] \\
    & = \E[ h_{\bbX, U}(\bbX, U)^2 \mid \bbX] - \E[ h_{\bbX, U}(\bbX, U) \mid \bbX]^2 + \E[\var(\epsilon_Y \mid \bbX, U) \mid \bbX],
\end{align*}
where the last equality holds by Assumption \ref{as:ind}. Hence, $ \sigma^2(Z, \bbX)$ does not depend on $Z$.

\clearpage

\section{Identification of the Conditional Counterfactual Distribution Function}
\phantomsection
\label{S-sec: counterfactual_distribution}

In this section, we show that the identification strategy in Theorem~\ref{th:identification} extends from the conditional mean effect to a broader class of counterfactual distribution functions. This extends the result in \citet{S-tchetgen2024nudge} for the nudgeable subgroup with a binary instrument.

Following \citet{S-tchetgen2024nudge}, we consider the identification of a more general function $ \E[\Upsilon(Y(d), \boldsymbol{V}) \mid \boldsymbol{V}  ]$, where $\Upsilon(Y(d), \boldsymbol{V})$ is any specified measurable function and $\boldsymbol{V}$ is a subset of measured covariates $\bbX$. When $\Upsilon(Y(d), \boldsymbol{V})=\mathbb{I}(Y(d) \leq y)$, $\E[\Upsilon(Y(d), \boldsymbol{V}) \mid \boldsymbol{V}  ]$ reduces to $\Pr(Y(d) \leq y \mid \boldsymbol{V})$, which is the conditional cumulative distribution function of the counterfactual outcome $Y(d)$.

Before presenting the identification result, we introduce some notation and smoothness conditions. For $d \in \{0, 1\}$, define $\mu_d(z, \bbx) = \E[\mathbb{I}(D=d) \mid Z=z, \bbX=\bbx]$ and    $\mu_d(z, \boldsymbol{v}) = \E[\mu_d(z, \bbX) \mid \boldsymbol{V}=\boldsymbol{v}]$. Further define
\begin{equation*}
   G_d(z, \boldsymbol{v}) \coloneqq \E[\mathbb{I}(D(z) =d) \Upsilon(Y(z), \boldsymbol{V}) \mid \boldsymbol{V} = \boldsymbol{v} ] = \int \E [\mathbb{I}(D=d) \Upsilon(Y, \boldsymbol{V}) \mid Z=z, \bbX=\bbx]\diff F(\bbx \mid \boldsymbol{v}),
  \end{equation*} where $Y(z) = Y(D(z))$.

We impose the following analogue of the no unmeasured common effect modifier condition.
\begingroup
\setprimeassumption{2}{assumption.5doubleprime}
\begin{assumption}\label{S-as:5e}
  For all $z \in \calZ$ and $d \in \{0, 1\}$, $\cov\{\E[\mathbb{I}(D=d) \mid Z=z, \bbX, U] - \E[\mathbb{I}(D=d) \mid \bbX, U] , \E[\Upsilon(Y(d), \boldsymbol{V}) \mid \bbX, U] \mid \boldsymbol{V} \} =0$ almost surely.
\end{assumption}
\endgroup

The formal identification result is presented in the following theorem, which is in parallel to Theorem~\ref{th:identification}. The proof is provided in Supplementary Material~\ref{S-proof: identification_functional_general}.

\begin{theorem}\label{S-th:identification_functional_general}
   Suppose that Assumptions \ref{as:relevance}--\ref{as:dsep} and \ref{S-as:5e} hold. Let
$\gamma(Z, \boldsymbol{V}) \in L^2(P_{Z,\boldsymbol{V}})$. If $\delta^D_{\gamma,d}(\boldsymbol{V}) \coloneqq
\E[\{ \gamma(Z, \boldsymbol{V}) - \E[\gamma(Z, \boldsymbol{V})  \mid \boldsymbol{V}]\}\mu_d(Z,\boldsymbol{V}) \mid \boldsymbol{V}]
\neq 0 $
almost surely, then
\begin{equation}\label{S-eqn:identification_functional}
\E[\Upsilon(Y(d),\boldsymbol{V}) \mid \boldsymbol{V}]
=\dfrac{\E[ \{ \gamma(Z, \boldsymbol{V}) - \E[\gamma(Z, \boldsymbol{V})  \mid \boldsymbol{V}]\} G_d(Z,\boldsymbol{V}) \mid \boldsymbol{V}] }{ \E[\{ \gamma(Z, \boldsymbol{V}) - \E[\gamma(Z, \boldsymbol{V})  \mid \boldsymbol{V}]\} \mu_d(Z,\boldsymbol{V}) \mid \boldsymbol{V}]}
\coloneqq
\dfrac{\delta^Y_{\gamma,d}(\boldsymbol{V})}{\delta^D_{\gamma,d}(\boldsymbol{V})}.
\end{equation}
\end{theorem}

\begin{remark}
 Setting $\boldsymbol{V} = \bbX$ and $\Upsilon(Y(d), \boldsymbol{V})=Y(d)$, taking the difference of \eqref{S-eqn:identification_functional} for $d=1$ and $d=0$ recovers \eqref{eq: ATE-generalIV}.
\end{remark}

\newpage

\section{Proof of Theorems}

\subsection{Proof of Theorem~\ref{th:identification}}
\phantomsection
\label{S-app:th:identification}

First note that
\begin{align}
    \mu^Y(Z, \bbX) = &\ \E_{ U \mid \bbX} \left [\E[ Y \mid Z, \bbX, U] \right]\quad\quad \left(Z \independent U \mid \bbX \right) \notag\\
    =&\ \E_{ U \mid \bbX} \left[ \sum_{d \in \{0, 1\}} \E[ Y \mid D=d, Z, \bbX, U]\Pr( D=d \mid Z, \bbX, U)  \right]  \notag \\
     =&\ \E_{ U \mid \bbX} \Big[\{ \E[ Y \mid D=1, \bbX, U]-\E[ Y \mid D=0, \bbX, U]\}\Pr( D=1 \mid Z, \bbX, U) \notag\\
     &\ +\E[ Y \mid D=0, \bbX, U] \Big] \quad\quad \left(Z \independent Y \mid (D, \bbX, U) \right) \notag\\
    =&\ \E_{ U \mid \bbX} \Big[\{ \E[ Y(1) \mid \bbX, U]-\E[ Y(0) \mid \bbX, U]\}\Pr( D=1 \mid Z, \bbX, U) + \E[ Y(0) \mid \bbX, U] \Big] \notag \\
    & \quad \left(Y(d) \independent D \mid (\bbX, U) \right) \notag \\
    =&\ \E_{ U \mid \bbX} \left[\E[ Y(1) -Y(0) \mid \bbX, U] \Pr( D=1 \mid Z, \bbX, U) \right] +\E\left[ Y(0) \mid \bbX \right]. \label{S-eqn: conditional exp Y}
\end{align}

Marginalizing over $Z$ yields
\begin{align*}
 \mu^Y(\bbX) =&\ \E_{ U \mid \bbX} \left[\E[ Y(1) -Y(0) \mid \bbX, U] \Pr( D=1 \mid \bbX, U) \right] +\E\left[ Y(0) \mid \bbX \right].
\end{align*}
Hence,
\begin{align*}
\mu^Y(Z, \bbX) - \mu^Y(\bbX)
 =&\ \E_{ U \mid \bbX} \left[\E[ Y(1) -Y(0) \mid \bbX, U] \{ \Pr( D=1 \mid Z, \bbX, U) - \Pr( D=1 \mid \bbX, U)\}\right] \\
 =&\ \E[ Y(1) -Y(0) \mid \bbX]  \{ \Pr( D=1 \mid Z, \bbX) - \Pr( D=1 \mid \bbX)\} \quad (\text{by Assumption~\ref{as:5d}}).
\end{align*}
Multiplying both sides by $\gamma(Z, \bbX)$ and taking the conditional expectation gives
\begin{align*}
    &\E[\gamma(Z, \bbX) \{\mu^Y(Z, \bbX) - \mu^Y(\bbX) \} \mid \bbX ] \\
    =& \ \E[ Y(1) -Y(0) \mid \bbX]\E[\gamma(Z, \bbX) \{\mu^D(Z, \bbX) - \mu^D(\bbX) \} \mid \bbX ].
\end{align*}
Equivalently,
\begin{align*}
    & \E[\{\gamma(Z, \bbX) - \E[\gamma(Z, \bbX) \mid \bbX]\} \mu^Y(Z, \bbX)  \mid \bbX ] \\
     =& \ \E[ Y(1) -Y(0) \mid \bbX]\E[\{\gamma(Z, \bbX) - \E[\gamma(Z, \bbX) \mid \bbX]\} \mu^D(Z, \bbX) \mid \bbX ].
\end{align*}
Thus,
\begin{align*}
    \E[ Y(1) -Y(0) \mid \bbX] = \dfrac{\E[\{\gamma(Z, \bbX) - \E[\gamma(Z, \bbX) \mid \bbX]\} \mu^Y(Z, \bbX)  \mid \bbX ]}{\E[\{\gamma(Z, \bbX) - \E[\gamma(Z, \bbX) \mid \bbX]\} \mu^D(Z, \bbX) \mid \bbX ]},
\end{align*}
provided that the denominator is nonzero almost surely.

This concludes our proof.

\subsection[Proof of weighted-derivative and categorical-weighted identification]{Proof of \eqref{identification:weighted-derivative} and \eqref{identification:categorical-weighted}}
\phantomsection
Although Section~\ref{S-app:th:identification} gives a unified proof of Theorem~\ref{th:identification} for both continuous and categorical instruments, our identification results in \eqref{eq: ATE-generalIV} were first motivated by equations \eqref{identification:weighted-derivative} and \eqref{identification:categorical-weighted}. Here we give a brief proof.
Given \eqref{S-eqn: conditional exp Y}, $\forall \ z_0 \neq z_1 \in \calZ$, we have
 \begin{align*}
    &\ \mu^Y(z_1, \bbX) - \mu^Y(z_0, \bbX) \\
    =&\ \E_{ U \mid \bbX} \left[ \E[ Y(1) -Y(0) \mid \bbX, U] \left\{ \Pr( D=1 \mid Z=z_1, \bbX, U) - \Pr( D=1 \mid Z=z_0, \bbX, U)  \right\} \right] \\
    =&\ \E_{ U \mid \bbX} [\E[ Y(1) -Y(0) \mid \bbX, U]] \E_{ U \mid \bbX} \left[  \Pr( D=1 \mid Z=z_1, \bbX, U) - \Pr( D=1 \mid Z=z_0, \bbX, U) \right]  \\
    & \quad\quad (\text{by Assumption~\ref{as:5d}})\\
    =&\ \E[ Y(1) - Y(0) \mid \bbX] \{\mu^D(z_1, \bbX) - \mu^D(z_0, \bbX)\}.
 \end{align*}

 When $Z$ is continuous, it follows that
 \begin{align*}
 \muyp (z, \bbX)  = &\ \lim_{z_1 \to z}\dfrac{ \mu^Y(z_1, \bbX) - \mu^Y(z, \bbX)}{z_1 - z}  \\
 = &\ \E[ Y(1) - Y(0) \mid \bbX] \lim_{z_1 \to z}\dfrac{ \mu^D(z_1, \bbX) - \mu^D(z, \bbX)}{z_1 - z} \\
 = &\ \E[ Y(1) - Y(0) \mid \bbX] \mudp(z, \bbX).
 \end{align*}

 So we have $\omega(Z, \bbX)\muyp(Z, \bbX)=\E[ Y(1) - Y(0) \mid \bbX] \omega(Z, \bbX)\mudp(Z, \bbX)$. Taking expectation over $Z$ on both sides yields
 \begin{equation*}
     \E[\omega(Z, \bbX)\muyp(Z, \bbX) \mid \bbX]=\E[ Y(1) - Y(0) \mid \bbX]  \E[\omega(Z, \bbX)\mudp(Z, \bbX) \mid \bbX].
 \end{equation*}
 Then we directly have
 \begin{equation*}
 \E[ Y(1) - Y(0) \mid \bbX]=\dfrac{\E[\omega(Z, \bbX)\muyp(Z, \bbX) \mid \bbX]}{\E[\omega(Z, \bbX)\mudp(Z, \bbX) \mid \bbX]},
 \end{equation*}
 provided that $\E[\omega(Z, \bbX)\mudp(Z, \bbX) \mid \bbX] \neq 0$.

 When $Z$ is categorical, it follows that
 \begin{align*}
 & \sum\limits_{0 \leq k <j \leq J-1}\omega_{jk}(\bbX) \{\mu^Y(j, \bbX) - \mu^Y(k, \bbX)\} \\
 & = \E[ Y(1) - Y(0) \mid \bbX] \sum\limits_{0 \leq k <j \leq J-1}\omega_{jk}(\bbX) \{\mu^D(j, \bbX) - \mu^D(k, \bbX)\}.
 \end{align*}
 Then we have
 \begin{equation*}
     \E[ Y(1) - Y(0) \mid \bbX] = \dfrac{\sum_{k <j}\omega_{jk}(\bbX) \{\mu^Y(j, \bbX) - \mu^Y(k, \bbX)\}}{\sum_{k <j}\omega_{jk}(\bbX) \{\mu^D(j, \bbX) - \mu^D(k, \bbX)\}},
 \end{equation*}
 provided that $\sum_{k <j}\omega_{jk}(\bbX) \{\mu^D(j, \bbX) - \mu^D(k, \bbX)\} \neq 0$.

\subsection{Proof of Theorem \ref{th:nonparametric}}
\phantomsection
\label{S-app:th:nonparametric}

\subsubsection{Proof of Statement (1)}
In the following proof, we will use the following identities repeatedly:
\begin{align}
    & \forall \, \bbx \mapsto f (\bbx), \E[f(\bbX)s( Y \mid \bbX)]=0, \label{S-eq:S1} \\
    & \forall \, (z, \bbx) \mapsto f (z, \bbx), \E[f(Z, \bbX)s( Y \mid Z, \bbX) \mid \bbX]=0,
    \label{S-eq:S2}
\end{align}
where $s (Y \mid \bbX)$ denotes the conditional score of $Y$ given $\bbX$ and $s (Y \mid Z, \bbX)$ is defined analogously.

Let $p_t(Y, D, Z, \bbX) =p_t(Y, D\mid Z, \bbX)p_t(Z \mid \bbX)p_t(\bbX)$ denote a one-dimensional path of $\calM_{\np}$ indexed by $t$, under which $\Delta_{\gamma, t}=\E_t\left[\dfrac{\delta_{\gamma,t}^Y(\bbX)}{\delta_{\gamma,t}^D(\bbX)}\right]$. The score function is
\begin{align*}
s_t(Y, D, Z, \bbX) &=\ s_t(Y, D \mid Z, \bbX) + s_t(Z \mid \bbX) + s_t(\bbX) \\
 &=\ \dfrac{\partial}{\partial t} \log p_t(Y, D \mid Z, \bbX) + \dfrac{\partial}{\partial t} \log p_t(Z \mid \bbX) + \dfrac{\partial}{\partial t} \log p_t(\bbX).
\end{align*}
To find the influence function in $\calM_{\np}$, we need to find a random variable $\Psi$ such that $\E(\Psi)=0$ and, for all one-dimensional paths, we have
\begin{equation*}
    \left.\dfrac{\partial}{\partial t} \Delta_{\gamma,t} \right|_{t=0}=\ \E[\Psi \cdot s_t(Y, D, Z, \bbX)]|_{t=0}.
\end{equation*}

First, note that
\begin{align*}
    &\ \left.\dfrac{\partial}{\partial t} \Delta_{\gamma,t} \right|_{t=0} \\
 = &\ \left.\dfrac{\partial}{\partial t} \E_t \left[ \dfrac{\delta_{\gamma,t}^Y(\bbX)}{\delta_{\gamma,t}^D(\bbX)}\right] \right|_{t=0}\\
 =& \ \E \left[ \dfrac{\delta_\gamma^Y(\bbX)}{\delta_\gamma^D(\bbX)} s(\bbX) \right] + \E \left. \left[ \dfrac{\dfrac{\partial}{\partial t}\delta_{\gamma,t}^Y(\bbX) \delta_\gamma^D(\bbX)- \delta_\gamma^Y(\bbX)\dfrac{\partial}{\partial t}\delta_{\gamma,t}^D(\bbX)}{(\delta_\gamma^D(\bbX))^2}  \right] \right|_{t=0} \\
 = &\ \E \left[ \dfrac{\delta_\gamma^Y(\bbX)}{\delta_\gamma^D(\bbX)} s(Y, D, Z, \bbX) \right] + \E \left. \left[ \dfrac{\dfrac{\partial}{\partial t}\delta_{\gamma,t}^Y(\bbX) \delta_\gamma^D(\bbX)- \delta_\gamma^Y(\bbX)\dfrac{\partial}{\partial t}\delta_{\gamma,t}^D(\bbX)}{(\delta_\gamma^D(\bbX))^2}  \right] \right|_{t=0}. \quad\quad (\text{by~\eqref{S-eq:S1}})
\end{align*}

Since $\gamma(Z, \bbX)$ is a given function, we have
\begin{align}
\left.\dfrac{\partial}{\partial t}\delta_{\gamma,t}^Y(\bbX)\right|_{t=0} = &\ \left. \dfrac{\partial}{\partial t} \E_t\left[  \left\{ \gamma(Z, \bbX) - \E_t[\gamma(Z, \bbX) \mid \bbX] \right\} Y \,\middle\vert\, \bbX\right]\right|_{t=0} \notag \\
     =&\ \left.\dfrac{\partial}{\partial t} \E\left[ \left\{ \gamma(Z, \bbX) - \E_t[\gamma(Z, \bbX) \mid \bbX] \right\} Y \,\middle\vert\, \bbX \right] \right|_{t=0}  \label{S-eqn: A}\\
     & + \E\left[ \left\{ \gamma(Z, \bbX) - \E[\gamma(Z, \bbX) \mid \bbX] \right\} Y S(Y, Z \mid \bbX) \,\middle\vert\, \bbX\right]. \label{S-eqn: B}
\end{align}
We first have
\begin{align*}
   \eqref{S-eqn: A}=&\   \left.\dfrac{\partial}{\partial t} \E\left[ \left\{ \gamma(Z, \bbX) - \E_t[\gamma(Z, \bbX) \mid \bbX] \right\} Y \,\middle\vert\, \bbX \right] \right|_{t=0}  \\
    =&\  - \E\left[ \E[\gamma(Z, \bbX) S(Z \mid \bbX) \mid \bbX ] Y \,\middle\vert\, \bbX \right]  \\
    =&\  -  \E \left[\gamma(Z, \bbX) S(Z \mid \bbX) \mid \bbX \right] \E[ Y \mid \bbX ] \\
    =&\  -  \E \left[\gamma(Z, \bbX) S(Y, D, Z \mid \bbX) \mid \bbX \right] \E[ Y \mid \bbX ]
  \quad\quad(\text{by~\eqref{S-eq:S2}}) \\
    =&\  -  \E \left[ \left\{ \gamma(Z, \bbX) - \E[\gamma(Z, \bbX) \mid \bbX] \right\}  S(Y, D, Z, \bbX) \mid \bbX \right] \E[ Y \mid \bbX ] \\
        =&\  -  \E \left[ \left\{ \gamma(Z, \bbX) - \E[\gamma(Z, \bbX) \mid \bbX] \right\} \E[ Y \mid \bbX ] S(Y, D, Z, \bbX) \mid \bbX \right] .
\end{align*}
And
\begin{align*}
  \eqref{S-eqn: B}=&\ \E\left[ \left\{ \gamma(Z, \bbX) - \E[\gamma(Z, \bbX) \mid \bbX] \right\} Y S(Y, Z \mid \bbX) \,\middle\vert\, \bbX\right]\\
    =&\ \E\left[ \left\{ \gamma(Z, \bbX) - \E[\gamma(Z, \bbX) \mid \bbX] \right\} Y S(Y, D, Z \mid \bbX) \,\middle\vert\, \bbX\right]   \quad\quad (\text{by~\eqref{S-eq:S2}}) \\
 = & \ \E\left[ \left\{ \gamma(Z, \bbX) - \E[\gamma(Z, \bbX) \mid \bbX] \right\} Y S(Y, D, Z, \bbX) \,\middle\vert\, \bbX\right] - \delta_{\gamma}^Y(\bbX) S(\bbX).
\end{align*}
Therefore, we have
\begin{align*}
    \left.\dfrac{\partial}{\partial t}\delta_{\gamma,t}^Y(\bbX)\right|_{t=0} = & \ \eqref{S-eqn: A} + \eqref{S-eqn: B} \\
    =& \  \E\left[\left\{ \gamma(Z, \bbX) - \E[\gamma(Z, \bbX) \mid \bbX] \right\} \left \{ Y - \E[ Y \mid \bbX] \right\} s(Y, D, Z, \bbX)  \,\middle\vert\, \bbX\right] - \delta_{\gamma}^Y(\bbX) S(\bbX)  .
\end{align*}

Similarly,
\begin{align*}
 \left.\dfrac{\partial}{\partial t}\delta_{\gamma,t}^D(\bbX)\right|_{t=0} = \E\left[\left\{ \gamma(Z, \bbX) - \E[\gamma(Z, \bbX) \mid \bbX] \right\} \left \{ D - \E[ D \mid \bbX] \right\} s(Y, D, Z, \bbX)  \,\middle\vert\, \bbX\right]  -\delta_{\gamma}^D(\bbX) S(\bbX)   .
\end{align*}

Define
\begin{align*}
\Tilde{\Psi}_{1}(\bbO; P) \coloneqq \dfrac{\gamma(Z, \bbX) - \E[\gamma(Z, \bbX) \mid \bbX]} {\delta_{\gamma}^D(\bbX)} \{Y- \mu^Y(\bbX)-\{D-  \mu^D(\bbX)\}\delta_{\gamma}(\bbX)\} +\delta_{\gamma}(\bbX).
\end{align*}
We then have $\left.\dfrac{\partial}{\partial t} \Delta_{\gamma,t} \right|_{t=0}  = \E[\Tilde{\Psi}_{1}(\bbO; P) s(Y, D, Z, \bbX)]$.
It is easy to show that $\E [\Tilde{\Psi}_{1}(\bbO; P)] = \Delta_\gamma$, and hence $\Psi_{\gamma}(\bbO; P) = \tilde{\Psi}_{1}(\bbO; P) - \Delta_\gamma$ is an influence function of $\Delta_\gamma$ given the weight function $\gamma(Z, \bbX)$ .

Under $\calM_{\np}$, the tangent space is the entire Hilbert space of mean-zero and finite variance. Therefore, $\Psi_\gamma(\bbO; P)$ is the unique influence function, or equivalently, the efficient influence function of $\Delta_\gamma$.

\subsubsection{Proof of Statement (2)}
Under the semiparametric model $\calM_{\sp}$,
\begin{equation*}
\mu^Y(z, \bbX) - \mu^Y(\bbX) = \delta(\bbX) \{ \mu^D(z, \bbX) - \mu^D(\bbX) \}, \forall \ z \in \calZ.
\end{equation*}
And we have $\delta_\gamma(\bbX) = \delta(\bbX)$, $\Delta_\gamma = \Delta$, and $\calE_\gamma = \calE$. Then
\begin{equation*}
\Psi_\gamma(\bbO; P) = \dfrac{\gamma(Z, \bbX) - \E[\gamma(Z, \bbX) \mid \bbX]}{\delta_{\gamma}^D(\bbX)} \calE +\delta(\bbX)  - \Delta,
\end{equation*}
is an IF of $\Delta$ under $\calM_{\sp}$ depending on $\gamma$. Hence, $\{ \Psi_\gamma(\bbO; P): \delta_\gamma^D(\bbX) \neq 0 \ \text{almost surely}\}$ is a class of IFs of $\Delta$.
\subsubsection{Proof of Statement (3)}

We now characterize the orthocomplement to the tangent space $\Lambda_{\sp}^{\perp}$ of the semiparametric model $\calM_{\sp}$, given as the lemma below.

\begin{lemma}
\phantomsection
\label{S-lem:tangent space}
The orthocomplement to the tangent space of the semiparametric model $\calM_{\sp}$ can be characterized as follows:
\begin{align*}
\Lambda_{\sp}^{\perp} = \left\{ \left( \dfrac{\bar{m} (Z, \bbX)}{\mu^D(Z, \bbX) - \mu^D(\bbX)} - \E \left[ \dfrac{\bar{m} (Z, \bbX)}{\mu^D(Z, \bbX) - \mu^D(\bbX)} \,\middle\vert\,  \bbX \right] \right) \calE: \E [\bar{m} (Z, \bbX) \mid \bbX] = 0 \ \text{almost surely} \right\}.
\end{align*}
\end{lemma}

\begin{proof}
The proof of Lemma~\ref{S-lem:tangent space} follows from constructing a particular one-dimensional path in $\calM_{\sp}$, which induces scores orthogonal to every element of $\Lambda_{\sp}^{\perp}$. To this end, recall that we denote the true observed-data distribution $P$ with probability density/mass function $p (y, d, z, \bbx)$. We perturb the observed-data distribution as follows:
\begin{align*}
p_{t} (y, d, z, \bbx) = p_{t} (z, \bbx) p_{t} (d \mid z, \bbx) p_{t} (y \mid d, z, \bbx),
\end{align*}
such that $p_{t = 0} (\bbo) \equiv p (\bbo)$ and
\begin{align}
& p_{t} (z, \bbx) = p (z, \bbx)  (1 + t  f (z, \bbx)), \text{ where } \E [f (Z, \bbX)] = 0, \nonumber \\
& p_{t} (d \mid z, \bbx) = p (d \mid z, \bbx)  (1 + t  g (d, z, \bbx)), \text{ where } \E [g (D, Z, \bbX) \mid Z, \bbX] = 0, \nonumber \\
& p_{t} (y \mid d, z, \bbx) = p (y \mid d, z, \bbx)  (1 + t  h_{1} (y, d, z, \bbx) + t^{2}  h_{2} (y, d, z, \bbx)), \label{S-path3} \\
& \text{where } \E [h_{1} (Y, D, Z, \bbX) \mid D, Z, \bbX] = \E [h_{2} (Y, D, Z, \bbX) \mid D, Z, \bbX] = 0. \nonumber
\end{align}
It is not difficult to verify that the perturbation $p_{t}$ just constructed induces the following score
\begin{align*}
s (Y, D, Z, \bbX) = f (Z, \bbX) + g (D, Z, \bbX) + h_{1} (Y, D, Z, \bbX).
\end{align*}
To simplify notation, we let $S \coloneqq s (Y, D, Z, \bbX)$, $H_{1} \coloneqq h_{1} (Y, D, Z, \bbX)$, $H_{2} \coloneqq h_{2} (Y, D, Z, \bbX)$, $G \coloneqq g (D, Z, \bbX)$, and $F \coloneqq f (Z, \bbX)$. Since we need to show that $S$ is orthogonal to every element of $\Lambda_{\sp}^{\perp}$, we also need to add the following restrictions to $S$: for any $\bar{m}$ such that $\E [\bar{m} (Z, \bbX) \mid \bbX] = 0$ almost surely,
\begin{align*}
\E \left[ S  \left( \dfrac{\bar{m} (Z, \bbX)}{\mu^D(Z, \bbX) - \mu^D(\bbX)} - \E \left[ \dfrac{\bar{m} (Z, \bbX)}{\mu^D(Z, \bbX) - \mu^D(\bbX)} \,\middle\vert\, \bbX \right] \right) \calE \right]=0.
\end{align*}
We then need to verify that $p_{t}$ is a valid path, that is, $p_{t} \in \calM_{\sp}$. Equivalently, we need to show that for any $z'$,
\begin{align}
\label{S-target}
\delta_{t} (z, \bbx) \coloneqq \dfrac{\mu_t^Y (z, \bbx) - \mu_t^Y (z', \bbx)}{\mu_t^D (z, \bbx) - \mu_t^D (z', \bbx)}
\end{align}
is a function of $\bbx$ but not of $z$. In particular, the numerator and denominator of the right-hand side of the above display can be further simplified as follows:
\begin{align}
&\ \mu_t^D (z, \bbx) - \mu_t^D (z', \bbx) \nonumber \\=&  \ \mu^D (z, \bbx) - \mu^D (z', \bbx)
+ t \left( \E [D  G \mid Z = z, \bbX = \bbx] - \E [D  G \mid Z = z', \bbX = \bbx] \right), \label{S-mu_D_perturb}
\end{align}
and there exists a function $\nu_{1}$ of $\bbx$ alone such that
\begin{align}
\phantomsection
& \ \mu_t^Y (z, \bbx) - \mu_t^Y (z', \bbx) \nonumber \\
= & \ \left\{\delta (\bbx) + t  \nu_{1} (\bbx)\right\}  \left\{\mu_t^D (z, \bbx) - \mu_t^D (z', \bbx)\right\} \nonumber \\
& + t^{2}  \left\{ \begin{array}{c}
\E [Y  (G  H_{1} + H_{2}) \mid Z = z, \bbX = \bbx] - \E [Y  (G  H_{1} + H_{2}) \mid Z = z', \bbX = \bbx] \\
- \, \nu_{1} (\bbx)  \left(\E [D  G \mid Z = z, \bbX = \bbx] - \E [D  G \mid Z = z', \bbX = \bbx]\right)
\end{array} \right\} \nonumber \\
& + t^{3}  \left(\E [Y  G  H_{2} \mid Z = z, \bbX = \bbx] - \E [Y  G  H_{2} \mid Z = z', \bbX = \bbx]\right), \label{S-mu_Y_perturb}
\end{align}
where the equality holds by applying Lemma~\ref{S-lem:id}, which is stated later.

Finally, it is noteworthy that $H_{2}$ does not appear as part of the score $S$ induced by the path $p_{t}$. As a consequence, $H_{2}$ does not need to be orthogonal to every element of $\Lambda_{\sp}^{\perp}$. This extra flexibility allows us to show that there must exist $H_{2} = h_{2} (Y, D, Z, \bbX)$ such that (1) $\E [H_{2} \mid D, Z, \bbX] = 0$ almost surely and (2) there exists a function $\nu_{2} (\cdot)$ of $\bbx$ alone satisfying
\begin{equation}
\phantomsection
\label{S-H2}
    \E \left[ Y  \left( \nu_{2} (\bbX) - \nu_{1} (\bbX)  G \right)  H_{2} \mid Z, \bbX \right]  = \nu_{2} (\bbX)^{2}  \E [D \mid Z, \bbX] - \nu_{2} (\bbX)  \E [Y  G  H_{1} \mid Z, \bbX].
\end{equation}

This is because we can first choose any other function $h_{2}' (\bbo)$ such that $\E [H_{2}' \mid D, Z, \bbX] = 0$ almost surely where $H_{2}' \coloneqq h_{2}' (\bbO)$ and then evaluate the LHS of \eqref{S-H2} with $H_{2}$ replaced by $H_{2}'$. We denote the resulting conditional mean (a function of $(z, \bbx)$) as $\kappa (z, \bbx)$. We can then simply find $H_{2}$ by solving the equation below:
\begin{align*}
H_{2} \cdot \kappa (Z, \bbX) = H_{2}' \cdot \left( \nu_{2} (\bbX)^{2}  \E [D \mid Z, \bbX] - \nu_{2} (\bbX)  \E [Y  G  H_{1} \mid Z, \bbX] \right).
\end{align*}

By elementary calculations, \eqref{S-H2} helps yield the following identity:
\begin{align*}
& \left\{ \begin{array}{c}
\E [Y  (G  H_{1} + H_{2}) \mid Z = z, \bbX = \bbx] - \E [Y  (G  H_{1} + H_{2}) \mid Z = z', \bbX = \bbx] \\
- \, \nu_{1} (\bbx)  \left( \E [D  G \mid Z = z, \bbX = \bbx] - \E [D  G \mid Z = z', \bbX = \bbx]\right)
\end{array} \right\} \\
& + t  \left(\E [Y  G  H_{2} \mid Z = z, \bbX = \bbx] - \E [Y  G  H_{2} \mid Z = z', \bbX = \bbx] \right) \\
& = \nu_{2} (\bbx)  \{\mu_t^D (z, \bbx) - \mu_t^D (z', \bbx)\}.
\end{align*}

Applying the above result to \eqref{S-mu_Y_perturb}, we have
\begin{align*}
\delta_{t} (z, \bbx) = \delta (\bbx) + t  \nu_{1} (\bbx) + t^{2}  \nu_{2} (\bbx),
\end{align*}
which completes the proof.
\end{proof}

Armed with Lemma~\ref{S-lem:tangent space}, we can find the efficient influence function of $\Delta$ under $\calM_{\sp}$ by projecting any influence function of $\Delta$ onto $\Lambda_{\sp}$:
\begin{align*}
\Psi_{\eff} = \Pi [\Psi_{\gamma} \mid \Lambda_{\sp}] = \Psi_{\gamma} - \Pi [\Psi_{\gamma} \mid \Lambda_{\sp}^{\perp}],
\end{align*}
where $\Pi$ is the projection operator. Elementary algebra then shows that $\Psi_{\eff}$ has the form given in Statement (3) of Theorem~\ref{th:nonparametric}.

Finally, we state Lemma~\ref{S-lem:id} used in the proof of Lemma~\ref{S-lem:tangent space}.

\begin{lemma}
\phantomsection
\label{S-lem:id}
Let $S \coloneqq s (Y, D, Z, \bbX)$ denote a joint score and $S$ permits the orthogonal decomposition into conditional scores:
\begin{align*}
S = H + G + F, H \coloneqq h (Y, D, Z, \bbX), G \coloneqq g (D, Z, \bbX), F \coloneqq f (Z, \bbX),
\end{align*}
where $\E [H \mid D, Z, \bbX] = 0$, $\E [G \mid Z, \bbX] = 0$, and $\E[F] = 0$ almost surely. If $S \in \Lambda_{\sp}$, then there exists a function $\bbx \mapsto \nu(\bbx)$ such that
\begin{align*}
& \ \E [Y  (G + H) \mid Z, \bbX] - \E [Y  (G + H) \mid \bbX] \\
= & \ \delta (\bbX)  \left(\E [D  G \mid Z, \bbX] - \E [D  G \mid \bbX]\right) + \nu (\bbX)  \{ \mu^D(Z, \bbX) - \mu^D(\bbX)\}, \text{ almost surely}.
\end{align*}
\end{lemma}

The proof of Lemma~\ref{S-lem:id} is straightforward by verifying that the desired constraint on $G + H$ suffices for $S$ to belong to $\Lambda_{\sp}$.

\begin{remark}
\phantomsection
\label{S-rem:non-convex}
One can heuristically derive $\Lambda_{\sp}^{\perp}$ by differentiating the following sole moment algebraic constraint induced by the semiparametric model $\calM_{\sp}$ with respect to $t \in \mathbb{R}$ at $t = 0$:
\begin{equation}
\label{S-moment_constraint}
\E_{t} \left[ \delta_{t} (Z, \bbX) \{m (Z, \bbX) - \E_{t} [m (Z, \bbX) \mid \bbX]\} \right] = 0,
\end{equation}
for any $(z, \bbx) \mapsto m (z, \bbx)$, where $\E_{t}$ denotes the mean with respect to a perturbed observed-data distribution $P_{t}$ with $P_{t = 0} = P$ and $\delta_{t}$ is the corresponding $\delta$ under $P_{t}$. The moment algebraic constraint \eqref{S-moment_constraint} encodes the restriction that $\delta_{t} (z, \bbx)$ is only a function of $\bbx$ not varying with $z$. Another natural approach to conjecturing the form of $\Lambda_{\sp}^{\perp}$ is inspecting the difference between any two IFs characterized in statement (2) of Theorem~\ref{th:nonparametric} (see Supplementary Material~\ref{S-app:rem:semiparametric}).

A rigorous proof demands construction of explicit paths (parametric submodels) in $\calM_{\sp}$ with the score orthogonal to $\Lambda_{\sp}^{\perp}$. A usual approach, as laid out in several textbooks such as \citet{S-laan2003unified}, is to postulate one-dimensional paths of the form $p_{t} (\bbo) = p (\bbo) (1 + t s (\bbo))$ where $t \in \mathbb{R}$ and $s (\bbo)$ is the corresponding score. However, without imposing further assumptions on the score $s (\bbo)$, such a first-order perturbation of $p (\bbo)$ will fail to ensure that $\delta_{t}$ does not depend on $z$, so $p_{t}$ is not a path by definition.

Second-order or even higher-order paths, to the best of our knowledge, first appeared in \citet{S-pfanzagl1983asymptotic}; also see \citet{S-van2014higher} for their application in defining higher-order tangent spaces. The motivation was to generalize the classical efficiency theory from first-order to higher-order. Our use of higher-order paths comes from a different motivation. Higher-order paths are constructed to overcome the difficulty of ensuring their validity, that is, strictly belonging to $\calM_{\sp}$. To see this, suppose that $p_{t} (y \mid d, z, \bbx)$ is perturbed from $p (y \mid d, z, \bbx)$ as in \eqref{S-path3} except that $h_{2} (\cdot)$ is set to 0, so the perturbation is only of first-order. The denominator of $\delta_{t} (z, \bbx)$ \eqref{S-target} stays the same as in \eqref{S-mu_D_perturb}, whereas the numerator will change from \eqref{S-mu_Y_perturb} to:
\begin{align*}
& \ \mu_t^Y (z, \bbx) - \mu_t^Y (z', \bbx) \\
= & \ \{\delta (\bbx) + t  \nu_{1} (\bbx)\}  \{\mu_t^D (z, \bbx) - \mu_t^D (z', \bbx)\} \\
& + t^{2}  \underbrace{\left\{ \begin{array}{c}
\E [Y  G  H_{1} \mid Z = z, \bbX = \bbx] - \E [Y  G  H_{1} \mid Z = z', \bbX = \bbx] \\
- \, \nu_{1} (\bbx)  \left( \E [D  G \mid Z = z, \bbX = \bbx] - \E [D  G \mid Z = z', \bbX = \bbx]\right)
\end{array} \right\}}_{(\ast) \, \coloneqq}.
\end{align*}
To show $\delta_{t} (z, \bbx)$ does not depend on $z$, the term $(\ast)$ should be equal to $\nu' (\bbx)  \{\mu_t^D (z, \bbx) - \mu_t^D (z', \bbx)\}$ for some function $x \mapsto \nu'(\bbx)$ or simply 0. In the first case, we have a contradiction as $(\ast)$ does not depend on $t$; in the second case, we need to impose extra restrictions on the score $S$, which has been demonstrated to be unnecessary in the proof.

We remark that the results of \citet{S-ai2012semiparametric}, in particular Theorem 2.1 therein, for establishing semiparametric efficiency bounds of semiparametric models with sequential moment restrictions may be applicable to computing the semiparametric efficiency bound of $\Delta$ under our semiparametric model $\calM_{\sp}$. Nonetheless, their results rely on several predicates summarized as Assumptions 1, 2, and A in \citet{S-ai2012semiparametric}, which assume the existence of paths in $\calM_{\sp}$, among other things. As we have seen, verifying these assumptions is not necessarily trivial and constructing valid paths as in our proof is a key step to applying the results or using the proof techniques of \citet{S-ai2012semiparametric}.
\end{remark}

\subsubsection{Some Further Comments on Theorem~\ref{th:nonparametric}}
\label{S-app:rem:semiparametric}

One heuristic approach to conjecturing $\Lambda_{\sp}^{\perp}$ is to take the difference between any two IFs characterized in statement (2) of Theorem~\ref{th:nonparametric}. There, the set $\{\Psi_\gamma(\bbO; P):\delta_\gamma^D(\bbX)\neq0\text{ almost surely}\}$, where
\begin{equation*}
\Psi_\gamma(\bbO; P)=\dfrac{\gamma(Z, \bbX)-\E[\gamma(Z, \bbX)\mid\bbX]}{\delta_\gamma^D(\bbX)}\calE+\delta(\bbX)-\Delta,
\end{equation*}
is a class of IFs for $\Delta$. By Theorem 4.2 of \citet{S-tsiatis2006semiparametric}, for any two IFs $\Psi_{\gamma_1}(\bbO; P)$ and $\Psi_{\gamma_2}(\bbO; P)$,
\begin{equation*}
\E\left[\{\Psi_{\gamma_1}-\Psi_{\gamma_2}\}s(\bbO)\right]\equiv 0.
\end{equation*}
This suggests that the orthocomplement of the semiparametric tangent space under $\calM_{\sp}$ is the linear span
\begin{equation*}
\Lambda_{\sp}^{\perp}=cl\{c(\Psi_{\gamma_1}-\Psi_{\gamma_2}):c\in\mathbb{R}\},
\end{equation*}
where $cl(\mathcal{A})$ denotes the mean-squared closure of $\mathcal{A}$. Note that
\begin{equation*}
\Psi_{\gamma_1}-\Psi_{\gamma_2}
=\left[\dfrac{\gamma_1(Z,\bbX)-\E\{\gamma_1(Z,\bbX)\mid\bbX\}}{\delta_{\gamma_1}^D(\bbX)}
-\dfrac{\gamma_2(Z,\bbX)-\E\{\gamma_2(Z,\bbX)\mid\bbX\}}{\delta_{\gamma_2}^D(\bbX)}\right]\calE.
\end{equation*}
By construction, $\E[\gamma(Z, \bbX)-\E[\gamma(Z, \bbX)\mid\bbX]\mid\bbX]=0$ and $\E[\{\gamma(Z, \bbX)-\E[\gamma(Z, \bbX)\mid\bbX]\}\mu^D(Z,\bbX)\mid\bbX]=\delta_\gamma^D(\bbX)$ for any $\gamma(Z, \bbX)$. It follows that
\begin{align*}
\E\left[\dfrac{\gamma_1(Z,\bbX)-\E\{\gamma_1(Z,\bbX)\mid\bbX\}}{\delta_{\gamma_1}^D(\bbX)}
-\dfrac{\gamma_2(Z,\bbX)-\E\{\gamma_2(Z,\bbX)\mid\bbX\}}{\delta_{\gamma_2}^D(\bbX)}\,\middle|\,\bbX\right]&=0,\\
\E\left[\left\{\dfrac{\gamma_1(Z,\bbX)-\E\{\gamma_1(Z,\bbX)\mid\bbX\}}{\delta_{\gamma_1}^D(\bbX)}
-\dfrac{\gamma_2(Z,\bbX)-\E\{\gamma_2(Z,\bbX)\mid\bbX\}}{\delta_{\gamma_2}^D(\bbX)}\right\}\mu^D(Z,\bbX)\,\middle|\,\bbX\right]&=0.
\end{align*}
This gives the conjectured orthocomplement in Lemma~\ref{S-lem:tangent space}.

\subsection{Proof of Theorem~\ref{th:localeff}}\label{S-sec:prooflocaleff}
Define
\begin{align*}
\Delta_{\gamma_0}
=\E\left[\dfrac{\E[\{\gamma_0(Z,\bbX)-\E[\gamma_0(Z,\bbX) \mid \bbX]\}Y\mid\bbX]}{\E[\{\gamma_0(Z,\bbX)-\E(\gamma_0(Z,\bbX) \mid \bbX)\}D\mid\bbX]}\right]
=\E\left[\dfrac{\delta_{\gamma_0}^Y(\bbX)}{\delta_{\gamma_0}^D(\bbX)}\right]
=\E\{\delta_{\gamma_0}(\bbX)\}.
\end{align*}
We first establish that the efficient influence function of $\Delta_{\gamma_0}$ under the nonparametric model $\calM_{\np}$ is
\begin{align*}
\Psi_{\gamma_0}(\bbO;P)
&=\dfrac{\{\gamma_0(Z,\bbX)-\E[\gamma_0(Z,\bbX)\mid\bbX]\}\{Y-\mu^Y(\bbX)-\delta_{\gamma_0}(\bbX)(D-\mu^D(\bbX))\}}{\delta_{\gamma_0}^D(\bbX)}
\\
&\quad+\dfrac{\{D-\mu^D(Z,\bbX)\}\{\mu^Y(Z,\bbX)-\delta_{\gamma_0}(\bbX)\mu^D(Z,\bbX)\}
-\{D-\mu^D(\bbX)\}\{\mu^Y(\bbX)-\delta_{\gamma_0}(\bbX)\mu^D(\bbX)\}}{\delta_{\gamma_0}^D(\bbX)} \\
&\quad +\delta_{\gamma_0}(\bbX)-\Delta_{\gamma_0}.
\end{align*}
It then follows that, under $\calM_{\sp}$, $\Delta_{\gamma_0}=\Delta$ and
\begin{equation*}
\Psi_{\gamma_0}(\bbO;P)=\dfrac{\mu^D(Z,\bbX)-\mu^D(\bbX)}{\delta_{\gamma_0}^D(\bbX)}\calE+\delta(\bbX)-\Delta
\end{equation*}
is an influence function of $\Delta$.

Following the proof of Theorem~\ref{th:nonparametric}, it suffices to find a random variable $\Psi_{\gamma_0}$ such that $\E(\Psi_{\gamma_0})=0$ and, for all one-dimensional paths,
\begin{align*}
\left.\dfrac{\partial}{\partial t}\Delta_{\gamma_0,t}\right|_{t=0}
=\left.\dfrac{\partial}{\partial t}\E_t\left[\dfrac{\delta_{\gamma_0,t}^Y(\bbX)}{\delta_{\gamma_0,t}^D(\bbX)}\right]\right|_{t=0}
=\left.\E[\Psi_{\gamma_0}s_t(Y,D,Z,\bbX)]\right|_{t=0}.
\end{align*}
First, note that
\begin{align*}
\left.\dfrac{\partial}{\partial t}\Delta_{\gamma_0,t}\right|_{t=0}
&=\E\left[\dfrac{\delta_{\gamma_0}^Y(\bbX)}{\delta_{\gamma_0}^D(\bbX)}s(\bbX)\right]
+\E\left[\left.\dfrac{\{\partial\delta_{\gamma_0,t}^Y(\bbX)/\partial t\}\delta_{\gamma_0}^D(\bbX)-\delta_{\gamma_0}^Y(\bbX)\{\partial\delta_{\gamma_0,t}^D(\bbX)/\partial t\}}{\{\delta_{\gamma_0}^D(\bbX)\}^2}\right|_{t=0}\right]\\
&=\E[\delta_{\gamma_0}(\bbX)s(Y,D,Z,\bbX)]
+\E\left[\left.\dfrac{\{\partial\delta_{\gamma_0,t}^Y(\bbX)/\partial t\}\delta_{\gamma_0}^D(\bbX)-\delta_{\gamma_0}^Y(\bbX)\{\partial\delta_{\gamma_0,t}^D(\bbX)/\partial t\}}{\{\delta_{\gamma_0}^D(\bbX)\}^2}\right|_{t=0}\right].
\end{align*}

Since
\begin{align*}
\left.\dfrac{\partial}{\partial t}\mu_t^D(Z,\bbX)\right|_{t=0}
&=\E[\{D-\mu^D(Z,\bbX)\}s(Y,D,Z,\bbX)\mid Z,\bbX],\\
\left.\dfrac{\partial}{\partial t}\mu_t^D(\bbX)\right|_{t=0}
&=\E[\{D-\mu^D(\bbX)\}s(Y,D,Z,\bbX)\mid \bbX],
\end{align*}
it follows that
\begin{align*}
\left.\dfrac{\partial}{\partial t}\delta_{\gamma_0,t}^Y(\bbX)\right|_{t=0}
&=\left.\dfrac{\partial}{\partial t}\E_t\left[\{\mu_t^D(Z,\bbX)-\mu_t^D(\bbX)\}Y\mid \bbX\right]\right|_{t=0}\\
&=\left.\dfrac{\partial}{\partial t}\E\left[\{\mu_t^D(Z,\bbX)-\mu_t^D(\bbX)\}Y\mid \bbX\right]\right|_{t=0}
+\E[\{\mu^D(Z,\bbX)-\mu^D(\bbX)\}Y s(Y,D,Z\mid\bbX)\mid\bbX]\\
&=\E[\{\gamma_0(Z,\bbX)-\E[\gamma_0(Z,\bbX)\mid\bbX]\}Y s(Y,D,Z,\bbX)\mid\bbX]-\delta_{\gamma_0}^Y(\bbX)s(\bbX)\\
&\quad+\E[\{D-\mu^D(Z,\bbX)\}\mu^Y(Z,\bbX)s(Y,D,Z,\bbX)\mid\bbX]\\
&\quad
-\E[\{D-\mu^D(\bbX)\}\mu^Y(\bbX)s(Y,D,Z,\bbX)\mid\bbX].
\end{align*}
Similarly,
\begin{align*}
\left.\dfrac{\partial}{\partial t}\delta_{\gamma_0,t}^D(\bbX)\right|_{t=0}
&=\E[\{\gamma_0(Z,\bbX)-\E[\gamma_0(Z,\bbX)\mid\bbX]\}D s(Y,D,Z,\bbX)\mid\bbX]-\delta_{\gamma_0}^D(\bbX)s(\bbX)\\
&\quad+\E[\{D-\mu^D(Z,\bbX)\}\mu^D(Z,\bbX)s(Y,D,Z,\bbX)\mid\bbX] \\
&\quad
-\E[\{D-\mu^D(\bbX)\}\mu^D(\bbX)s(Y,D,Z,\bbX)\mid\bbX].
\end{align*}
Therefore,
\begin{align*}
\left.\dfrac{\partial}{\partial t}\Delta_{\gamma_0,t}\right|_{t=0}
&=\E[\delta_{\gamma_0}(\bbX)s(Y,D,Z,\bbX)]\\
&\quad+\E\left[\dfrac{1}{\delta_{\gamma_0}^D(\bbX)}\E[\{\gamma_0(Z,\bbX)-\E[\gamma_0(Z,\bbX)\mid\bbX]\}\{Y-\delta_{\gamma_0}(\bbX)D\}s(Y,D,Z,\bbX)\mid\bbX]\right]\\
&\quad+\E\left[\dfrac{1}{\delta_{\gamma_0}^D(\bbX)}\E[\{D-\mu^D(Z,\bbX)\}\{\mu^Y(Z,\bbX)-\delta_{\gamma_0}(\bbX)\mu^D(Z,\bbX)\}s(Y,D,Z,\bbX)\mid\bbX]\right]\\
&\quad-\E\left[\dfrac{1}{\delta_{\gamma_0}^D(\bbX)}\E[\{D-\mu^D(\bbX)\}\{\mu^Y(\bbX)-\delta_{\gamma_0}(\bbX)\mu^D(\bbX)\}s(Y,D,Z,\bbX)\mid\bbX]\right]\\
&=\E[\delta_{\gamma_0}(\bbX)s(Y,D,Z,\bbX)] \\
&\quad
+\E\left[\dfrac{\gamma_0(Z,\bbX)-\E[\gamma_0(Z,\bbX)\mid\bbX]}{\delta_{\gamma_0}^D(\bbX)}\{Y-\delta_{\gamma_0}(\bbX)D\}s(Y,D,Z,\bbX)\right]\\
&\quad+\E\left[\dfrac{\{D-\mu^D(Z,\bbX)\}\{\mu^Y(Z,\bbX)-\delta_{\gamma_0}(\bbX)\mu^D(Z,\bbX)\}}{\delta_{\gamma_0}^D(\bbX)}s(Y,D,Z,\bbX)\right]\\
&\quad-\E\left[\dfrac{\{\gamma_0(Z,\bbX)+(D-\mu^D(Z,\bbX))\}\{\mu^Y(\bbX)-\delta_{\gamma_0}(\bbX)\mu^D(\bbX)\}}{\delta_{\gamma_0}^D(\bbX)}s(Y,D,Z,\bbX)\right]\\
&=\E[\Tilde{\Psi}_2(\bbO; P)\,s(Y,D,Z,\bbX)],
\end{align*}
where
\begin{align*}
\Tilde{\Psi}_2(\bbO; P)
&=\dfrac{\{\gamma_0(Z,\bbX)-\E[\gamma_0(Z,\bbX)\mid\bbX]\}\{Y-\mu^Y(\bbX)-\delta_{\gamma_0}(\bbX)(D-\mu^D(\bbX))\}}{\delta_{\gamma_0}^D(\bbX)}\\
&\quad+\dfrac{\{D-\mu^D(Z,\bbX)\}\{\mu^Y(Z,\bbX)-\mu^Y(\bbX)-\delta_{\gamma_0}(\bbX)(\mu^D(Z,\bbX)-\mu^D(\bbX))\}}{\delta_{\gamma_0}^D(\bbX)} \\
& \quad +\delta_{\gamma_0}(\bbX).
\end{align*}
It is straightforward to verify that $\E[\Tilde{\Psi}_2(\bbO; P)]=\Delta_{\gamma_0}$. Hence $\Psi_{\gamma_0}(\bbO; P)=\Tilde{\Psi}_2(\bbO; P)-\Delta_{\gamma_0}$ is an influence function of $\Delta_{\gamma_0}$. Since $\calM_{\np}$ is the nonparametric model, $\Psi_{\gamma_0}(\bbO; P)$ is the efficient influence function of $\Delta_{\gamma_0}$.

Further under $\calM_{\sp}$, $\mu^Y(Z,\bbX)-\mu^Y(\bbX)=\delta_{\gamma_0}(\bbX)\{\mu^D(Z,\bbX)-\mu^D(\bbX)\}$, and hence
\begin{align}\label{S-eqn: IF1}
\dfrac{\mu^D(Z,\bbX)-\mu^D(\bbX)}{\E[\{\mu^D(Z,\bbX)-\mu^D(\bbX)\}D\mid\bbX]}\calE+\delta(\bbX)-\Delta
\end{align}
is an influence function of $\Delta$.
The local efficiency follows because $\E[\{\mu^D(Z,\bbX)-\mu^D(\bbX)\}D\mid\bbX]= \var\{\mu^D(Z,\bbX) \mid \bbX \}$ and hence \eqref{S-eqn: IF1} $=$ \eqref{eq:eif_conhom}. Therefore, \eqref{S-eqn: IF1} coincides with $\Psi_{\eff}(\bbO; P)$ on $\calM_{\ch}$.

\subsection{Proof of Theorem \ref{th:Delta_1}}\label{S-app:th:Delta_1}
\subsubsection{Proof of Theorem \ref{th:Delta_1}}
First note that
\begin{equation*}
\widehat{\Delta}_{\mathrm{RM}}=\dfrac{1}{n}\sum_{k=1}^K\sum_{i\in\mathcal I_k}\psi_{\gamma_0}(\bbO_i;\widehat P_{(-k)})= \sum_{k=1}^K \dfrac{n_k}{n}\mathbb{P}_{\mathcal{I}_{k}} \left\{   \psi_{\gamma_0}(\bbO; \widehat{P}_{(-k)})\right\},
\end{equation*}
where $n_k = |\mathcal{I}_k|$ and $\mathbb{P}_{\mathcal{I}_{k}} \left\{ \psi_{\gamma_0}(\bbO; \widehat{P}_{(-k)})\right\} = \dfrac{1}{n_k} \sum_{i \in \mathcal{I}_k} \psi_{\gamma_0}(\bbO_i; \widehat{P}_{(-k)})$.
Denote $\widehat{\Delta}_{k}
=\mathbb{P}_{\mathcal{I}_{k}}\psi_{\gamma_0}(\bbO;\widehat P_{(-k)})$. We establish below that
\begin{equation*}
\sqrt{n_k}\big(\widehat{\Delta}_{k}-\Delta\big)
=\dfrac{1}{\sqrt{n_k}}\sum_{i\in\mathcal I_k}\{\psi_{\gamma_0}(\bbO_i;P)-\Delta\}+o_p(1).
\end{equation*}
Then
\begin{align*}
\sqrt n(\widehat{\Delta}_{\mathrm{RM}}-\Delta)
&=\sqrt n\left\{\dfrac{1}{n}\sum_{k=1}^K\sum_{i\in\mathcal I_k}\psi_{\gamma_0}(\bbO_i;\widehat P_{(-k)})-\Delta\right\}\\
&=\dfrac{1}{\sqrt n}\sum_{i=1}^n\{\psi_{\gamma_0}(\bbO_i;P)-\Delta\}+o_p(1).
\end{align*}
Let $\E_{(-k)}[f(\bbO)] = \E[f(\bbO) \mid \mathcal I_k^c]$ be the conditional expectation that $f(\bbO)$ is a fixed function after conditional on the data from $\mathcal I_k^c$. Let $\mathbb{G}_{\mathcal{I}_k} f(\bbO)=\dfrac{1}{\sqrt{n_k}} \sum_{i \in \mathcal{I}_k}\{ f(\bbO_i) - \E_P[f(\bbO)]\}$ be the empirical process of $f(\bbO)$ and $\mathbb{G}_{\mathcal{I}_k, (-k)} \widehat{f}(\bbO) = \dfrac{1}{\sqrt{n_k}} \sum_{i \in \mathcal{I}_k}\{\widehat{f}(\bbO_i) - \E_{(-k)}[\widehat{f}(\bbO)]\}$ be the empirical process of $\widehat{f}(\bbO)$. We can show that
\begin{align}
\sqrt{n_k}\{\widehat{\Delta}_{k}-\Delta\}
&=\sqrt{n_k}\{\mathbb{P}_{\mathcal{I}_{k}}\psi_{\gamma_0}(\bbO;\widehat P_{(-k)})-\Delta\} \notag\\
&=\dfrac{1}{\sqrt{n_k}}\sum_{i\in\mathcal I_k}\{\psi_{\gamma_0}(\bbO_i;\widehat P_{(-k)})-\Delta\} \notag\\
&=\dfrac{1}{\sqrt{n_k}}\sum_{i\in\mathcal I_k}\{\psi_{\gamma_0}(\bbO_i;P)-\Delta\} \label{S-term:if}\\
&\quad+\sqrt{n_k}\,\E_{(-k)}\{\psi_{\gamma_0}(\bbO;\widehat P_{(-k)})-\psi_{\gamma_0}(\bbO;P)\} \label{S-term1}\\
&\quad+\mathbb{G}_{\mathcal{I}_k,(-k)}\{\psi_{\gamma_0}(\bbO;\widehat P_{(-k)})-\psi_{\gamma_0}(\bbO;P)\}. \label{S-term2}
\end{align}
Note that \eqref{S-term:if} is asymptotically normal by the central limit theorem. If we can show \eqref{S-term1} and \eqref{S-term2} are $o_p(1)$, then we will have
\begin{align*}
    \sqrt{n}(\widehat{\Delta}_{\mathrm{RM}}  - \Delta) = \dfrac{1}{\sqrt{n}} \sum_{i=1}^n \Psi_{\gamma_0} (\bbO_i; P) + o_p(1) \to_d N(0, \mathcal{V}),
\end{align*}
where $\Psi_{\gamma_0}(\bbO;P)=\psi_{\gamma_0}(\bbO;P)-\Delta$ and $\mathcal{V} = \var\{\Psi_{\gamma_0}(\bbO; P)\}$.

For simplicity, we omit the random variable $\bbX$ in the following proof. We first show \eqref{S-term1} is $o_p(1)$ by showing that $\dfrac{1}{\sqrt{n_k}}$ \eqref{S-term1} is $o_p(n^{-1/2})$.
\begin{align*}
&\quad\ \E_{(-k)}\{\psi_{\gamma_0}(\bbO;\widehat P_{(-k)})-\psi_{\gamma_0}(\bbO;P)\}\\
&=\E_{(-k)}
\Bigg[
\widehat{\xi}^{D}_{(-k)}(Z)\widehat{\tau}_{(-k)}
\left\{Y-\widehat{\mu}^{Y}_{(-k)}
-\widehat{\delta}_{(-k)}\big(D-\widehat{\mu}^{D}_{(-k)}\big)\right\}
+\widehat{\delta}_{(-k)}  -\xi^D(Z)\tau
\left\{Y-\mu^Y-\delta\big(D-\mu^D\big)\right\}
-\delta \Bigg]\\
& = \E_{(-k)}\Bigg[ \widehat{\xi}^{D}_{(-k)}(Z)\widehat{\tau}_{(-k)} \left\{  -\big(\widehat{\mu}^{Y}_{(-k)}-\mu^Y\big) + \widehat{\delta}_{(-k)}\big(\widehat{\mu}^{D}_{(-k)}-\mu^D\big) - \big(\widehat{\delta}_{(-k)}-\delta\big)\xi^D(Z) \right\} + \widehat{\delta}_{(-k)}-\delta \Bigg] \\
&=\E_{(-k)} \Bigg[  \{\widehat{\xi}^{D}_{(-k)}(Z) - \xi^D(Z)\} \widehat{\tau}_{(-k)} \left\{ -\big(\widehat{\mu}^{Y}_{(-k)}-\mu^Y\big) + \widehat{\delta}_{(-k)}\big(\widehat{\mu}^{D}_{(-k)}-\mu^D\big) - \big(\widehat{\delta}_{(-k)}-\delta\big)\xi^D(Z)\right\} \Bigg] \\
& \quad +  \E_{(-k)} \Bigg[ \xi^D(Z) \{\widehat{\tau}_{(-k)} - \tau\} \left\{ -\big(\widehat{\mu}^{Y}_{(-k)}-\mu^Y\big) + \widehat{\delta}_{(-k)}\big(\widehat{\mu}^{D}_{(-k)}-\mu^D\big) - \big(\widehat{\delta}_{(-k)}-\delta\big)\xi^D(Z) \right\}\Bigg] \\
&=\E_{(-k)} \Bigg[  \{\widehat{\xi}^{D}_{(-k)}(Z) - \xi^D(Z)\} \widehat{\tau}_{(-k)} \left\{ -\big(\widehat{\mu}^{Y}_{(-k)}-\mu^Y\big) + \widehat{\delta}_{(-k)}\big(\widehat{\mu}^{D}_{(-k)}-\mu^D\big) - \big(\widehat{\delta}_{(-k)}-\delta\big)\xi^D(Z)\right\} \Bigg] \\
& \quad - \E_{(-k)} \Bigg[ \{\xi^D(Z)\}^2 \{\widehat{\tau}_{(-k)} - \tau\} \big(\widehat{\delta}_{(-k)}-\delta\big)  \Bigg]
\end{align*}
Since $D$ is binary, $\mu^D(Z)$ and $\mu^D$ belong to $[0,1]$. It follows that $\xi^D(Z) \in [-1, 1]$ and $\{\xi^D(Z)\}^2 \in [0, 1]$. By Assumptions~\ref{as:Boundedness} and \ref{as:Rate},
\begin{align*}
& \E_{(-k)}\{\psi_{\gamma_0}(\bbO;\widehat P_{(-k)})-\psi_{\gamma_0}(\bbO;P)\}\\
&\lesssim \left(r_{D,Z}^{(-k)}+r_D^{(-k)}\right)\left(r_D^{(-k)}+r_Y^{(-k)}+r_\delta^{(-k)}\right)+r_\tau^{(-k)}r_\delta^{(-k)}\\
&= o_p(n^{-1/2}).
\end{align*}

We then show that \eqref{S-term2} is $o_p(1)$. Note that $\E_{(-k)}\left[\mathbb{G}_{\mathcal{I}_k, (-k)} \left\{ \psi_{\gamma_0}(\bbO; \widehat{P}_{(-k)}) - \psi_{\gamma_0}(\bbO; P) \right\}\right]=0$. The variance of \eqref{S-term2} is
\begin{align*}
&\var_{(-k)}\left(\mathbb{G}_{\mathcal{I}_k,(-k)}
\left\{\psi_{\gamma_0}(\bbO;\widehat P_{(-k)})-\psi_{\gamma_0}(\bbO;P)\right\}\right) \\
&\lesssim
\var_{(-k)}\left\{\psi_{\gamma_0}(\bbO;\widehat P_{(-k)})-\psi_{\gamma_0}(\bbO;P)\right\} \\
& \leq
\E_{(-k)}\left[\left\{\psi_{\gamma_0}(\bbO;\widehat P_{(-k)})-\psi_{\gamma_0}(\bbO;P)\right\}^2\right] \\
&= \E_{(-k)}
\left[
\left\{
\left( \widehat{\xi}^{D}_{(-k)}(Z)\widehat{\tau}_{(-k)} - \xi^D(Z) \tau  \right)\calE
+\widehat{\xi}^{D}_{(-k)}(Z) \widehat{\tau}_{(-k)} \left(\widehat{\calE}_{(-k)}-\calE\right)
+  \widehat{\delta}_{(-k)}-\delta
\right\}^2
\right] \\
&\lesssim \E_{(-k)}\left[ \left\{ \left[\left( \widehat{\xi}^{D}_{(-k)}(Z) - \xi^D(Z)\right)\widehat{\tau}_{(-k)} + \xi^D(Z) \left( \widehat{\tau}_{(-k)} - \tau\right) \right] \calE \right\}^2\right] \\
& \quad + \E_{(-k)}\left[ \left\{ \widehat{\xi}^{D}_{(-k)}(Z) \widehat{\tau}_{(-k)} \left[ -\left(\widehat{\mu}^{Y}_{(-k)}-\mu^Y\right)
-\left(\widehat{\delta}_{(-k)}-\delta\right)D
+\widehat{\delta}_{(-k)}\left(\widehat{\mu}^{D}_{(-k)}-\mu^D\right)
+\left(\widehat{\delta}_{(-k)}-\delta\right)\mu^D
 \right] \right\}^2\right] \\
& \quad + \E_{(-k)}
\left[\left(\widehat{\delta}_{(-k)}-\delta\right)^2\right] \\
&\lesssim \left(r_{D,Z}^{(-k)}\right)^2+\left(r_D^{(-k)}\right)^2+\left(r_\tau^{(-k)}\right)^2+\left(r_Y^{(-k)}\right)^2+\left(r_\delta^{(-k)}\right)^2 \\
&= o_p(1).
\end{align*}

It remains to show that the variance estimator $\widehat{\mathcal{V}}$ is consistent. Write
\begin{align*}
\widehat{\mathcal{V}}=\sum_{k=1}^K \dfrac{n_k}{n}\widehat{\mathcal{V}}_k, \qquad
\widehat{\mathcal{V}}_k=\mathbb{P}_{\mathcal{I}_{k}}\left[\left\{\psi_{\gamma_0}(\bbO;\widehat P_{(-k)})-\widehat{\Delta}_{\mathrm{RM}}\right\}^2\right].
\end{align*}
Let $\Tilde{\mathcal{V}}_k=\mathbb{P}_{\mathcal{I}_{k}}\left[\left\{\psi_{\gamma_0}(\bbO;P)-\Delta\right\}^2\right]$ and $\Tilde{\mathcal{V}}=\sum_{k=1}^K \dfrac{n_k}{n}\Tilde{\mathcal{V}}_k$. By the law of large numbers, $\Tilde{\mathcal{V}}-\mathcal{V}=o_p(1)$. It is enough to prove $\widehat{\mathcal{V}}_k-\Tilde{\mathcal{V}}_k=o_p(1)$ for each $k$. By $a^2-b^2=(a-b)^2+2b(a-b)$ and Cauchy--Schwarz inequality,
\begin{align*}
\left|\widehat{\mathcal{V}}_k-\Tilde{\mathcal{V}}_k\right|
\leq&\ \mathbb{P}_{\mathcal{I}_{k}}\left[\left\{\psi_{\gamma_0}(\bbO;\widehat P_{(-k)})-\widehat{\Delta}_{\mathrm{RM}}-\psi_{\gamma_0}(\bbO;P)+\Delta\right\}^2\right] \\
& +2\left\{\mathbb{P}_{\mathcal{I}_{k}}\left[\left\{\psi_{\gamma_0}(\bbO;\widehat P_{(-k)})-\widehat{\Delta}_{\mathrm{RM}}-\psi_{\gamma_0}(\bbO;P)+\Delta\right\}^2\right]\right\}^{1/2}
\left\{\mathbb{P}_{\mathcal{I}_{k}}\left[\left\{\psi_{\gamma_0}(\bbO;P)-\Delta\right\}^2\right]\right\}^{1/2}.
\end{align*}
Since $\mathbb{P}_{\mathcal{I}_{k}}\left[\left\{\psi_{\gamma_0}(\bbO;P)-\Delta\right\}^2\right]=\mathcal{V}+o_p(1)=O_p(1)$, it suffices to show that the first term on the right-hand side is $o_p(1)$. By $(a+b)^2\lesssim a^2+b^2$ and the law of large numbers,
\begin{align*}
& \mathbb{P}_{\mathcal{I}_{k}}\left[\left\{\psi_{\gamma_0}(\bbO;\widehat P_{(-k)})-\widehat{\Delta}_{\mathrm{RM}}-\psi_{\gamma_0}(\bbO;P)+\Delta\right\}^2\right] \\
\lesssim&\ \mathbb{P}_{\mathcal{I}_{k}}\left[\left\{\psi_{\gamma_0}(\bbO;\widehat P_{(-k)})-\psi_{\gamma_0}(\bbO;P)\right\}^2\right]+(\widehat{\Delta}_{\mathrm{RM}}-\Delta)^2 \\
= & \ \E_{(-k)}\left[\left\{\psi_{\gamma_0}(\bbO;\widehat P_{(-k)})-\psi_{\gamma_0}(\bbO;P)\right\}^2\right] +(\widehat{\Delta}_{\mathrm{RM}}-\Delta)^2 + o_p(1) \\
= & \ o_p(1).
\end{align*}
Therefore $\widehat{\mathcal{V}}_k-\Tilde{\mathcal{V}}_k=o_p(1)$ for each $k$, and hence $\widehat{\mathcal{V}}-\Tilde{\mathcal{V}}=o_p(1)$. Together with $\Tilde{\mathcal{V}}-\mathcal{V}=o_p(1)$, this gives $\widehat{\mathcal{V}}-\mathcal{V}=o_p(1)$.

This concludes our proof.

\subsubsection{Role of Additional Sample Splitting in Error Control}\label{S-app:samplesplit}
To illustrate the role of the additional split, suppose that $\widehat \xi^D_{(-k,c_j)}(Z,\bbX)$ is constructed using observations in $\mathcal I_k^{c_j}$, whereas $\widehat\tau_{(-k,c_{3-j})}(\bbX)$ is fitted using observations in $\mathcal I_k^{c_{j'}}$ for $j' = 3-j$, $j\in\{1,2\}$. Then
\begin{align*}
\widehat\tau_{(-k,c_{j'})}(\bbX)-\tau(\bbX)
=\{\widehat\tau_{(-k,c_{j'})}(\bbX)-\tau^o_{(-k,c_{j})}(\bbX)\}+\{\tau^o_{(-k,c_{j})}(\bbX)-\tau(\bbX)\},
\end{align*}
where
\begin{align*}
\tau^o_{(-k,c_{j})}(\bbX)
=\argmin_{g \in L^2(P_{\bbX})}\E\left[\widehat \xi^D_{(-k,c_j)}(Z,\bbX)^2\left\{\dfrac{1}{\widehat \xi^D_{(-k,c_j)}(Z,\bbX)^2}-g(\bbX)\right\}^2 \,\middle|\, \mathcal I_k^{c_j} \right]
\end{align*}
is the conditional population target induced by $\widehat \xi^D_{(-k,c_j)}(Z,\bbX)$. The $L^2(P_{\bbX})$-norm of the second term, $\tau^o_{(-k,c_{j})}(\bbX)-\tau(\bbX)$, is controlled by $\|\widehat\mu^D_{(-k,c_j)}(Z,\bbX)-\mu^D(Z,\bbX)\|_{P,2}$ and $\|\widehat\mu^D_{(-k,c_j)}(\bbX)-\mu^D(\bbX)\|_{P,2}$ under boundedness and positivity conditions. The first term, $\widehat\tau_{(-k,c_{j'})}(\bbX)-\tau^o_{(-k,c_{j})}(\bbX)$, is the oracle learning error of the second-stage regression. Conditional on $\mathcal I_k^{c_j}$, $\widehat \xi^D_{(-k,c_j)}$ is fixed and independent of the observations in $\mathcal I_k^{c_{j'}}$, so the $L^2(P_{\bbX})$-norm of this term can be controlled by the $L^2(P_{\bbX})$ learning rate of the second-stage regression. Without additional sample splitting, one often still needs to impose Donsker-type empirical-process conditions to control the error rate of the first term, which can be difficult to justify with high-dimensional covariates or highly adaptive learning methods.

\subsection[Derivation of the Riesz representer]{Proof of Lemma \ref{S-le:rr}}
\phantomsection
\label{S-app:le:rr}

When $Z$ is categorical, under Assumption~\ref{S-A1}, we have
\begin{align*}
&\ \E[m_\omega(Z,\bbX; h)\mid\bbX=\bbx]\\
=&\ \sum_{0 \leq k < j \leq J-1}\omega_{jk}(\bbx)\{h(j,\bbx)-h(k,\bbx)\}\\
=&\ \sum_{z=0}^{J-1}\left\{\sum_{k<z}\omega_{zk}(\bbx)-\sum_{k>z}\omega_{kz}(\bbx)\right\}h(z,\bbx)\\
=&\ \sum_{z=0}^{J-1}p(z\mid\bbx)\gamma_\omega(z,\bbx)h(z,\bbx)\\
=&\ \E[\gamma_\omega(Z,\bbX)h(Z,\bbX)\mid\bbX=\bbx],
\end{align*}
where $\gamma_\omega(Z,\bbX)= \dfrac{\sum_{k<z}\omega_{zk}(\bbx)-\sum_{k>z}\omega_{kz}(\bbx)}{p(z \mid \bbX)}$.
Taking expectations over $\bbX$ gives
\begin{align*}
    \E[m_\omega(Z,\bbX;h)] =    \E[\gamma_\omega(Z,\bbX)h(Z,\bbX)].
\end{align*}
This proves the conditional and marginal identities in the categorical case.

When $Z$ is continuous, let $\Omega(\bbx)=(s(\bbx),t(\bbx))$, where $s(\bbx) = \inf \Omega(\bbx)$ and $t(\bbx) = \sup \Omega(\bbx)$. Under Assumption~\ref{S-A2}, we have
\begin{align*}
&\ \E[\omega(Z,\bbX)\dot h(Z,\bbX)\mid\bbX=\bbx]\\
=&\ \int_{s(\bbx)}^{t(\bbx)}
\omega(z,\bbx)\dot h(z,\bbx)p(z\mid\bbx)\diff z\\
=&\ \left[
\omega(z,\bbx)p(z\mid\bbx)h(z,\bbx)
\right]_{s(\bbx)}^{t(\bbx)}
-\int_{s(\bbx)}^{t(\bbx)}
h(z,\bbx)
\dfrac{\partial\{\omega(z,\bbx)p(z\mid\bbx)\}}{\partial z}
\diff z\\
=&\ -\int_{s(\bbx)}^{t(\bbx)}
h(z,\bbx)
\dfrac{\partial\{\omega(z,\bbx)p(z\mid\bbx)\}}{\partial z}
\diff z\\
=&\ \int_{s(\bbx)}^{t(\bbx)}
\gamma_\omega(z,\bbx)h(z,\bbx)p(z\mid\bbx)
\diff z\\
=&\ \E[\gamma_\omega(Z,\bbX)h(Z,\bbX)\mid\bbX=\bbx],
\end{align*}
where $\left[
\omega(z,\bbx)p(z\mid\bbx)h(z,\bbx)
\right]_{s(\bbx)}^{t(\bbx)}=0$ by Assumption~\ref{S-A2} and $h$ is bounded, and
\[
    \gamma_\omega(z,\bbx)
    =
    -\dfrac{\partial \omega(z,\bbx)}{\partial z}
    -\omega(z,\bbx)
    \dfrac{\partial \log p(z\mid\bbx)}{\partial z}.
\]
Taking expectations over $\bbX$ gives
\[
    \E[\omega(Z,\bbX)\dot h(Z,\bbX)]
    =
    \E[\gamma_\omega(Z,\bbX)h(Z,\bbX)].
\]
This proves the conditional and marginal identities in the continuous case.

\subsection[Proof of the conditional linear-functional identity]{Proof of \eqref{S-eqn: conditionallf}}
\label{S-app:le:crrr}

According to \eqref{omega},
\begin{equation}\label{S-eqn:ode}
    \gamma(z, \bbx) p(z \mid \bbx) = - \dfrac{\partial \{\omega(z, \bbx)p(z \mid \bbx) \}}{\partial z}.
\end{equation}
Integrating \eqref{S-eqn:ode} from $s(\bbx) = \inf \Omega(\bbx)$ to $z$ yields
\begin{equation*}
\omega(z, \bbx)p(z \mid \bbx)  =- \int_{s(\bbx)}^{z} \gamma(\tilde{z}, \bbx) p(\tilde{z} \mid \bbx) \diff \tilde{z}
\end{equation*}
Since $\E[\gamma(Z, \bbX) \mid \bbX]=0$, it follows that $\lim_{z \to  \inf\Omega(\bbx)}\omega(z,\bbx)p(z\mid\bbx) = \lim_{z \to \sup\Omega(\bbx)}\omega(z,\bbx)p(z\mid\bbx)=0$.
For every bounded $h$ such that $z\mapsto h(z,x)$ is absolutely continuous, integration by parts gives
\begin{align*}
\E[\omega(Z, \bbX)\dot h(Z, \bbX)\mid  \bbX= \bbx]
&= \int_{s(\bbx)}^{t(\bbx)} \omega(z, \bbx)p(z \mid \bbx) \dot h(z,  \bbx)\diff z  \\
&= \left[\omega(z, \bbx)p(z \mid \bbx) h(z, \bbx)\right]_{s(x)}^{t(x)} - \int_{s(\bbx)}^{t(\bbx)} h(z,\bbx) \dfrac{\partial \omega(z, \bbx)p(z \mid \bbx)}{\partial z}\diff z \\
&= \int_{s(\bbx)}^{t(\bbx)} h(z, \bbx)\gamma(z, \bbx)p(z \mid \bbx)\diff z  \\
&= \E[\gamma(Z, \bbX)h(Z, \bbX)\mid \bbX=\bbx].
\end{align*}

This proves \eqref{S-eqn: conditionallf}.

\subsection{Proof of Theorem~\ref{S-th:Delta_eff}}\label{S-prof:Delta_eff}
The proof of Theorem~\ref{S-th:Delta_eff} follows the proof of Theorem~\ref{th:Delta_1}. Note that
\begin{align*}
    \widehat{\Delta}^{\eff}_{\mathrm{DS}} = \dfrac{1}{n}\sum_{k=1}^K\sum_{i \in \mathcal{I}_{k}} \psi(\bbO_i; \widehat{P}_{(-k)}) = \sum_{k=1}^K \dfrac{n_k}{n}\mathbb{P}_{\mathcal{I}_{k}} \left\{   \psi(\bbO; \widehat{P}_{(-k)})\right\}.
\end{align*}

Denote $\widehat{\Delta}_{\,\eff,k}= \mathbb{P}_{\mathcal{I}_{k}} \left\{  \psi(\bbO; \widehat{P}_{(-k)})\right\} $. We establish below that
\begin{align*}
    \sqrt{n_k} \left\{\widehat{\Delta}_{\,\eff,k} - \Delta \right\} = \dfrac{1}{\sqrt{n_k}} \sum_{i \in \mathcal{I}_k} \Psi_{\eff}(\bbO_i; P) +  o_p(1).
\end{align*}
It then follows that
\begin{align*}
    \sqrt{n}( \widehat{\Delta}^{\eff}_{\mathrm{DS}}  - \Delta) & =  \sqrt{n}\left(\sum_{k=1}^K \dfrac{n_k}{n} \widehat{\Delta}_{\,\eff,k} - \Delta  \right) \\
    & = \dfrac{1}{\sqrt{n}} \sum_{i=1}^n \Psi_{\eff} (\bbO_i; P) + o_p(1).
\end{align*}
Using the notation $\E_{(-k)}$ and $\mathbb{G}_{\mathcal{I}_k,(-k)}$ from the proof of Theorem~\ref{th:Delta_1}, we can show that
\begin{align}
     \sqrt{n_k} \left\{\widehat{\Delta}_{\,\eff,k}  - \Delta \right\} & = \sqrt{n_k} \left\{  \mathbb{P}_{\mathcal{I}_{k}} \left\{  \psi(\bbO; \widehat{P}_{(-k)})\right\} - \Delta \right\} \notag \\
     & = \dfrac{1}{\sqrt{n_k}} \sum_{i \in \mathcal{I}_{k}} \left\{\psi(\bbO_i; \widehat{P}_{(-k)}) - \Delta \right\} \notag  \\
     & = \dfrac{1}{\sqrt{n_k}} \sum_{i \in \mathcal{I}_{k}} \Psi_{\eff} (\bbO_i; P) \label{S-term:eif} \\
     & \quad + \sqrt{n_k} \E_{(-k)} \left\{ \psi(\bbO; \widehat{P}_{(-k)}) - \psi(\bbO; P)\right\} \label{S-term:remaindereif} \\
     & \quad  + \mathbb{G}_{\mathcal{I}_k, (-k)} \left\{ \psi(\bbO; \widehat{P}_{(-k)}) - \psi(\bbO; P) \right\}. \label{S-term:empiricaleif}
\end{align}
Note that \eqref{S-term:eif} is asymptotically normal by the central limit theorem. If we can show \eqref{S-term:remaindereif} and \eqref{S-term:empiricaleif} are $o_p(1)$, then we will have
\begin{align*}
    \sqrt{n}(\widehat{\Delta}^{\eff}_{\mathrm{DS}}  - \Delta) = \dfrac{1}{\sqrt{n}} \sum_{i=1}^n \Psi_{\eff} (\bbO_i; P) + o_p(1) \to_d N(0, \mathcal{V}_{\eff}),
\end{align*}
where $\psi(\bbO;P)=\gamma_{\opt}(Z)\calE+\delta$, $\Psi_{\eff}(\bbO;P)=\psi(\bbO;P)-\Delta$ and $\mathcal{V}_{\eff} = \var\{\Psi_{\eff}(\bbO; P)\}$.

For simplicity, we omit the random variable $\bbX$ in the following proof. Write $\gamma_{\rm opt}(Z)=N(Z)/M$ and $\widehat{\gamma}_{{\rm opt}, (-k)}(Z)=\widehat{N}_{(-k)}(Z)/\widehat{M}_{(-k)}$, where $N$ and $M$ are the numerator and denominator of $\gamma_{\rm opt}$, respectively. We first show \eqref{S-term:remaindereif} is $o_p(1)$ by showing that $\dfrac{1}{\sqrt{n_k}}$\eqref{S-term:remaindereif} is $o_p(n^{-1/2})$. Note that
\begin{align*}
& \E_{(-k)} \left\{ \psi(\bbO; \widehat{P}_{(-k)}) - \psi(\bbO; P)\right\} \\
= \ & \E_{(-k)} \Bigg[ \dfrac{\widehat{\kappa}_{1,(-k)}\big(\widehat{\mu}^D_{(-k)}(Z)-\mu^D(Z)\big)} {\widehat{M}_{(-k)}\,\widehat{\sigma}^2_{(-k)}(Z)}
\Big\{ \widehat{\delta}_{(-k)} \big(\widehat{\mu}^D_{(-k)}-\mu^D\big)
-\big(\widehat{\mu}^Y_{(-k)}-\mu^Y\big)
-\big(\widehat{\delta}_{(-k)}-\delta\big)\big(\mu^D(Z)-\mu^D\big)
\Big\} \Bigg] \\
  & + \E_{(-k)} \Bigg[ \dfrac{\mu^D(Z)\big(\widehat{\kappa}_{1,(-k)}-\kappa_1\big)} {\widehat{M}_{(-k)}\,\widehat{\sigma}^2_{(-k)}(Z)}
\Big\{ \widehat{\delta}_{(-k)} \big(\widehat{\mu}^D_{(-k)}-\mu^D\big)
-\big(\widehat{\mu}^Y_{(-k)}-\mu^Y\big)
-\big(\widehat{\delta}_{(-k)}-\delta\big)\big(\mu^D(Z)-\mu^D\big)
\Big\} \Bigg]\\
& - \E_{(-k)} \Bigg[ \dfrac{\big(\widehat{\kappa}_{2,(-k)}-\kappa_2\big)} {\widehat{M}_{(-k)}\,\widehat{\sigma}^2_{(-k)}(Z)}
\Big\{ \widehat{\delta}_{(-k)} \big(\widehat{\mu}^D_{(-k)}-\mu^D\big)
-\big(\widehat{\mu}^Y_{(-k)}-\mu^Y\big)
-\big(\widehat{\delta}_{(-k)}-\delta\big)\big(\mu^D(Z)-\mu^D\big)
\Big\} \Bigg] \\
& - \E_{(-k)} \Bigg[ \dfrac{N(Z)\big(\widehat{\sigma}^2_{(-k)}(Z)-\sigma^2(Z)\big)} {\widehat{M}_{(-k)}\,\widehat{\sigma}^2_{(-k)}(Z)}
\Big\{ \widehat{\delta}_{(-k)} \big(\widehat{\mu}^D_{(-k)}-\mu^D\big)
-\big(\widehat{\mu}^Y_{(-k)}-\mu^Y\big)
-\big(\widehat{\delta}_{(-k)}-\delta\big)\big(\mu^D(Z)-\mu^D\big)
\Big\} \Bigg] \\
& + \E_{(-k)} \Bigg[ \dfrac{ \widehat{\delta}_{(-k)}-\delta} {\widehat{M}_{(-k)}} \Big\{ \widehat{\kappa}_{1,(-k)} \big(\widehat{\kappa}_{3,(-k)}-\kappa_3\big) + \kappa_3 \big( \widehat{\kappa}_{1,(-k)} - \kappa_1 \big)- \big(\widehat{\kappa}_{2,(-k)} + \kappa_2 \big)\big(\widehat{\kappa}_{2,(-k)}-\kappa_2\big)  \Big\} \Bigg].
\end{align*}
Since $D$ is binary, $\mu^D(Z) \in [0, 1]$ and $\mu^D \in [0, 1]$, together with the Assumptions in Theorem \ref{S-th:Delta_eff}, we can show that
\begin{align*}
& \E_{(-k)} \left\{ \psi(\bbO; \widehat{P}_{(-k)}) - \psi(\bbO; P)\right\} \\
&\lesssim \big(\left\|\widehat{\mu}^D_{(-k)}(Z)-\mu^D(Z)\right\|_{P,2}+\left\|\widehat{\kappa}_{1,(-k)}-\kappa_1\right\|_{P,2}+\left\|\widehat{\kappa}_{2,(-k)}-\kappa_2\right\|_{P,2}+\left\|\widehat{\sigma}^2_{(-k)}(Z)-\sigma^2(Z)\right\|_{P,2}\big) \\
& \quad\times \big(\left\|\widehat{\mu}^D_{(-k)}-\mu^D\right\|_{P,2}+\left\|\widehat{\mu}^Y_{(-k)}-\mu^Y\right\|_{P,2}+\left\|\widehat{\delta}_{(-k)}-\delta\right\|_{P,2}\big) \\
& \quad+\big(\left\|\widehat{\kappa}_{1,(-k)}-\kappa_1\right\|_{P,2}+\left\|\widehat{\kappa}_{2,(-k)}-\kappa_2\right\|_{P,2}+\left\|\widehat{\kappa}_{3,(-k)}-\kappa_3\right\|_{P,2}\big)\left\|\widehat{\delta}_{(-k)}-\delta\right\|_{P,2} \\
&= o_p(n^{-1/2}).
\end{align*}

We then show that \eqref{S-term:empiricaleif} is $o_p(1)$. First note that $\E_{(-k)}\left[\mathbb{G}_{\mathcal{I}_k, (-k)} \left\{ \psi(\bbO; \widehat{P}_{(-k)}) - \psi(\bbO; P) \right\}\right]=0$. The variance of \eqref{S-term:empiricaleif} is
\begin{align*}
    &  \var_{(-k)}\left(\mathbb{G}_{\mathcal{I}_k, (-k)} \left\{ \psi(\bbO; \widehat{P}_{(-k)}) - \psi(\bbO; P) \right\} \right) \\
    \lesssim \ & \var_{(-k)}\left(  \psi(\bbO; \widehat{P}_{(-k)}) - \psi(\bbO; P) \right) \\
    \leq \ &  \E_{(-k)}\left[ \left\{ \psi(\bbO; \widehat{P}_{(-k)}) - \psi(\bbO; P) \right\}^2 \right] \\
  = \ &  \E_{(-k)}\left[ \left\{ \widehat{\gamma}_{\opt,(-k)}(Z) \widehat{\calE}_{(-k)} - \gamma_{\opt}(Z) \calE + \widehat{\delta}_{(-k)} - \delta \right\}^2 \right]\\
  = \ & \E_{(-k)}\left[ \left\{  (\widehat{\gamma}_{\opt, (-k)}(Z) - \gamma_{\opt}(Z))\calE + \widehat{\gamma}_{\opt,(-k)}(Z)(\widehat{\calE}_{(-k)} - \calE) + \widehat{\delta}_{(-k)} - \delta \right\}^2 \right]\\
  = \ & \E_{(-k)} \, \Bigg[ \, \Bigg\{  \dfrac{ \widehat{N}_{(-k)}(Z) - N(Z) - \gamma_{\opt}(Z) ( \widehat{M}_{(-k)} -M)}{\widehat{M}_{(-k)}}  \left\{Y - \mu^Y - \delta (D - \mu^D)\right\}  \\
  & + \widehat{\gamma}_{\opt, (-k)}(Z) \left\{ -( \widehat{\mu}^Y_{(-k)} - \mu^Y) - (\widehat{\delta}_{(-k)} - \delta)D + \widehat{\delta}_{(-k)} (\widehat{\mu}^D_{(-k)} - \mu^D) + (\widehat{\delta}_{(-k)} - \delta)\mu^D\right\} + \widehat{\delta}_{(-k)} - \delta  \, \Bigg\}^2 \, \Bigg] \\
  \lesssim \ & \E_{(-k)} \, \Bigg[ \, \Bigg\{  \dfrac{ \widehat{N}_{(-k)}(Z) - N(Z) - \gamma_{\opt}(Z) ( \widehat{M}_{(-k)} -M)}{\widehat{M}_{(-k)}} \left\{Y - \mu^Y - \delta (D - \mu^D)\right\} \, \Bigg\}^2 \, \Bigg]  \\
   & + \E_{(-k)} \, \Bigg[ \, \Bigg\{  \widehat{\gamma}_{\opt, (-k)}(Z) \left\{ -( \widehat{\mu}^Y_{(-k)} - \mu^Y) - (\widehat{\delta}_{(-k)} - \delta)D + \widehat{\delta}_{(-k)} (\widehat{\mu}^D_{(-k)} - \mu^D) + (\widehat{\delta}_{(-k)} - \delta)\mu^D\right\} + \widehat{\delta}_{(-k)} - \delta  \, \Bigg\}^2 \, \Bigg] \\
  \lesssim \ & \left\|\widehat{\mu}^D_{(-k)}(Z) - \mu^D(Z)\right\|_{P,2}^2 + \left\|\widehat{\kappa}_{1,(-k)} - \kappa_1\right\|_{P,2}^2 + \left\|\widehat{\kappa}_{2,(-k)} - \kappa_2\right\|_{P,2}^2 + \left\|\widehat{\kappa}_{3,(-k)} - \kappa_3\right\|_{P,2}^2  \\
  & + \left\|\widehat{\mu}^D_{(-k)} - \mu^D\right\|_{P,2}^2 + \left\|\widehat{\mu}^Y_{(-k)} - \mu^Y\right\|_{P,2}^2 + \left\|\widehat{\delta}_{(-k)} - \delta\right\|_{P,2}^2 + \left\|\widehat{\sigma}^2_{(-k)}(Z) - \sigma^2(Z)\right\|_{P,2}^2 \\
  = \ & o_p(1).
\end{align*}

It remains to verify consistency of $\widehat{\mathcal{V}}_{\eff}$. Write
\begin{align*}
\widehat{\mathcal{V}}_{\eff}=\sum_{k=1}^K \dfrac{n_k}{n}\widehat{\mathcal{V}}_{\eff,k}, \qquad
\widehat{\mathcal{V}}_{\eff,k}=\mathbb{P}_{\mathcal{I}_{k}}\left[\left\{\psi(\bbO;\widehat P_{(-k)})-\widehat{\Delta}^{\eff}_{\mathrm{DS}}\right\}^2\right].
\end{align*}
Let $\Tilde{\mathcal{V}}_{\eff,k}=\mathbb{P}_{\mathcal{I}_{k}}\left[\left\{\psi(\bbO;P)-\Delta\right\}^2\right]$ and $\Tilde{\mathcal{V}}_{\eff}=\sum_{k=1}^K \dfrac{n_k}{n}\Tilde{\mathcal{V}}_{\eff,k}$. By the law of large numbers, $\Tilde{\mathcal{V}}_{\eff}-\mathcal{V}_{\eff}=o_p(1)$. It suffices to prove $\widehat{\mathcal{V}}_{\eff,k}-\Tilde{\mathcal{V}}_{\eff,k}=o_p(1)$ for each $k$. As above, $a^2-b^2=(a-b)^2+2b(a-b)$ and Cauchy--Schwarz inequality yield
\begin{align*}
\left|\widehat{\mathcal{V}}_{\eff,k}-\Tilde{\mathcal{V}}_{\eff,k}\right|
\leq&\ \mathbb{P}_{\mathcal{I}_{k}}\left[\left\{\psi(\bbO;\widehat P_{(-k)})-\widehat{\Delta}^{\eff}_{\mathrm{DS}}-\psi(\bbO;P)+\Delta\right\}^2\right] \\
& +2\left\{\mathbb{P}_{\mathcal{I}_{k}}\left[\left\{\psi(\bbO;\widehat P_{(-k)})-\widehat{\Delta}^{\eff}_{\mathrm{DS}}-\psi(\bbO;P)+\Delta\right\}^2\right]\right\}^{1/2}
\left\{\mathbb{P}_{\mathcal{I}_{k}}\left[\left\{\psi(\bbO;P)-\Delta\right\}^2\right]\right\}^{1/2}.
\end{align*}
Also $\mathbb{P}_{\mathcal{I}_{k}}\left[\left\{\psi(\bbO;P)-\Delta\right\}^2\right]=\mathcal{V}_{\eff}+o_p(1)=O_p(1)$. Hence it suffices to show that the first term on the right-hand side is $o_p(1)$. By the same argument as in the proof of Theorem~\ref{th:Delta_1}, we have
\begin{align*}
& \mathbb{P}_{\mathcal{I}_{k}}\left[\left\{\psi(\bbO;\widehat P_{(-k)})-\widehat{\Delta}^{\eff}_{\mathrm{DS}}-\psi(\bbO;P)+\Delta\right\}^2\right] \\
\lesssim&\ \mathbb{P}_{\mathcal{I}_{k}}\left[\left\{\psi(\bbO;\widehat P_{(-k)})-\psi(\bbO;P)\right\}^2\right]+(\widehat{\Delta}^{\eff}_{\mathrm{DS}}-\Delta)^2=o_p(1).
\end{align*}
Therefore, $\widehat{\mathcal{V}}_{\eff,k}-\Tilde{\mathcal{V}}_{\eff,k}=o_p(1)$ for each $k$, and hence $\widehat{\mathcal{V}}_{\eff}-\Tilde{\mathcal{V}}_{\eff}=o_p(1)$. Combining with $\Tilde{\mathcal{V}}_{\eff}-\mathcal{V}_{\eff}=o_p(1)$ gives $\widehat{\mathcal{V}}_{\eff}-\mathcal{V}_{\eff}=o_p(1)$.

This concludes our proof.

\subsection{Proof of Theorem \ref{S-th:identification_functional_general}}\label{S-proof: identification_functional_general}
First, note that, for each $z$ and $\boldsymbol{v}$,
\begin{align*}
G_d(z, \boldsymbol{v})
= &\ \E \left[ \mathbb{I}(D(z)=d)\Upsilon(Y(z), \boldsymbol{V}) \mid \boldsymbol{V}=\boldsymbol{v} \right] \\
= &\ \E \left[ \mathbb{I}(D(z)=d)\Upsilon(Y(d), \boldsymbol{V}) \mid \boldsymbol{V}=\boldsymbol{v} \right] \\
= &\ \E \left[
\E \left\{\mathbb{I}(D(z)=d)\Upsilon(Y(d), \boldsymbol{V}) \mid \bbX, U \right\}
\mid \boldsymbol{V}=\boldsymbol{v}
\right] \\
= &\ \E \left[
\E\{\mathbb{I}(D(z)=d) \mid \bbX, U\}
\E\{\Upsilon(Y(d), \boldsymbol{V}) \mid \bbX, U\}
\mid \boldsymbol{V}=\boldsymbol{v}
\right] \quad\quad \left(Y(d) \independent D \mid \bbX, U\right)\\
= &\ \E \left[
\Pr(D=d \mid Z=z, \bbX, U)
\E\{\Upsilon(Y(d), \boldsymbol{V}) \mid \bbX, U\}
\mid \boldsymbol{V}=\boldsymbol{v}
\right].
\end{align*}
Similarly,
\begin{align*}
\mu_d(z,\boldsymbol{v})
= &\ \E \left[\mathbb{I}(D(z)=d) \mid \boldsymbol{V}=\boldsymbol{v}\right] \\
= &\ \E \left[
\E\{\mathbb{I}(D(z)=d) \mid \bbX, U\}
\mid \boldsymbol{V}=\boldsymbol{v}
\right] \\
= &\ \E \left[
\Pr(D=d \mid Z=z, \bbX, U)
\mid \boldsymbol{V}=\boldsymbol{v}
\right].
\end{align*}
Therefore, for each $z$,
\begin{align*}
G_d(z, \boldsymbol{V})
= &\ \E \left[ \Pr(D=d \mid Z=z, \bbX, U)\E\{\Upsilon(Y(d), \boldsymbol{V}) \mid \bbX, U\} \mid \boldsymbol{V}\right] \\
= &\ \E \left[\{\Pr(D=d \mid Z=z, \bbX, U)-\Pr(D=d \mid \bbX, U)\}
\E\{\Upsilon(Y(d), \boldsymbol{V}) \mid \bbX, U\} \mid \boldsymbol{V}\right] \\
& + \E \left[ \Pr(D=d \mid \bbX, U) \E\{\Upsilon(Y(d), \boldsymbol{V}) \mid \bbX, U\} \mid \boldsymbol{V} \right] \\
= &\ \E \left[ \Pr(D=d \mid Z=z, \bbX, U)-\Pr(D=d \mid \bbX, U) \mid \boldsymbol{V} \right]
\E \left[\E\{\Upsilon(Y(d), \boldsymbol{V}) \mid \bbX, U\}\mid \boldsymbol{V} \right] \\
& + \E \left[ \Pr(D=d \mid \bbX, U) \E\{\Upsilon(Y(d), \boldsymbol{V}) \mid \bbX, U\}
\mid \boldsymbol{V} \right] \quad\quad (\text{by Assumption~\ref{S-as:5e}})\\
= &\ \mu_d(z,\boldsymbol{V})
\E\{\Upsilon(Y(d), \boldsymbol{V}) \mid \boldsymbol{V}\} -
\E\{\Pr(D=d \mid \bbX, U) \mid \boldsymbol{V}\}\E\{\Upsilon(Y(d), \boldsymbol{V}) \mid \boldsymbol{V}\} \\
& + \E \left[\Pr(D=d \mid \bbX, U)\E\{\Upsilon(Y(d), \boldsymbol{V}) \mid \bbX, U\} \mid \boldsymbol{V}\right].
\end{align*}
Multiplying both sides by
$\gamma(Z,\boldsymbol{V})-\E\{\gamma(Z,\boldsymbol{V}) \mid \boldsymbol{V}\}$
and taking the conditional expectation given $\boldsymbol{V}$ gives
\begin{align*}
&\ \E \left[
\{\gamma(Z,\boldsymbol{V})-\E[\gamma(Z,\boldsymbol{V}) \mid \boldsymbol{V}]\}
G_d(Z,\boldsymbol{V})
\mid \boldsymbol{V}
\right] \\
= &\ \E \left[
\{\gamma(Z,\boldsymbol{V})-\E[\gamma(Z,\boldsymbol{V}) \mid \boldsymbol{V}]\}
\mu_d(Z,\boldsymbol{V})
\mid \boldsymbol{V}
\right]
\E\{\Upsilon(Y(d), \boldsymbol{V}) \mid \boldsymbol{V}\},
\end{align*}
because
\[
\E \left[
\gamma(Z,\boldsymbol{V})-\E\{\gamma(Z,\boldsymbol{V}) \mid \boldsymbol{V}\}
\mid \boldsymbol{V}
\right]=0.
\]
Thus,
\begin{align*}
\E\{\Upsilon(Y(d),\boldsymbol{V}) \mid \boldsymbol{V}\}
= \dfrac{
\E \left[
\{\gamma(Z,\boldsymbol{V})-\E[\gamma(Z,\boldsymbol{V}) \mid \boldsymbol{V}]\}
G_d(Z,\boldsymbol{V})
\mid \boldsymbol{V}
\right]
}{
\E \left[
\{\gamma(Z,\boldsymbol{V})-\E[\gamma(Z,\boldsymbol{V}) \mid \boldsymbol{V}]\}
\mu_d(Z,\boldsymbol{V})
\mid \boldsymbol{V}
\right]
},
\end{align*}
given that the denominator is nonzero almost surely.

This concludes our proof.

\newpage

\section{Additional Simulation Details and Results}
\phantomsection

\begin{table}[ht]
\centering
\caption{Simulation results of the semiparametric locally efficient debiased estimators under the treatment mechanism without interaction}
\label{S-tab:datagen-binary-if}
\begin{tabular}{@{}c l r r r r r r r@{}}
n & Estimator & Bias & Median bias & MCSD & Avg. SE & SD acc.$^*$ & RMSE & Coverage \\[2pt]
\multirow{2}{*}{\centering $500$} & $\widehat{\Delta}_{\mathrm{DS}}$ & 0.045 & 0.002 & 0.050 & 0.224 & 0.099 & 2.252 & 0.949 \\
 & $\widehat{\Delta}_{\mathrm{RM}}$ & 0.002 & 0.005 & 0.002 & 0.102 & 1.035 & 0.099 & 0.963 \\[4pt]
\multirow{2}{*}{\centering $1000$} & $\widehat{\Delta}_{\mathrm{DS}}$ & 0.001 & 0.002 & 0.002 & 0.066 & 0.854 & 0.077 & 0.947 \\
 & $\widehat{\Delta}_{\mathrm{RM}}$ & 0.001 & 0.002 & 0.001 & 0.067 & 1.042 & 0.064 & 0.955 \\[4pt]
\multirow{2}{*}{\centering $2000$} & $\widehat{\Delta}_{\mathrm{DS}}$ & -0.001 & 0.000 & 0.001 & 0.044 & 0.965 & 0.045 & 0.940 \\
 & $\widehat{\Delta}_{\mathrm{RM}}$ & 0.001 & 0.001 & 0.001 & 0.046 & 0.999 & 0.047 & 0.954 \\[4pt]
\multirow{2}{*}{\centering $5000$} & $\widehat{\Delta}_{\mathrm{DS}}$ & -0.001 & -0.001 & 0.001 & 0.027 & 0.981 & 0.028 & 0.946 \\
 & $\widehat{\Delta}_{\mathrm{RM}}$ & 0.001 & 0.000 & 0.001 & 0.028 & 1.021 & 0.028 & 0.958 \\[2pt]
\multicolumn{9}{@{}p{0.92\textwidth}@{}}{\footnotesize MCSD, Monte Carlo standard deviation; Avg. SE, average estimated standard error; SD acc., SD accuracy, the ratio of the average estimated standard error to the Monte Carlo standard deviation; RMSE, root mean squared error.}
\end{tabular}
\end{table}

\begin{table}[ht]
\centering
\caption{Simulation results of the semiparametric locally efficient debiased estimators under the treatment mechanism with interaction}
\label{S-tab:datagen2-binary-if}
\begin{tabular}{@{}c l r r r r r r r@{}}
n & Estimator & Bias & Median bias & MCSD & Avg. SE & SD acc. & RMSE & Coverage \\[2pt]
\multirow{2}{*}{\centering $500$} & $\widehat{\Delta}_{\mathrm{DS}}$ & -0.160 & 0.007 & 0.129 & 0.373 & 0.065 & 5.762 & 0.945 \\
 & $\widehat{\Delta}_{\mathrm{RM}}$ & 0.008 & 0.011 & 0.003 & 0.110 & 0.750 & 0.147 & 0.948 \\[4pt]
\multirow{2}{*}{\centering $1000$} & $\widehat{\Delta}_{\mathrm{DS}}$ & 0.019 & 0.005 & 0.008 & 0.087 & 0.244 & 0.356 & 0.935 \\
 & $\widehat{\Delta}_{\mathrm{RM}}$ & 0.006 & 0.005 & 0.002 & 0.070 & 0.962 & 0.073 & 0.944 \\[4pt]
\multirow{2}{*}{\centering $2000$} & $\widehat{\Delta}_{\mathrm{DS}}$ & 0.002 & 0.003 & 0.002 & 0.049 & 0.527 & 0.094 & 0.933 \\
 & $\widehat{\Delta}_{\mathrm{RM}}$ & 0.002 & 0.002 & 0.001 & 0.049 & 0.970 & 0.050 & 0.946 \\[4pt]
\multirow{2}{*}{\centering $5000$} & $\widehat{\Delta}_{\mathrm{DS}}$ & 0.002 & 0.001 & 0.001 & 0.029 & 0.817 & 0.036 & 0.935 \\
 & $\widehat{\Delta}_{\mathrm{RM}}$ & 0.002 & 0.002 & 0.001 & 0.030 & 0.954 & 0.032 & 0.942 \\[2pt]
\multicolumn{9}{@{}p{0.92\textwidth}@{}}{\footnotesize MCSD, Monte Carlo standard deviation; Avg. SE, average estimated standard error; SD acc., SD accuracy, the ratio of the average estimated standard error to the Monte Carlo standard deviation; RMSE, root mean squared error.}
\end{tabular}
\end{table}

\begin{table}[ht]
\centering
\caption{Simulation results of the semiparametric locally efficient debiased estimators under the treatment mechanism without interaction, with the instrument dichotomized at the $20\%$ quantile}
\label{S-tab:datagen-quantile-binary-q20-if1}
\begin{tabular}{@{}c l r r r r r r r@{}}
n & Estimator & Bias & Median bias & MCSD & Avg. SE & SD acc. & RMSE & Coverage \\[2pt]
\multirow{2}{*}{\centering $500$} & $\widehat{\Delta}_{\mathrm{DS}}$ & -95.724 & -0.028 & 96.969 & 111 & 0.026 & 4337 & 0.979 \\
 & $\widehat{\Delta}_{\mathrm{RM}}$ & -0.045 & -0.032 & 0.009 & 0.321 & 0.797 & 0.405 & 0.970 \\[2pt]
\multirow{2}{*}{\centering $1000$} & $\widehat{\Delta}_{\mathrm{DS}}$ & 3.375 & -0.025 & 3.114 & 10.696 & 0.077 & 139 & 0.980 \\
 & $\widehat{\Delta}_{\mathrm{RM}}$ & -0.016 & -0.012 & 0.004 & 0.167 & 0.987 & 0.169 & 0.976 \\[2pt]
\multirow{2}{*}{\centering $2000$} & $\widehat{\Delta}_{\mathrm{DS}}$ & 0.793 & -0.019 & 2.444 & 8.484 & 0.078 & 109 & 0.984 \\
 & $\widehat{\Delta}_{\mathrm{RM}}$ & -0.012 & -0.012 & 0.002 & 0.103 & 0.965 & 0.107 & 0.971 \\[2pt]
\multirow{2}{*}{\centering $5000$} & $\widehat{\Delta}_{\mathrm{DS}}$ & -0.721 & -0.043 & 0.441 & 3.143 & 0.159 & 19.744 & 0.980 \\
 & $\widehat{\Delta}_{\mathrm{RM}}$ & -0.009 & -0.009 & 0.001 & 0.057 & 1.148 & 0.050 & 0.977 \\[2pt]
\multicolumn{9}{@{}p{0.92\textwidth}@{}}{\footnotesize MCSD, Monte Carlo standard deviation; Avg. SE, average estimated standard error; SD acc., SD accuracy, the ratio of the average estimated standard error to the Monte Carlo standard deviation; RMSE, root mean squared error.}
\end{tabular}
\end{table}

\begin{table}[ht]
\centering
\caption{Simulation results of the semiparametric locally efficient debiased estimators under the treatment mechanism without interaction, with the instrument dichotomized at the $50\%$ quantile}
\label{S-tab:datagen-quantile-binary-q50-if1}
\begin{tabular}{@{}c l r r r r r r r@{}}
n & Estimator & Bias & Median bias & MCSD & Avg. SE & SD acc.$^*$ & RMSE & Coverage \\[2pt]
\multirow{2}{*}{\centering $500$} & $\widehat{\Delta}_{\mathrm{DS}}$ & 0.292 & 0.001 & 0.231 & 0.456 & 0.044 & 10.313 & 0.950 \\
 & $\widehat{\Delta}_{\mathrm{RM}}$ & -0.003 & -0.004 & 0.003 & 0.138 & 1.072 & 0.129 & 0.973 \\[2pt]
\multirow{2}{*}{\centering $1000$} & $\widehat{\Delta}_{\mathrm{DS}}$ & 0.004 & 0.006 & 0.003 & 0.083 & 0.710 & 0.117 & 0.958 \\
 & $\widehat{\Delta}_{\mathrm{RM}}$ & 0.002 & 0.004 & 0.002 & 0.086 & 1.096 & 0.078 & 0.969 \\[2pt]
\multirow{2}{*}{\centering $2000$} & $\widehat{\Delta}_{\mathrm{DS}}$ & -0.001 & -0.002 & 0.001 & 0.054 & 0.969 & 0.056 & 0.946 \\
 & $\widehat{\Delta}_{\mathrm{RM}}$ & 0.001 & -0.000 & 0.001 & 0.058 & 1.014 & 0.057 & 0.954 \\[2pt]
\multirow{2}{*}{\centering $5000$} & $\widehat{\Delta}_{\mathrm{DS}}$ & -0.001 & -0.001 & 0.001 & 0.034 & 1.019 & 0.033 & 0.953 \\
 & $\widehat{\Delta}_{\mathrm{RM}}$ & 0.001 & 0.001 & 0.001 & 0.035 & 1.070 & 0.033 & 0.968 \\[2pt]
\multicolumn{9}{@{}p{0.92\textwidth}@{}}{\footnotesize MCSD, Monte Carlo standard deviation; Avg. SE, average estimated standard error; SD acc., SD accuracy, the ratio of the average estimated standard error to the Monte Carlo standard deviation; RMSE, root mean squared error.}
\end{tabular}
\end{table}

\begin{table}[ht]
\centering
\caption{Simulation results of the semiparametric locally efficient debiased estimators under the treatment mechanism without interaction, with the instrument dichotomized at the $80\%$ quantile}
\label{S-tab:datagen-quantile-binary-q80-if1}
\begin{tabular}{@{}c l r r r r r r r@{}}
n & Estimator & Bias & Median bias & MCSD & Avg. SE & SD acc.$^*$ & RMSE & Coverage \\[2pt]
\multirow{2}{*}{\centering $500$} & $\widehat{\Delta}_{\mathrm{DS}}$ & 0.501 & 0.029 & 0.805 & 6.835 & 0.190 & 35.995 & 0.981 \\
 & $\widehat{\Delta}_{\mathrm{RM}}$ & -0.020 & -0.018 & 0.012 & 0.403 & 0.741 & 0.544 & 0.972 \\[2pt]
\multirow{2}{*}{\centering $1000$} & $\widehat{\Delta}_{\mathrm{DS}}$ & 1.789 & 0.014 & 0.985 & 6.211 & 0.141 & 44.069 & 0.983 \\
 & $\widehat{\Delta}_{\mathrm{RM}}$ & -0.009 & -0.007 & 0.004 & 0.189 & 1.000 & 0.189 & 0.976 \\[2pt]
\multirow{2}{*}{\centering $2000$} & $\widehat{\Delta}_{\mathrm{DS}}$ & -0.390 & 0.030 & 0.437 & 3.339 & 0.171 & 19.541 & 0.981 \\
 & $\widehat{\Delta}_{\mathrm{RM}}$ & 0.006 & 0.009 & 0.003 & 0.118 & 1.043 & 0.113 & 0.972 \\[2pt]
\multirow{2}{*}{\centering $5000$} & $\widehat{\Delta}_{\mathrm{DS}}$ & 0.239 & 0.034 & 0.595 & 2.849 & 0.107 & 26.590 & 0.976 \\
 & $\widehat{\Delta}_{\mathrm{RM}}$ & 0.002 & 0.003 & 0.001 & 0.066 & 1.106 & 0.060 & 0.976 \\[2pt]
\multicolumn{9}{@{}p{0.92\textwidth}@{}}{\footnotesize MCSD, Monte Carlo standard deviation; Avg. SE, average estimated standard error; SD acc., SD accuracy, the ratio of the average estimated standard error to the Monte Carlo standard deviation; RMSE, root mean squared error.}
\end{tabular}
\end{table}

\clearpage
\renewcommand{\refname}{Supplementary Material References}


\begin{thebibliography}{}

\bibitem[Angrist et~al., 1996]{angrist1996identification}
Angrist, J.~D., Imbens, G.~W., and Rubin, D.~B. (1996).
\newblock Identification of causal effects using instrumental variables.
\newblock {\em Journal of the American Statistical Association}, 91(434):444--455.

\bibitem[Berrie et~al., 2024]{berrie2024does}
Berrie, L., Feng, Z., Rice, D., Clemens, T., Williamson, L., and Dibben, C. (2024).
\newblock Does cycle commuting reduce the risk of mental ill-health? {A}n instrumental variable analysis using distance to nearest cycle path.
\newblock {\em International Journal of Epidemiology}, 53(1):dyad153.

\bibitem[Burgess et~al., 2011]{burgess2011avoiding}
Burgess, S., Thompson, S.~G., and {Crp Chd Genetics Collaboration} (2011).
\newblock Avoiding bias from weak instruments in {M}endelian randomization studies.
\newblock {\em International Journal of Epidemiology}, 40(3):755--764.

\bibitem[Chen et~al., 2025]{chen2025identification}
Chen, S., Zhang, P., and Cui, Y. (2025).
\newblock Identification and debiased learning of causal effects with general instrumental variables.
\newblock {\em arXiv preprint arXiv:2510.20404}.

\bibitem[Chen and Santos, 2018]{chen2018overidentification}
Chen, X. and Santos, A. (2018).
\newblock Overidentification in regular models.
\newblock {\em Econometrica}, 86(5):1771--1817.

\bibitem[Chen et~al., 2026]{chen2026equivalence}
Chen, Y., Kennedy, E.~H., and Balakrishnan, S. (2026).
\newblock On the equivalence between {N}eyman orthogonality and pathwise differentiability.
\newblock {\em arXiv preprint arXiv:2603.15817}.

\bibitem[Chernozhukov et~al., 2018]{chernozhukov_doubledebiased_2018}
Chernozhukov, V., Chetverikov, D., Demirer, M., Duflo, E., Hansen, C., Newey, W., and Robins, J. (2018).
\newblock Double/debiased machine learning for treatment and structural parameters.
\newblock {\em The Econometrics Journal}, 21(1):C1--C68.

\bibitem[Chernozhukov et~al., 2024]{chernozhukov2024conditional}
Chernozhukov, V., Newey, W.~K., and Syrgkanis, V. (2024).
\newblock Conditional influence functions.
\newblock {\em arXiv preprint arXiv:2412.18080}.

\bibitem[Conway, 1990]{conway2019course}
Conway, J.~B. (1990).
\newblock {\em A Course in Functional Analysis}, volume~96.
\newblock New York: Springer.

\bibitem[Cui and Tchetgen~Tchetgen, 2021]{cui2021semiparametric}
Cui, Y. and Tchetgen~Tchetgen, E. (2021).
\newblock A semiparametric instrumental variable approach to optimal treatment regimes under endogeneity.
\newblock {\em Journal of the American Statistical Association}, 116(533):162--173.

\bibitem[Darolles et~al., 2011]{darolles2011nonparametric}
Darolles, S., Fan, Y., Florens, J.-P., and Renault, E. (2011).
\newblock Nonparametric instrumental regression.
\newblock {\em Econometrica}, 79(5):1541--1565.

\bibitem[Davies et~al., 2015]{davies2015many}
Davies, N.~M., von Hinke Kessler~Scholder, S., Farbmacher, H., Burgess, S., Windmeijer, F., and Smith, G.~D. (2015).
\newblock The many weak instruments problem and {M}endelian randomization.
\newblock {\em Statistics in Medicine}, 34(3):454--468.

\bibitem[Didelez and Sheehan, 2007]{didelez2007mendelian}
Didelez, V. and Sheehan, N. (2007).
\newblock Mendelian randomization as an instrumental variable approach to causal inference.
\newblock {\em Statistical Methods in Medical Research}, 16(4):309--330.

\bibitem[Hartford et~al., 2017]{hartford2017deep}
Hartford, J., Lewis, G., Leyton-Brown, K., and Taddy, M. (2017).
\newblock Deep {IV}: A flexible approach for counterfactual prediction.
\newblock In {\em Proceedings of the 34th International Conference on Machine Learning}, pages 1414--1423. PMLR.

\bibitem[Hartwig et~al., 2023]{hartwig2023average}
Hartwig, F.~P., Wang, L., Smith, G.~D., and Davies, N.~M. (2023).
\newblock Average causal effect estimation via instrumental variables: The no simultaneous heterogeneity assumption.
\newblock {\em Epidemiology}, 34(3):325--332.

\bibitem[Hines et~al., 2026]{hines2026parameterizing}
Hines, O.~J., Diaz-Ordaz, K., and Vansteelandt, S. (2026).
\newblock Parameterizing the effect of a continuous treatment using average derivative effects.
\newblock {\em Biometrika}, 113(2):asag012.

\bibitem[K{\'e}dagni and Mourifi{\'e}, 2020]{kedagni2020generalized}
K{\'e}dagni, D. and Mourifi{\'e}, I. (2020).
\newblock Generalized instrumental inequalities: Testing the instrumental variable independence assumption.
\newblock {\em Biometrika}, 107(3):661--675.

\bibitem[Kennedy et~al., 2019]{kennedy2019robust}
Kennedy, E.~H., Lorch, S., and Small, D.~S. (2019).
\newblock Robust causal inference with continuous instruments using the local instrumental variable curve.
\newblock {\em Journal of the Royal Statistical Society Series B: Statistical Methodology}, 81(1):121--143.

\bibitem[Lee et~al., 2025]{lee2025inference}
Lee, Y., Yu, M., Liu, J., Park, C., Zhang, Y., Robins, J.~M., and Tchetgen~Tchetgen, E.~J. (2025).
\newblock Inference on nonlinear counterfactual functionals under a multiplicative {IV} model.
\newblock {\em arXiv preprint arXiv:2507.15612}.

\bibitem[Leigh and Schembri, 2004]{leigh2004instrumental}
Leigh, J.~P. and Schembri, M. (2004).
\newblock Instrumental variables technique: Cigarette price provided better estimate of effects of smoking on {SF}-12.
\newblock {\em Journal of Clinical Epidemiology}, 57(3):284--293.

\bibitem[Levis et~al., 2024]{levis2024nonparametric}
Levis, A.~W., Kennedy, E.~H., and Keele, L. (2024).
\newblock Nonparametric identification and efficient estimation of causal effects with instrumental variables.
\newblock {\em arXiv preprint arXiv:2402.09332}.

\bibitem[Liu et~al., 2025]{liu2025multiplicative}
Liu, J., Park, C., Lee, Y., Zhang, Y., Yu, M., Robins, J.~M., and Tchetgen, E. J.~T. (2025).
\newblock The multiplicative instrumental variable model.
\newblock {\em arXiv preprint arXiv:2507.09302}.

\bibitem[Newey and Powell, 2003]{newey2003instrumental}
Newey, W.~K. and Powell, J.~L. (2003).
\newblock Instrumental variable estimation of nonparametric models.
\newblock {\em Econometrica}, 71(5):1565--1578.

\bibitem[Okui et~al., 2012]{okui2012doubly}
Okui, R., Small, D.~S., Tan, Z., and Robins, J.~M. (2012).
\newblock Doubly robust instrumental variable regression.
\newblock {\em Statistica Sinica}, 22(1):173--205.

\bibitem[Park et~al., 2018]{park2018plausibility}
Park, Y., Peterson, L.~L., and Colditz, G.~A. (2018).
\newblock The plausibility of obesity paradox in cancer—point.
\newblock {\em Cancer Research}, 78(8):1898--1903.

\bibitem[Pearl, 2009]{pearl2009causality}
Pearl, J. (2009).
\newblock {\em Causality}.
\newblock Cambridge University Press.

\bibitem[Pfanzagl, 1983]{pfanzagl1983asymptotic}
Pfanzagl, J. (1983).
\newblock {\em Asymptotic Expansions for General Statistical Models}, volume~31 of {\em Lecture Notes in Statistics}.
\newblock Springer Science \& Business Media.

\bibitem[Richardson and Robins, 2013]{richardson2013single}
Richardson, T.~S. and Robins, J.~M. (2013).
\newblock Single world intervention graphs ({SWIG}s): A unification of the counterfactual and graphical approaches to causality.
\newblock {\em Center for the Statistics and the Social Sciences, University of Washington Series. Working Paper}, 128(30):2013.

\bibitem[Richardson et~al., 2017]{richardson2017modeling}
Richardson, T.~S., Robins, J.~M., and Wang, L. (2017).
\newblock On modeling and estimation for the relative risk and risk difference.
\newblock {\em Journal of the American Statistical Association}, 112(519):1121--1130.

\bibitem[Robins, 1994]{robins1994correcting}
Robins, J.~M. (1994).
\newblock Correcting for non-compliance in randomized trials using structural nested mean models.
\newblock {\em Communications in Statistics-Theory and Methods}, 23(8):2379--2412.

\bibitem[Sanderson et~al., 2022]{sanderson2022mendelian}
Sanderson, E., Glymour, M.~M., Holmes, M.~V., Kang, H., Morrison, J., Munaf{\`o}, M.~R., Palmer, T., Schooling, C.~M., Wallace, C., Zhao, Q., and Smith, G.~D. (2022).
\newblock Mendelian randomization.
\newblock {\em Nature Reviews Methods Primers}, 2(1):6.

\bibitem[Saunders et~al., 2022]{saunders2022genetic}
Saunders, G.~R., Wang, X., Chen, F., Jang, S.-K., Liu, M., Wang, C., Gao, S., Jiang, Y., Khunsriraksakul, C., Otto, J.~M., et~al. (2022).
\newblock Genetic diversity fuels gene discovery for tobacco and alcohol use.
\newblock {\em Nature}, 612(7941):720--724.

\bibitem[Scheidegger et~al., 2026]{scheidegger2026inference}
Scheidegger, C., Guo, Z., and B{\"u}hlmann, P. (2026).
\newblock Inference for heterogeneous treatment effects with efficient instruments and machine learning.
\newblock {\em Electronic Journal of Statistics}, 20(1):718--770.

\bibitem[Shepshelovich et~al., 2019]{shepshelovich2019body}
Shepshelovich, D., Xu, W., Lu, L., Fares, A., Yang, P., Christiani, D., Zhang, J., Shiraishi, K., Ryan, B.~M., Chen, C., et~al. (2019).
\newblock Body mass index ({BMI}), {BMI} change, and overall survival in patients with {SCLC} and {NSCLC}: A pooled analysis of the {International Lung Cancer Consortium}.
\newblock {\em Journal of Thoracic Oncology}, 14(9):1594--1607.

\bibitem[Swanson et~al., 2018]{swanson2018partial}
Swanson, S.~A., Hern{\'a}n, M.~A., Miller, M., Robins, J.~M., and Richardson, T.~S. (2018).
\newblock Partial identification of the average treatment effect using instrumental variables: Review of methods for binary instruments, treatments, and outcomes.
\newblock {\em Journal of the American Statistical Association}, 113(522):933--947.

\bibitem[Tchetgen~Tchetgen, 2025]{tchetgen2024nudge}
Tchetgen~Tchetgen, E.~J. (2025).
\newblock {The Nudge Average Treatment Effect}.
\newblock {\em arXiv preprint arXiv:2410.23590}.
\newblock Version 3, revised 12 August 2025.

\bibitem[Tchetgen~Tchetgen and Vansteelandt, 2013]{tchetgen2013alternative}
Tchetgen~Tchetgen, E.~J. and Vansteelandt, S. (2013).
\newblock Alternative identification and inference for the effect of treatment on the treated with an instrumental variable.
\newblock Working Paper 166, Harvard University Biostatistics Working Paper Series.

\bibitem[van~der Laan et~al., 2007]{van2007super}
van~der Laan, M.~J., Polley, E.~C., and Hubbard, A.~E. (2007).
\newblock Super learner.
\newblock {\em Statistical Applications in Genetics and Molecular Biology}, 6(1):Article 25.

\bibitem[{van der Vaart}, 2014]{van2014higher}
{van der Vaart}, A. (2014).
\newblock Higher order tangent spaces and influence functions.
\newblock {\em Statistical Science}, 29(4):679--686.

\bibitem[Wang and Tchetgen~Tchetgen, 2018]{wang2018bounded}
Wang, L. and Tchetgen~Tchetgen, E. (2018).
\newblock Bounded, efficient and multiply robust estimation of average treatment effects using instrumental variables.
\newblock {\em Journal of the Royal Statistical Society Series B: Statistical Methodology}, 80(3):531--550.

\bibitem[Wang et~al., 2023]{wang2023instrumental}
Wang, L., Tchetgen~Tchetgen, E., Martinussen, T., and Vansteelandt, S. (2023).
\newblock Instrumental variable estimation of the causal hazard ratio.
\newblock {\em Biometrics}, 79(2):539--550.

\bibitem[Wooldridge, 2010]{wooldridge2010econometric}
Wooldridge, J.~M. (2010).
\newblock {\em Econometric Analysis of Cross Section and Panel Data}.
\newblock The MIT Press.

\bibitem[Yengo et~al., 2018]{yengo2018meta}
Yengo, L., Sidorenko, J., Kemper, K.~E., Zheng, Z., Wood, A.~R., Weedon, M.~N., Frayling, T.~M., Hirschhorn, J., Yang, J., Visscher, P.~M., et~al. (2018).
\newblock Meta-analysis of genome-wide association studies for height and body mass index in 700000 individuals of {E}uropean ancestry.
\newblock {\em Human Molecular Genetics}, 27(20):3641--3649.

\bibitem[Young and Shah, 2024]{young2024rose}
Young, E.~H. and Shah, R.~D. (2024).
\newblock {ROSE} random forests for robust semiparametric efficient estimation.
\newblock {\em arXiv preprint arXiv:2410.03471}.

\bibitem[Zeng et~al., 2025]{zeng2025nonparametric}
Zeng, Z., Levis, A.~W., Lee, J., Kennedy, E.~H., and Keele, L. (2025).
\newblock Nonparametric estimation of local treatment effects with continuous instruments.
\newblock {\em arXiv preprint arXiv:2504.03063}.

\bibitem[Zhang et~al., 2017]{zhang2017obesity}
Zhang, X., Liu, Y., Shao, H., and Zheng, X. (2017).
\newblock Obesity paradox in lung cancer prognosis: Evolving biological insights and clinical implications.
\newblock {\em Journal of Thoracic Oncology}, 12(10):1478--1488.

\end{thebibliography}

\begin{thebibliography}{}

\bibitem[Ai and Chen, 2012]{S-ai2012semiparametric}
Ai, C. and Chen, X. (2012).
\newblock The semiparametric efficiency bound for models of sequential moment restrictions containing unknown functions.
\newblock {\em Journal of Econometrics}, 170(2):442--457.

\bibitem[Chernozhukov et~al., 2018]{S-chernozhukov_doubledebiased_2018}
Chernozhukov, V., Chetverikov, D., Demirer, M., Duflo, E., Hansen, C., Newey, W., and Robins, J. (2018).
\newblock Double/debiased machine learning for treatment and structural parameters.
\newblock {\em The Econometrics Journal}, 21(1):C1--C68.

\bibitem[Hastie and Tibshirani, 1990]{S-hastie1990generalized}
Hastie, T.~J. and Tibshirani, R.~J. (1990).
\newblock {\em Generalized Additive Models}, volume~43 of {\em Monographs on Statistics and Applied Probability}.
\newblock Chapman\&Hall/CRC.

\bibitem[Hines et~al., 2026]{S-hines2026parameterizing}
Hines, O.~J., Diaz-Ordaz, K., and Vansteelandt, S. (2026).
\newblock Parameterizing the effect of a continuous treatment using average derivative effects.
\newblock {\em Biometrika}, 113(2):asag012.

\bibitem[Pfanzagl, 1983]{S-pfanzagl1983asymptotic}
Pfanzagl, J. (1983).
\newblock {\em Asymptotic Expansions for General Statistical Models}, volume~31 of {\em Lecture Notes in Statistics}.
\newblock Springer Science \& Business Media.

\bibitem[Stein, 1972]{S-stein1972bound}
Stein, C. (1972).
\newblock A bound for the error in the normal approximation to the distribution of a sum of dependent random variables.
\newblock In {\em Proceedings of the Sixth Berkeley Symposium on Mathematical Statistics and Probability, Volume 2: Probability Theory}, volume~6, pages 583--603. University of California Press.

\bibitem[Tchetgen~Tchetgen, 2025]{S-tchetgen2024nudge}
Tchetgen~Tchetgen, E.~J. (2025).
\newblock {The Nudge Average Treatment Effect}.
\newblock {\em arXiv preprint arXiv:2410.23590}.
\newblock Version 3, revised 12 August 2025.

\bibitem[Tsiatis, 2006]{S-tsiatis2006semiparametric}
Tsiatis, A.~A. (2006).
\newblock {\em Semiparametric Theory and Missing Data}, volume~4.
\newblock New York: Springer.

\bibitem[{van der Laan} and Robins, 2003]{S-laan2003unified}
{van der Laan}, M.~J. and Robins, J.~M. (2003).
\newblock {\em Unified Methods for Censored Longitudinal Data and Causality}.
\newblock New York: Springer.

\bibitem[{van der Vaart}, 2014]{S-van2014higher}
{van der Vaart}, A. (2014).
\newblock Higher order tangent spaces and influence functions.
\newblock {\em Statistical Science}, 29(4):679--686.

\end{thebibliography}
\end{document}